\documentclass[superscriptaddress,twocolumn,prb,showkeys,showpacs]{revtex4-1}

\usepackage[utf8]{inputenc}
\usepackage{amsmath}   
\usepackage{graphicx}   
\usepackage{verbatim}  
\usepackage{color}     
\usepackage{subfigure}  
\usepackage{hyperref}   
\usepackage{natbib}
\usepackage{gensymb}
\usepackage{epstopdf}
\usepackage{multirow}
\usepackage{xfrac}

\usepackage{listings}
\usepackage{cancel}
\usepackage{soul}
\usepackage{color}
\usepackage{cleveref}

\usepackage{amsmath,amssymb,graphics,epsfig,epstopdf,color,verbatim,ulem,braket,tabularx}

\begin{document}

\title{Ultrafast Dynamics of the Surface Photovoltage in Potassium Doped Black Phosphorus}

\author{G. Kremer}
\altaffiliation{Corresponding author.\\ geoffroy.kremer@unifr.ch}
\affiliation{D{\'e}partement de Physique and Fribourg Center for Nanomaterials, Universit{\'e} de Fribourg, CH-1700 Fribourg, Switzerland}

\author{M.~Rumo}
\affiliation{D{\'e}partement de Physique and Fribourg Center for Nanomaterials, Universit{\'e} de Fribourg, CH-1700 Fribourg, Switzerland}

\author{C.~Yue}
\affiliation{D{\'e}partement de Physique and Fribourg Center for Nanomaterials, Universit{\'e} de Fribourg, CH-1700 Fribourg, Switzerland}

\author{A.~Pulkkinen}
\affiliation{D{\'e}partement de Physique and Fribourg Center for Nanomaterials, Universit{\'e} de Fribourg, CH-1700 Fribourg, Switzerland}

\author{C.W.~Nicholson}
\affiliation{D{\'e}partement de Physique and Fribourg Center for Nanomaterials, Universit{\'e} de Fribourg, CH-1700 Fribourg, Switzerland}

\author{T.~Jaouen}
\affiliation{D{\'e}partement de Physique and Fribourg Center for Nanomaterials, Universit{\'e} de Fribourg, CH-1700 Fribourg, Switzerland}
\affiliation{Univ Rennes, CNRS, (IPR Institut de Physique de Rennes) - UMR 6251, F-35000 Rennes, France}

\author{F.O.~von~Rohr}
\affiliation{Department of Chemistry, University of Zurich, CH-8000 Zurich, Switzerland}

\author{P.~Werner}
\affiliation{D{\'e}partement de Physique and Fribourg Center for Nanomaterials, Universit{\'e} de Fribourg, CH-1700 Fribourg, Switzerland}

\author{C.~Monney}
\affiliation{D{\'e}partement de Physique and Fribourg Center for Nanomaterials, Universit{\'e} de Fribourg, CH-1700 Fribourg, Switzerland}

\begin{abstract}

Black phosphorus is a quasi-two-dimensional layered semiconductor with a narrow direct band gap of 0.3 eV. A giant surface Stark effect can be produced by the potassium doping of black phosphorus, leading to a semiconductor to semimetal phase transition originating from the creation of a strong surface dipole and associated band bending. By using time- and angle-resolved photoemission spectroscopy, we report the partial photoinduced screening of this band bending by the creation of a compensating surface photovoltage. We further resolve the detailed dynamics of this effect at the pertinent timescales and the related evolution of the band structure near the Fermi level. We demonstrate that after a fast rise time, the surface photovoltage exhibits a plateau over a few tens of picoseconds before decaying on the nanosecond timescale. We support our experimental results with simulations based on drift-diffusion equations.

\end{abstract}
\date{\today}
\maketitle

%%%%%%%%%%%%%%%%%% INTRODUCTION %%%%%%%%%%%%%%%%%%%%%

\begin{center}
\textbf{I. INTRODUCTION}
\end{center}

Semiconducting materials are the building block of modern technologies such as electronics, optoelectronics or computing devices.  Their band gap is a fundamental property governing their optical and electrical abilities, leading in recent decades to an increasing amount of research aiming to enhance the control of both its size and its nature (direct or indirect). In this context, quasi two-dimensional (2D) materials have appeared as a new promising emergent family\cite{chaves2020bandgap}. Indeed, 2D semiconducting materials form a  rich class of compounds whose electronic band gap can vary from 0 to more than 7 eV \textit{i.e.} ranging from the terahertz  to the infrared in monolayer Xenes\cite{ak2007rise,vogt2012silicene,li2014buckled,ji2016two,zhu2017multivalency,kochat2018atomically,deng2018epitaxial,wu2019large,bihlmayer2020plumbene}, to the visible with transition metal dichalcogenides (TMDCs)\cite{manzeli20172d} and finally to the ultraviolet with hexagonal boron nitride\cite{elias2019direct} and 2D oxides \cite{zhang2020epitaxial,lichtenstein2012probing,kremer2019electronic,kremer2020dispersing}. In addition to covering a large part of the electromagnetic spectrum, the main advantage of quasi 2D materials is the possibility to tune their band gap with numerous approaches such as strain\cite{guo2014tuning,rodin2014strain}, electric fields\cite{ramasubramaniam2011tunable}, surface gating\cite{kang2017universal}, chemical substitution\cite{gong2014band,ma2014postgrowth}, dimensionality reduction\cite{villaos2019thickness,lin2020dimensionality}, or by the creation of complex heterostructures\cite{novoselov20162d}. Besides the band gap tunability, this opens a broad range of possibilities to create and engineer new quantum materials with unprecedented physical properties. 

Among all these materials, black phosphorus (BP) is one of the most prominent because of its high carrier mobilities\cite{morita1986semiconducting,li2014black} and its tunable band gap over the whole infrared range, giving it a singular place between graphene and the TMDCs\cite{ling2015renaissance}.  BP is a layered semiconductor composed of only phosphorus atomic planes which are weakly interacting via out-of-plane van der Waals forces. Its band gap is highly dependent on the thickness, ranging from 0.3 eV in the bulk (BP) to 1.5 eV in the monolayer case (phosphorene).  In addition to this intrinsic variety, external stimuli such as mechanical strain or surface chemical gating have been demonstrated as powerful tools to modify its electronic properties\cite{mu2019two}. 

Chemical gating has attracted a lot of attention since the BP surface is very sensitive to external electric fields\cite{dolui2015quantum,liu2017gate} and induced surface dipoles\cite{kim2015observation,kim2017microscopic}. Indeed, by doping the surface of BP with alkali atoms, it is possible to create a strong dipole generating an electric field responsible for the continuous reduction of the band gap at the surface. At a sufficient alkali coverage the band gap closes, as recently demonstrated by angle resolved photoemission spectroscopy (ARPES), creating a remarkable semiconductor to semimetal transition\cite{kim2015observation}. The electric field of alkali doped BP is related to the large electron transfer from the alkali atoms to the BP. It creates an excess of free carriers, leading to the formation of a space charge region (SCR) and band bending (BB), a generic effect in semiconductors physics\cite{zhang2012band}. There is also a charges redistribution between the valence band (VB) and the conduction band (CB) whose wave functions are respectively pushed to the bulk and confined at the surface of the material\cite{dolui2015quantum,kim2015observation,liu2017gate,kim2017microscopic}. In particular, the CB forms a two-dimensional electrons gas (2DEG)\cite{kim2015observation,kiraly2019anisotropic} and  experiences a larger BB, which can be decomposed in the same BB feeled by the VB ($\phi_{BB}$) and a second one we name Stark potential ($\phi_{S}$), localized at the extreme surface. At a sufficient coverage, the total CB bending is enough to close the surface band gap but the bulk one remains the same. In analogy with the atomic Stark effect created by an electric field, this is called the surface giant Stark effect.

In recent works\cite{chen2020spectroscopy,hedayat2020non}, it has been shown that $\phi_{BB}$ in alkali doped BP at this critical coverage equals a few hundreds of meV and can be compensated via the illumination of the surface. Under the preexisting electric field, photo-generated electrons-holes are redistributed in the SCR and create a compensating field called surface photovoltage (SPV). Hedayat \textit{et al.} reported that this is felt as an energy shift of the whole surface photoemission spectrum but also as a spectral shape modification of the VB \cite{hedayat2020non}. However, their work mainly focussed on the transient dynamics up to \hbox{5 ps} after the photoexcitation.

Here, we report on the temporal evolution of the SPV and the band structure near the Fermi level (E\textsubscript{F}) of potassium (K) doped BP, from a few ps before to a few hundreds ps after photoexcitation. To do so, we use time-resolved ARPES (Tr-ARPES) which is the ideal tool to study the ultrafast dynamics of the electronic structure of materials \cite{bovensiepen2012elementary,smallwood2016ultrafast}. \hbox{Tr-ARPES} is particularly well suited since it allows the generation and the study of the SPV \cite{widdra2003time,tokudomi2008ultrafast,tanaka2012utility,yang2014electron} with energy and time resolutions. Whereas it appears quasi instantaneously, the SPV is remarkably stable over 60 ps and decreases on the nanosecond (ns) timescale. We demonstrate that the characteristic timescales of the SPV  are correlated to the spectral shape modification of the VB, which is highly affected by the potential profile in the SCR and as a consequence by the transient screening of  $\phi_{BB}$. Furthermore, we solve the drift-diffusion equations (DDEs) for K doped BP and  compare the simulated temporal evolution of the BB to the measured SPV dynamics. Our simulations show a good qualitative agreement and unambigously confirm the ultrafast transient evolution of $\phi_{BB}$ after the photoexcitation of the pump. Finally, the calculations demonstrate the importance of the electron-hole recombination processes for the long-time evolution.

\bigskip

%%%%%%%%%%%%%%%%%% METHODS %%%%%%%%%%%%%%%%%%%%%

\begin{center}
\textbf{II. METHODS}
\end{center}

Large single-crystals of BP were synthesized by a modified chemical vapor transport technique similar to the method first reported in Ref. \onlinecite{lange2007au3snp7}. Dry red phosphorus (150 mg, purity 99.999 $\%$), dry SnI$_{4}$ (\hbox{5 mg,} purity \hbox{99.999 $\%$}), tin shots (40 mg, purity \hbox{99.98 $\%$}-purity), and gold powder (20 mg, purity \hbox{99.999 $\%$}) were sealed in an evacuated quartz glass tube. The mixture was placed in a box furnace at 680$^{\circ}$C and held at this temperature for 24 h. It was then cooled with 0.1 K/h to 500$^{\circ}$C and quenched to room temperature (RT) by removal from the furnace. The crystals were manually removed from the metals. Residual SnI$_{4}$ was removed from the crystals by sealing them in an evacuated quartz tube and heating them to 400$^{\circ}$C for 5 h in a tube furnace. BP samples were cleaved at RT in a pressure of \hbox{1 $\times$ 10$^{-8}$ mbar}. ARPES measurements were performed using a ScientaOmicron DA30 photoelectron analyser with energy and angular resolutions better than 10 meV and 0.1°, respectively. The base pressure in the photoemission chamber was better than 3 $\times$ 10$^{-11}$ mbar. For the static measurements, we used monochromatized HeI$_{\alpha}$ radiation (\hbox{$h \nu$ = 21.2 eV}, SPECS UVLS with TMM 304 monochromator). For the Tr-ARPES measurements, half of the power of a femtosecond laser (Pharos, Light Conversion, operating at 1030 nm) is converted into 780 nm light with an optical parametric amplifier, which is then frequency-quadrupled to 6.3 eV in $\beta$-BaB$_{2}$O$_{4}$ crystals to generate probe pulses\cite{faure2012full}. The pump-probe measurements were performed at \hbox{200 kHz} and the pump fluence was \hbox{160 $\mu$J/cm$^{2}$.} The intrinsic resolution of the probe laser is 47 meV, as determined by the fit of the Fermi edge of a polycrystalline metal.  The remaining half of the fundamental laser power is directed into a collinear optical parametric amplifier (Orpheus, Light Conversion) to generate pump pulses at 780 nm (1.55 eV) with a duration of about 50 fs. The temporal resolution was estimated to be better than 100 fs by measuring the width of the cross-correlation between the pump and the probe pulses. We used  $s$ and $p$ polarization for the probe and an incident angle of $55^{\circ}$ with respect to the normal of the surface to measure along the armchair (AC) and zigzag (ZZ) directions, respectively. The Tr-ARPES data have been acquired with a negative bias of -5 V. All the photoemission measurements were realized at 30 K if not further specified. The K evaporation was performed at \hbox{30 K} from a SAES getter source in a pressure better than \hbox{2 $\times$ 10$^{-10}$ mbar}. 

Density functional theory (DFT) calculations were performed using the projector augmented wave (PAW) method implemented in the Vienna ab-initio software package (VASP~\cite{kresse1993,kresse1994,kresse1996a,kresse1996b,blochl1994,kresse1999}), with kinetic energy cutoff 400 eV and k-point spacing $<$ 0.2 \AA$^{-1}$. Exchange and correlation effects were included within the strongly constrained and appropriately normed (SCAN~\cite{sun2015}) meta-generalized gradient approximation. The structure of BP was relaxed until the residual forces on atoms were smaller than \hbox{1 meV/\AA}. The lattice parameters obtained after the relaxation of the structure with DFT calculations in the SCAN approximation are \hbox{a = 4.53 \AA} , \hbox{b = 3.29 \AA} \thinspace \thinspace and \hbox{c = 10.91 \AA} , \thinspace \thinspace in good agreement with x-ray diffraction measurements \cite{brown1965refinement}.

\bigskip

%%%%%%%%%%%%%%%%%%%%% DICUSSSIONS %%%%%%%%%%%%%%%%%%%

\bigskip

\begin{center}
\textbf{III. RESULTS}
\end{center}

\begin{center}
\textbf{A. Static Characterization of Black Phosphorus}
\end{center}

\begin{figure}[t]
\includegraphics[scale=0.52]{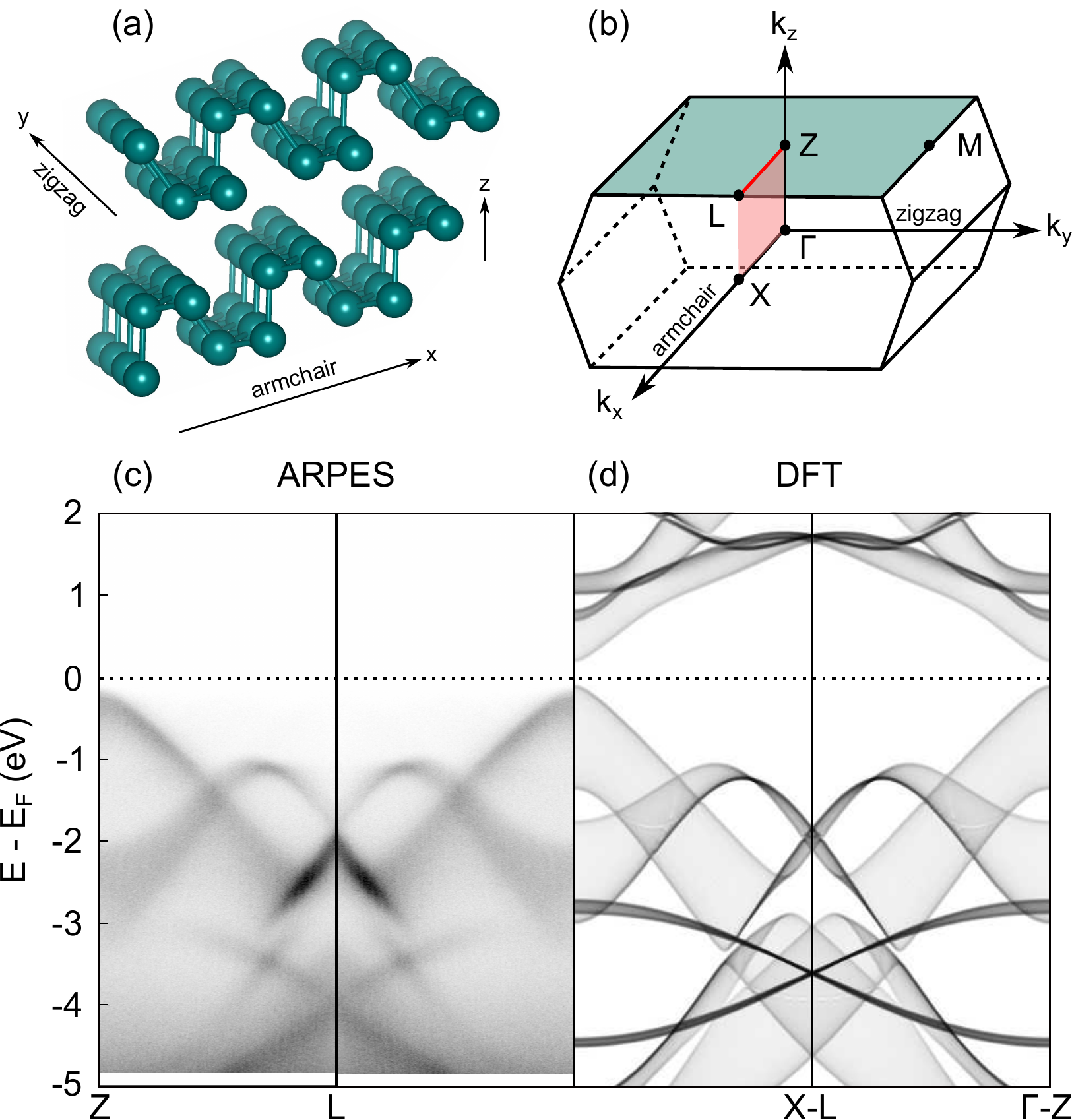}
\caption{(a) Crystal structure of BP plotted with the VESTA visualization program\cite{momma2011vesta} showing the two inequivalent in-plane directions, namely the ZZ and AC directions. (b) Corresponding three dimensional BZ and its 2D projection on the (001) plane (green). (c) ARPES spectrum along the $Z-L-Z$ high-symmetry line (thick red line in (b)) measured at \hbox{T = 300 K} and \hbox{h$\nu$ = 21.2 eV}. (d) Bulk projected electronic band structure (red plane in (b)) calculated from DFT in the SCAN approximation.}
\label{fig1}
\end{figure}

The crystal structure of BP is shown in Fig. \ref{fig1}(a). It crystallises in an orthorhombic crystal structure (space group \textit{Cmca}) with superimposed buckled layers of phosphorus atoms along the $z$ axis. The layered structure of BP and the weak interlayer van der Waals interactions favor mechanical cleaving of the material with the scotch-tape technique. The strongly anisotropic nature of BP in the basal plane is visible in its two high-symmetry directions along $x$ and $y$, namely the AC and ZZ directions.  The corresponding three dimensional Brillouin zone (BZ) and its 2D projection on the (001) plane (green), \textit{i.e.} the cleave plane, are shown in \hbox{Fig. \ref{fig1}(b)}.\\

\begin{figure*}[t]
\includegraphics[scale=0.32]{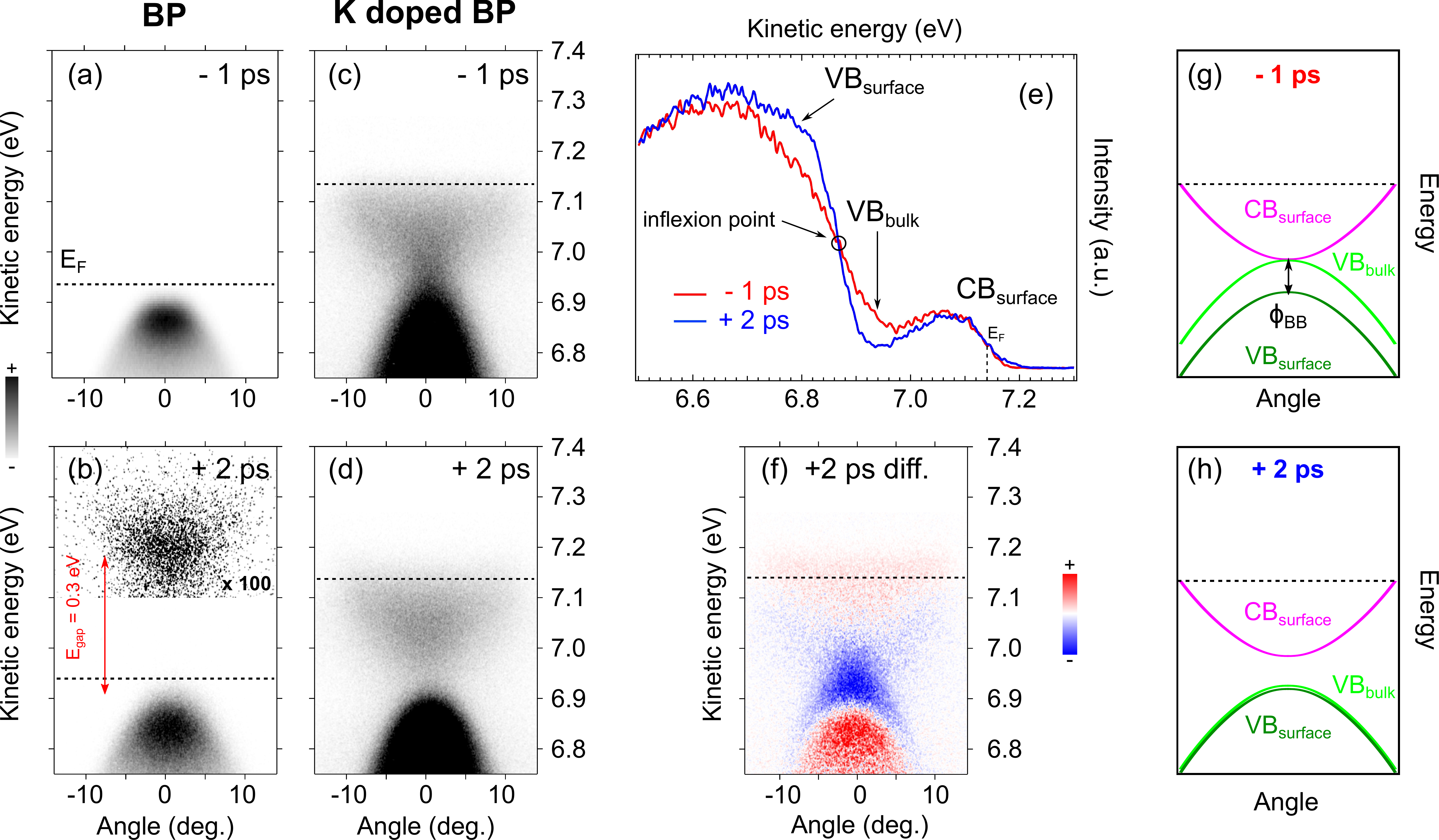}
\caption{ARPES spectra recorded at pump-probe delays of (a,c) -1 ps and (b,d) +2 ps for bare BP and K doped BP. In panel (b), the intensity at high KE has been multiplied by a factor 100 with respect to the low KE part in order to visualize the CB and the VB on the same linear color scale. The data have been acquired along the ZZ direction \textit{i.e.} the $M-Z-M$ high symmetry direction of the BZ. (e) EDCs at -1 ps (red) and +2 ps (blue) integrated $\pm 2^{\circ}$ around the normal emission for K doped BP. We highlight the position of the leading edge inflexion point of the VB, which is the marker we use in the following to obtain the evolution of the energy shift induced by the SPV. (f) Difference between the ARPES spectra of panels (c) and (d). Red and blue colors respectively correspond to an increase and a depletion of photoemission intensity. (g) and (h) are sketches of K doped BP photoemission spectra from panels (c,d) at negative and positive pump-probe delays.}
\label{fig2}
\end{figure*}

The static electronic band structure of pristine BP along the AC direction measured by ARPES is shown in \hbox{Fig. \ref{fig1}(c)} (note that the signal has been symmetrised with respect to the $L$ point). At this particular photon energy, we almost probe the in-plane $Z-L-Z$ high-symmetry line of the BZ\cite{taka1986} (thick red line in Fig. \ref{fig1}(b)). The ARPES intensity exhibits all the spectroscopic signatures of a clean BP surface\cite{golias2016disentangling}. In these energy and momentum ranges, the band structure is composed of six dispersive bands with the VB top presenting a maximum at the $Z$ point and located about \hbox{0.1 eV} below E\textsubscript{F}. In  \hbox{Fig. \ref{fig1}(d)}, we compare our experimental data with bulk projected DFT calculations \textit{i.e.} the superposition of the calculated band structure of BP along the $k_{z}$ direction, from the $\Gamma$ to the $Z$ point of the 3D BZ (red plane in Fig. \ref{fig1}(b)). The overall agreement between ARPES and DFT is excellent. Furthermore, the intensity variations  and the energy broadening of the bands are also well reproduced.

\begin{center}
\textbf{B. Dynamics of the Surface Photovoltage in K Doped Black Phosphorus}
\end{center}

Since the static ARPES only gives access to the occupied part of the band structure, we now turn our attention to Tr-ARPES measurements in order to obtain the band gap of BP. Figure \ref{fig2}(a,b) show the corresponding data taken at a pump-probe delay of -1 ps and +2 ps along the ZZ direction, respectively. We are using this orientation in the following because of detrimental matrix element effects along the AC direction. At negative delays, we only see the occupied part of the band structure as in the static case (shifted by the SPV as discussed below). At positive delay, the CB is now transiently populated and can be probed. Thus, we can directly estimate the band gap amplitude of the material to be 0.3 eV, in good agreement with previous measurements \cite{kiraly2017probing,qiu2017resolving} and our DFT calculations from Fig. \ref{fig1}(d) which give a band gap value of 307 meV.   

\begin{figure}
\begin{center}
\includegraphics[scale=0.105]{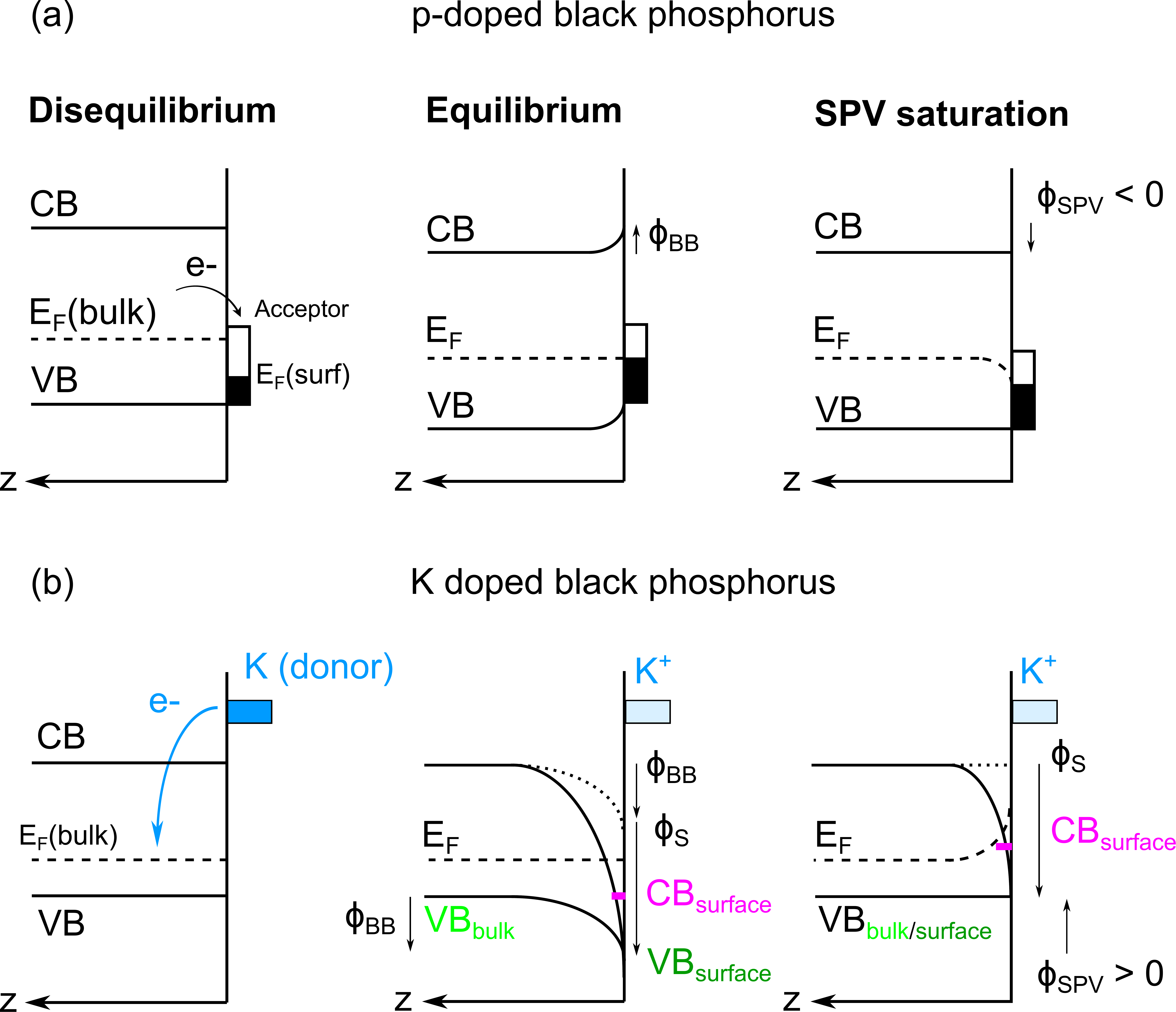} 
\end{center}
\caption{\label{fig7} Schematic diagram of electron energy levels near the surface of (a) p-doped BP sample and (b) K doped BP with and without illumination. For the p-doped (K doped) BP surface, an upward (downward) long-range BB and consequently a negative (positive) SPV are expected. In the K doped case, the CB is confined at the surface and experiences an additional Stark potential which is not screened by the pump excitation in our fluence regime. This Stark potential is responsible of the progressive collapse of the surface band gap as a function of K coverage. Note that the scales in panels (a) and (b) are not respected.}
\end{figure}

As visible in Fig. \ref{fig2}(b), E\textsubscript{F} is closer to the VB than to the CB. Since the effective masses of these bands are almost the same in pure BP\cite{morita1986semiconducting}, it means that our sample is p-doped. Recent scanning tunneling microscopy (STM) experiments have demonstrated that p-doped BP samples have an important density of surface acceptor states\cite{qiu2017resolving}. This is confirmed by our measurements showing that photoexcitation of the pump induces a global shift of -13 meV of the photoemission spectrum (see \hbox{Fig. \ref{fig6})}, \textit{i.e.} a shift to low kinetic energy (KE), originating from a negative SPV as expected by the existence of an upward BB at the surface of our sample\cite{zhang2012band}.  This negative SPV compensates the initial upward BB at the surface of our p-doped BP sample. We illustrate the situation in \hbox{Fig. \ref{fig7}(a)}, where we plot the energy band diagrams of the p-doped BP surface. We consider the nonequilibrium, the equilibrium and finally the photoexcited configurations.

At a negative pump-probe delay, the photoelectrons travelling from the surface to the photoemission analyser are decelerated/accelerated (depending on the initial BB curvature) by the electric field generated by the SPV due to the pump pulse arriving afterwards\cite{tanaka2012utility,yang2014electron}. The experimental consequence is a rigid KE shift of the entire photoemission spectrum. In other words, at negative delays we probe the spatial distribution of the electric field generated outside the sample by the SPV, but not the photoexcited band structure of the material. At positive delays, we are sensitive to the modification of the band structure induced by the SPV. \\

Subsequently, we have deposited K atoms on the  BP surface until reaching the semiconductor to semimetal transition, as depicted in Fig. \ref{fig2}(c). This panel shows the Tr-ARPES intensity taken at -1 ps pump-probe delay. Compared to the pristine BP surface, the VB has shifted down by 200 meV relative to E\textsubscript{F} whereas the CB is now visible in the occupied states and touches the top of the VB. By doping the surface with electrons from donor atoms, we switch from an upward to a strongly downward BB configuration. This is the reason why we now observe a positive SPV at the origin of a global shift to high KE of the whole photoemission spectrum (see \hbox{Fig. \ref{fig8}} for measurements with and without pump). Both VB and CB feel this common BB, namely  $\phi_{BB}$, but the CB also experiences a Stark potential at the extreme surface and forms a 2DEG. As a consequence, its total bending is larger than the VB one and, at a suffisient K coverage, lead to a collapse of the surface band gap. At the low fluence used in our experiment, we do not screen this strong Stark potential which is confined in the first layers of BP and, consequently, the 2DEG of the CB persists. The coexistence of a long range BB and a strong surface potential appears to be a common effect in layered semiconducting materials \cite{papalazarou2018unraveling}. Similarly to the VB, the 2DEG is only energy shifted by the SPV and the associated screening of  $\phi_{BB}$ (see Fig. \ref{fig7}(b)).

Let us now discuss the measurement taken at +2 ps and presented in Fig. \ref{fig2}(d). The most striking effect relates to the top of the VB which is shifted to low KE: the gap between the CB and the VB is apparently reopened. This is even more evident by looking at the energy-distribution curves (EDCs) taken at -1 ps and \hbox{+2 ps} (see Fig. \ref{fig2}(e)) or the difference in ARPES intensity between these two pump-probe delays (see \hbox{Fig. \ref{fig2}(f))}. It is worth to mention that this effect is independent of the crystal orientation, as shown in Fig. \ref{fig9}. Nevertheless, it is more convenient to work along the ZZ line because the CB has no spectral weight at the normal emission along the AC direction due to matrix element effects \cite{chen2018band}.

The EDCs in Fig. \ref{fig2}(e) clearly show that the spectral shape of the VB is unambiguously affected by the photoexcitation at positive pump-probe delay. Indeed, the leading edge of the VB is strongly sharpened, leading to a depletion of intensity at 6.95 eV and an increase of intensity at 6.8 eV KE, visible as a blue/white/red contrast in the difference ARPES intensity plot in \hbox{Fig. \ref{fig2}(f)}. In contrast, the CB spectral shape is marginally affected. By consequence, we confirm that the effect of the SPV on the VB is not limited to a global energy shift of the band structure; it also modifies its spectral shape\cite{hedayat2020non}. \\

\begin{figure*}[t]
\includegraphics[scale=0.113]{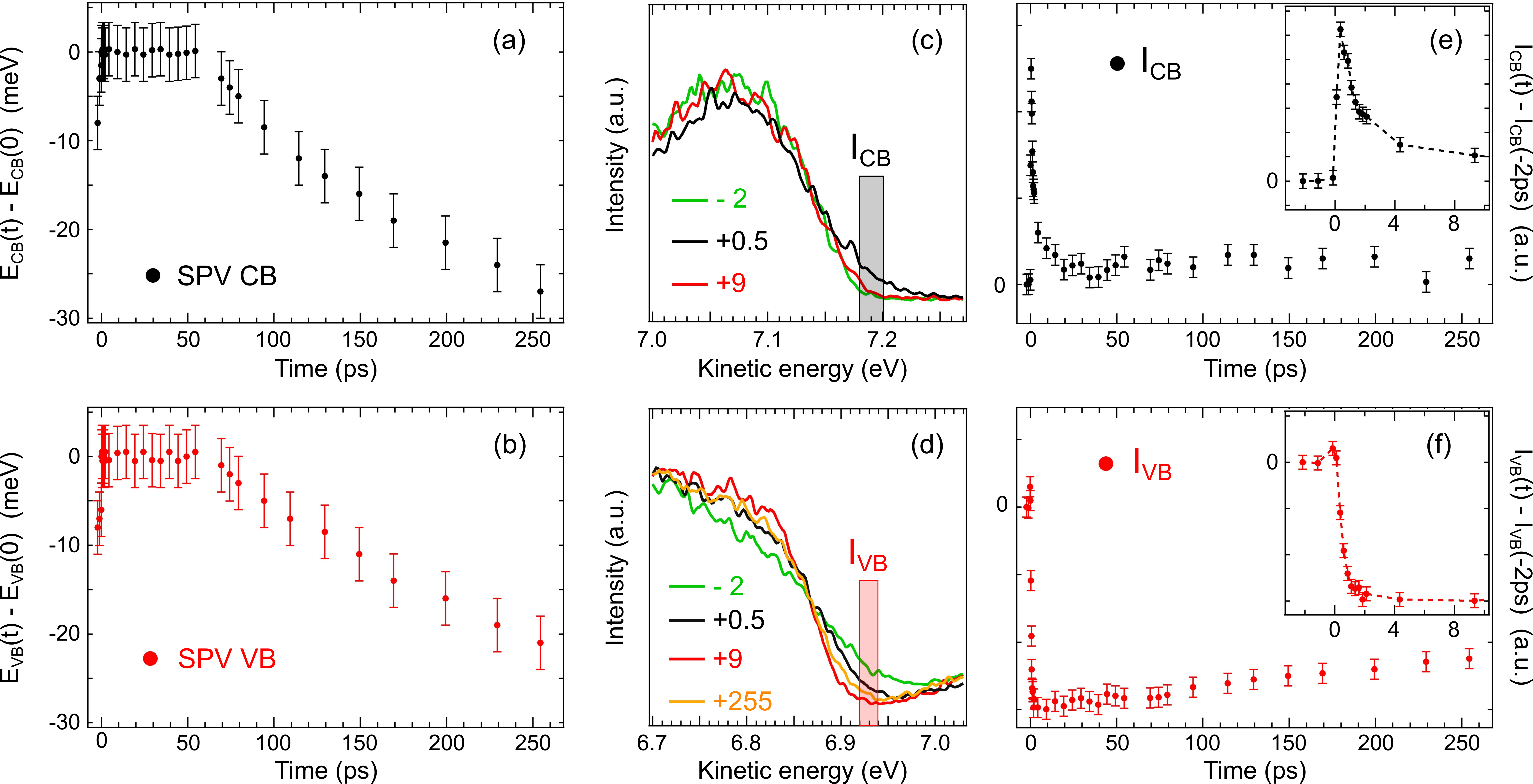}
\caption{Temporal evolution of the surface photovoltage (SPV) at the surface of K doped BP for (a) the CB and (b) the VB. (c) and (d) represent EDCs taken at normal emission, respectively for the CB and the VB, corrected by the SPV determined from panels (a,b) for a few pump-probe delays. (e) Temporal evolution of the integrated intensity I\textsubscript{CB} under the EDCs in panel (c) (grey area). The inset corresponds to a zoom over the first ps. (f) Same for  I\textsubscript{VB}  (red area in panel (d)).}
\label{fig3}
\end{figure*}

At negative delay, we probe the equilibrium state of the system for which the common BB,  $\phi_{BB}$,  is not yet screened. Since the VB is delocalised in the bulk of the material and the 6.3 eV laser is bulk sensitive (several nm) due to the extraction of very low kinetic energy electrons \cite{seah1979quantitative}, we integrate both the contributions of the VB from the surface and the bulk which are energy separated by the amplitude of  $\phi_{BB}$, \textit{i.e.} almost 200 meV (see Fig. \ref{fig7}(b)). The corresponding sketch of the band structure is represented in Fig. \ref{fig2}(g). It explains why the VB width is larger at negative delay, as observed in \hbox{Fig. \ref{fig2}(e)}. On the contrary, the CB forms a 2DEG confined at the surface and is visible as a single band we call CB\textsubscript{surface}, which aligns with the bulk contribution of the VB.

At positive delay, we now probe the photoexcited system at a pump fluence of 160 $\mu$J/cm$^{2}$. At this fluence, the SPV is saturated \cite{chen2020spectroscopy} and we recover a flat band regime for the VB: surface and bulk contributions of the VB are now aligned (see Fig. \ref{fig7}(b)). In contrast, the 2DEG persists because the Stark potential is not screened at this fluence. The consequence is that the bulk contribution of the VB shifts relatively to the surface one and to E\textsubscript{F}, leading to a decrease of the VB width and to a sharpening of its  leading edge, as visible in Fig. \ref{fig2}(e). This is also the reason why CB\textsubscript{surface} and the bulk contribution of the VB are now energy separated and why we observe an apparent transient band gap reopening. This second situation is schematized in Fig. \ref{fig2}(h). It is apparent because, in reality, the surface band gap was not closed since CB\textsubscript{surface} and VB\textsubscript{surface} are energy separated by 200 meV, as revealed once $\phi_{BB}$ is compensated by the SPV (Fig. \ref{fig2}(d)).  With our K coverage,  the surface band gap has not yet collapsed and has only been reduced by 100 meV with respect to the bulk band gap of BP. 

Nevertheless, it is an additionnal evidence that the surface band gap can be reduced with respect to the bulk by alkali doping. By supressing the photoemission artefact related to the integration of the VB along the out-of-place direction, we access the true surface band gap amplitude. For this reason, the claim of a semiconductor to semimetal transition has to be used carefully since it is only true (i) in the first layers of BP where the Stark potential exists and (ii) it is dependent on the depth integration of the technique, which varies with the photon energy in ARPES. \\

Thus, in this scenario the spectral shape evolution of the VB is directly correlated to the photoinduced SPV at positive delay. To confirm this, it is important to show that the SPV dynamics is unambiguously related to the dynamics of the spectral shape of the VB. First, we characterized the energy shift induced by the SPV both at negative and positive delays over a few hundreds of ps. To do so, we estimated the energy shift of the photoemission spectra as extracted from the EDCs analysis as a function of time (see a few EDCs at different pump-probe delays in \hbox{Fig. \ref{fig10}(a))}. The energy shift has been evaluated as the evolution of the position of the leading edge inflexion point of both the VB and the CB. The corresponding results are presented in Fig. \ref{fig3}(a,b). It is important to mention that the energy position of the inflexion point is independent of the evolution of the spectral shape and consequently only reflects the evolution of the experimentally measured SPV (in other words the position of CB\textsubscript{surface} and VB\textsubscript{surface}). 

In both cases, the SPV is increasing at negative delays up to a maximum around time zero. Then, it exhibits a remarkable plateau over 60 ps before slowly decaying over a few hundreds of ps. We have fitted the recovery of the SPV with an exponential fit. The extracted time constants for the CB and the VB are $>$ 1 ns (see \hbox{Fig. \ref{fig11}(a,b))}. The temporal evolution of the photoemission cutoff position shows the same features (see \hbox{Fig. \ref{fig10}(b)}). Thus, the energy shift induced by the SPV is the same for all the photoemission spectrum and not restricted to a particular KE range.  

Now, we address the temporal evolution of the spectral shape of the bands. In order to remove the influence of the energy shift induced by the SPV, we have plotted the EDCs corrected by the above determined SPV for both the CB and the VB, as shown in Fig. \ref{fig3}(c,d) for selected pump-probe delays. To understand the temporal evolution of the spectral shape, we have plotted the variation of the photoemission intensity above the energy position of the inflexion points for both the CB and the VB (it would be the complementary of taking an integration region below these points). The integration regions are highlighted by the grey and red rectangles. These plots are represented in Fig. \ref{fig3}(e,f) and respectively called I\textsubscript{CB} and I\textsubscript{VB}. 

The temporal evolution of I\textsubscript{CB} shows a strong increase around time zero before decaying over a few ps to its equilibrium value (see inset in panel (e)). It can be reproduced by an exponential fit with a time constant of 2 ps (see Fig. \ref{fig11}(c)) which is expected for the transient dynamics of the electronic temperature \cite{sobota2012ultrafast}. Thus, the effect of the SPV on the CB is limited to an energy shift and its spectral shape is not modified as expected by the strong surface confinement of electrons from the CB in the above mentioned scenario. 

On the contrary, the evolution of I\textsubscript{VB} is more complex. Over the first ps after photoexcitation, I\textsubscript{VB }is drastically reduced: it corresponds to the intensity depletion we discussed in Fig. \ref{fig2}. This decrease can be reproduced by an exponential fit with a time constant of 520 fs (see \hbox{Fig. \ref{fig11}(e)}). During the next 60 ps, I\textsubscript{VB} shows a plateau before slowly recovering with a time constant of 1.1 ns (see Fig. \ref{fig11}(d)). The plateau and the recovery timescales of I\textsubscript{VB} are in very good agreement with the ones observed for the SPV.  Thus, it unambiguously demonstrates that the spectral shape evolution of the VB in K doped BP is directly related to the magnitude of the SPV in the material. We can further note that, as expected and contrary to the experimentally measured SPV, I\textsubscript{VB} is not affected at negative delays because we probe the band structure of the unexcited system, \textit{i.e.} before the pump pulse arrival.

\begin{center}
\textbf{C. Discussion and Simulations}
\end{center}

\begin{figure}[t]
\includegraphics[scale=0.12]{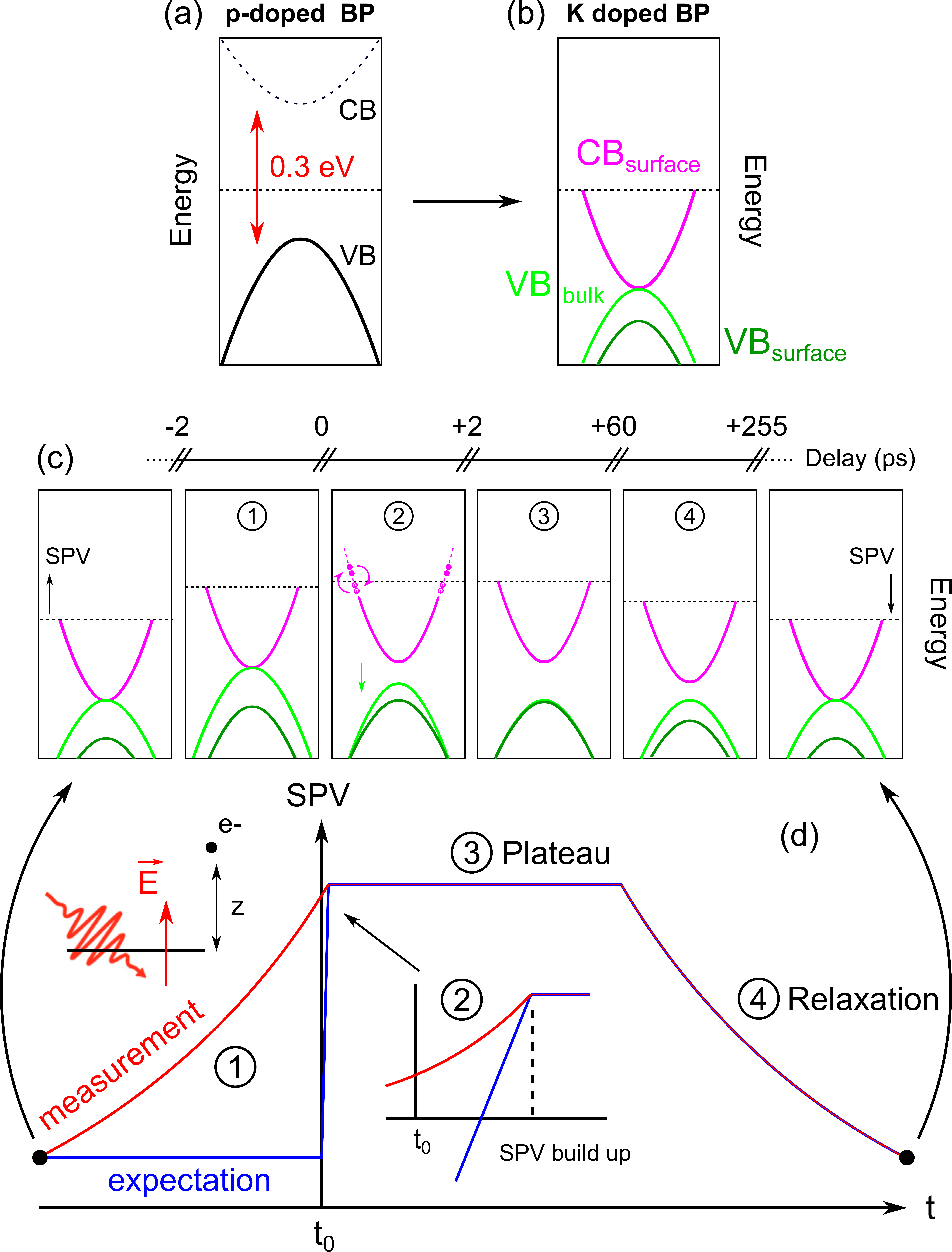}
\caption{(a,b) Cartoon representation of the band structure before and after K deposition on BP as measured by ARPES. The strong surface dipole originating from alkali deposition leads to a progressive surface semiconductor to semimetal transition named giant surface Stark effect. (c) Cartoon of the temporal evolution of K doped BP band structure at the pertinent timescales as measured by Tr-ARPES: \textcircled{1} Photoelectrons acceleration before time zero \textcircled{2} SPV build up, band structure modification and excited electrons  thermalization \textcircled{3} Plateau \textcircled{4} SPV relaxation and band structure recovery. (d) Schematic representation of the evolution of the SPV only, as experimentaly measured with Tr-ARPES  (red) and as expected from theory (blue).}
\label{fig4}
\end{figure}

We now summarize the above discussed evolution of the band structure, both in the static and dynamic cases. By doping the surface of a p-doped BP sample with K, we create a strong surface dipole which is the origin of the surface confinement and the shift of the CB down to the occupied states. At a sufficient coverage, the CB\textsubscript{surface} and the bulk contribution of the buried VB superimpose, visible as a band gap closing in static ARPES \hbox{(Fig. \ref{fig4}(a,b))}. By photoexciting the system with an intense infrared laser pulse, we induce and reveal the dynamics of the SPV, which is responsible for the shift of the whole surface photoemission spectrum and for the transient modification of the VB spectral shape.

The pertinent timescales are given in Fig. \ref{fig4}(c) and can be decomposed into four main steps. (i) At negative delays, since the probe induces photoemission before the pump  arrival, we are sensitive to the non-photoexcited band structure of the system. Nevertheless, the photoemission spectrum is shifted to higher KE by the electric field generated afterwards by the pump pulse. The smaller the pump-probe delay, the larger the field experienced by the outgoing photoelectrons. At a sufficient time before the pump excitation, the photoelectrons are not sensitive to this electric field anymore and the band structure looks the same as in the static case (see the extreme left panel). (ii) During the first picoseconds after photoexcitation, \textit{i.e.}  at positive delays, two processes are occuring. Firstly, a SPV quickly builds up and modifies the energy level diagram of the system. The VB is largely affected because it is delocalized in the bulk of BP while the surface confined CB\textsubscript{surface} is not altered. The consequence is the transient modification of the VB spectral shape and the apparent reopening of the band gap. Secondly, the pump pulse drives the system to a highly excited state. The electronic temperature increases, progressively equilibrates with the lattice and decreases: the consequence is a transient broadening of the Fermi-Dirac distribution in the first ps after the photoexcitation. (iii) Once the electronic temperature equilibrates and the band gap apparently reopens, the system exhibits a remarkable quasi steady state over \hbox{60 ps.} (iv) Then, the SPV progressively recovers with a time constant of the order of ns. The consequence is a progressive shift to lower KE this time and a recovery of the VB spectral shape. At a sufficient time after the pump excitation, the SPV has fully recovered and the band structure is similar to the one observed a long time before the photoexcitation or in the static case (see the extreme right panel).

\begin{figure*}[t]
\includegraphics[scale=0.31]{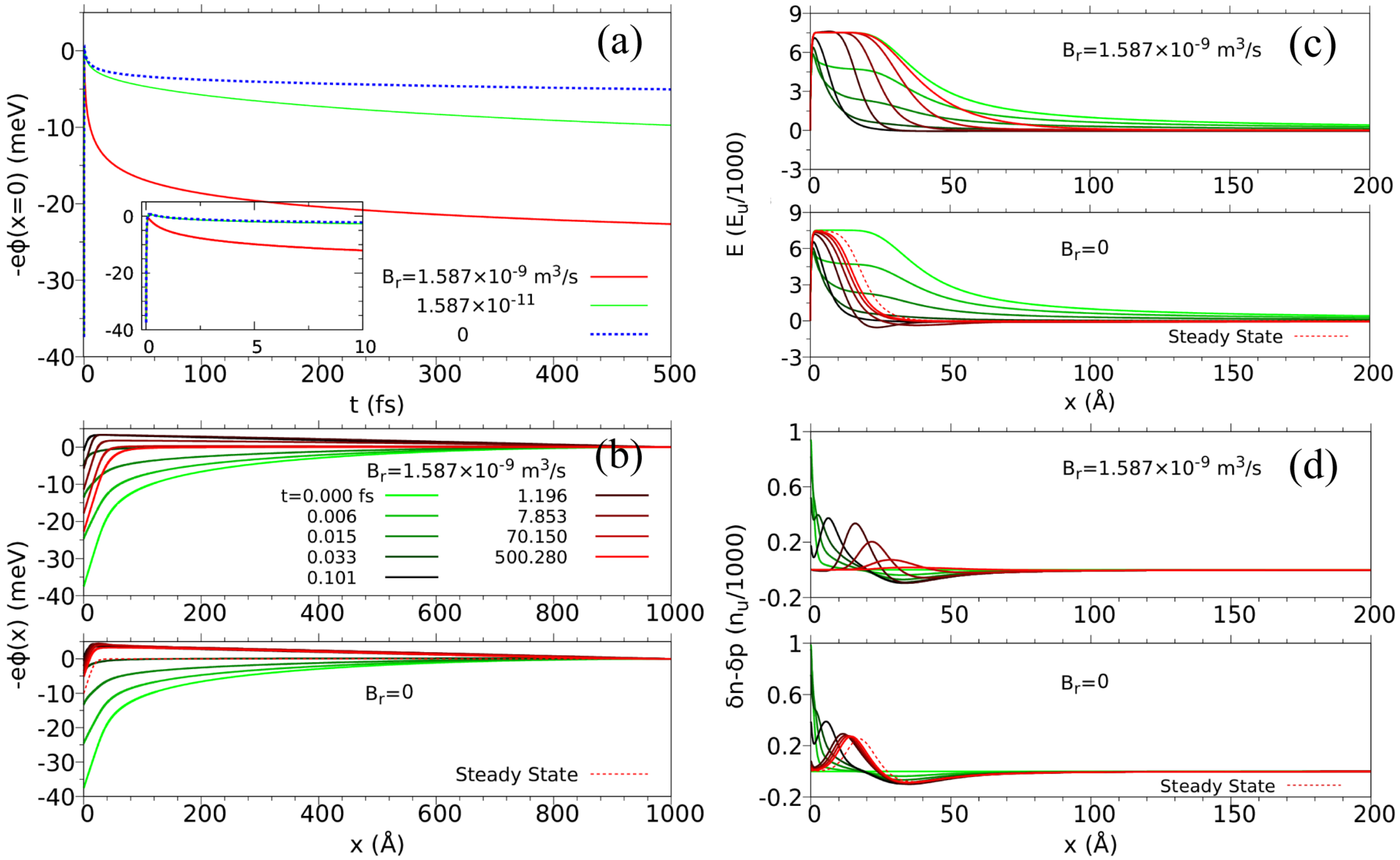}
\caption{Non-equilibrium dynamics of photo-excited carriers obtained by solving the DDEs.
(a) The SPV as a function of time after the pump for different recombination rates $B_r$. When $B_r=0$
(blue dashed-line), there is no recombination. The inset shows the fast initial dynamics in the range $0\le t<10$ fs.
(b) The band energy as a function of $x$ at different times. The upper panel is for $B_r\ne 0$ and the lower panel is for $B_r=0$ (similarly in panels (c) and (d)). The red-dashed line in the lower panel shows the solution of the steady-state after the pump when $B_r=0$.
(c) The internal electric field as a function of $x$ for different $t$.
(d) The net excess carrier density ($\delta n - \delta p$) as a function of $x$ for different $t$.
The parameters used in the simulation are shown in Tab.~\ref{tab:table_para}.
}
\label{fig:bending_largwin}
\end{figure*}

In Fig. \ref{fig4}(d), we give a schematic representation of the SPV time evolution as experimentally measured with Tr-ARPES (red curve), and as we would expect from theory (blue curve). Before time zero, the progressive increase of the experimentally extracted SPV is an artefact of the pump-probe approach, as discussed above.  As schematized in the step 1, we are theoretically expecting a zero SPV in this time regime (see blue curve). At time zero, the system is photoexcited and the SPV builds up (step 2). As discussed below, it occurs on the timescale of the pump pulse ($\approx$ 50 fs). In theory, the SPV starts from zero before building up to its maximum after this characteristic time (without step 1). This is in contrast with the experimentally measured SPV which is already at an almost maximum value around time zero due to the step 1, further complicating its experimental resolution.  Steps 3 and 4 are similar in both cases with a plateau and relaxation processes. \\

In order to simulate the non-equilibrium dynamics of photo-excited carriers in K doped BP at $T=30$ K we solve  the DDEs with realistic parameters (see appendix). Since we are looking at the effect of the photoexcitation of the pump on the common long-range BB, \textit{i.e.} $\phi_{BB}$, we assume that the depth profile potential of the CB and the VB are the same. The results are shown in Fig. \ref{fig:bending_largwin}. While electron-hole recombination can be neglected on the sub-ps timescales of the simulation, we also show results with an artificially short life time of the photocarriers. In these simulations, the band-to-band recombination \cite{Semi_cond_elect_book} between excess electron-hole pairs results from a term $B_r n\delta p$ ($n=n_0+\delta n$ with $n_0$ the static free electron
distribution) in the DDEs, where $B_r$ is the  recombination rate and $\delta n$ ($\delta p$) is the excess density of electrons (holes) generated by the laser pump.  The SPV amplitude $\phi_\text{SPV}=-e\phi(x=0)$ is plotted as a function of time in Fig. \ref{fig:bending_largwin}(a). An ultrafast increase of $\phi_\text{SPV}$ is observed for $t<0.1$ fs, as a result of two main factors. On the one hand, there is a large discrepancy between the high mobility of the holes, \hbox{$\mu_p \approx 6000$ $\mathrm{cm^2/Vs}$}, and the smaller  mobility of electrons \hbox{$\mu_n \approx 540$ $\mathrm{cm^2/Vs}$} at $T=30$ K\cite{morita1986semiconducting}.  On the other hand, a strong internal electric field ($E\sim 10^7$ $\mathrm{V/m}$) exists in the surface region due to the spatial separation between the K ions and the free electrons (see Fig. \ref{fig:static}).  The drift velocity of the holes and electrons can reach $v_p=\mu_p E \sim 56$ \AA$/ \mathrm{fs}$ and $v_n=\mu_nE\sim 5$ \AA$/\mathrm{fs}$, respectively \footnote{We caution here that the sub-fs timescale is obtained by assuming that the experimental mobilities are still valid for such a very high electric field. In practice, this might not be the case, resulting potentially a longer rise-time. As long as it remains shorter than the pump duration, our conclusions are not qualitatively affected.}. As a result, the holes flow rapidly into the bulk (the direction of the net field), while the electrons flow towards the surface. The time evolution of the excess carrier density, $\delta n -\delta p$, plotted in Fig. \ref{fig:bending_largwin}(d), illustrates how excess electrons rapidly accumulate near the surface while an excess of holes rapidly builds up slightly away from  the surface. The effect of this reshuffling of charge is a rapid weakening of the internal electric field, as shown in \hbox{Fig. \ref{fig:bending_largwin}(c)}, and  a rapid flattening of the energy band $-e\phi(x)$, \textit{i.e.} $\phi_{BB}$, as shown in Fig. \ref{fig:bending_largwin}(b).

The diffusion term becomes important when the density profile is sharp. It causes particles to flow from the high density region to the lower density region. As we can see from Fig. \ref{fig:bending_largwin}(d), the surface peak of $\delta n -\delta p$ decreases with time  and a broader peak appears which moves toward the bulk. At $t=0.1$ fs, the internal electric field $E$ stops weakening and $\phi_\text{SPV}$ reaches its maximum value as shown in Fig. \ref{fig:bending_largwin}(a). For $t>0.1$ fs, the surface peak in the net charge density begins to disappear slowly and shifts further into the bulk region. As a result, the internal electric field $E$ starts to increase again and some BB reappears, as shown in Fig. \ref{fig:bending_largwin}(b). We should note that these simulations are for an instantaneous photo-doping at $t=0_+$. In our experiment, the pump pulse has a duration of about 50 fs. Hence, in the experiment, the flattening of the band should occur on a timescale controlled by the pump pulse.  

Without recombination between electrons and holes ($B_r=0$), the system will relax to a new steady state controlled by the different electron and hole mobilities. As shown by the lower panel of Fig. \ref{fig:bending_largwin}(d), the time-evolution of $\delta n - \delta p$ approaches this steady state solution (red dashed line) rather slowly after $t=0.1$ fs. This solution corresponds to an almost flat band, which is distinctly different from the initial equilibrium solution (green line in Fig. \ref{fig:bending_largwin}(b)). Hence, as shown by the blue-dashed line in  Fig. \ref{fig:bending_largwin}(a), $\phi_\text{SPV}$ will eventually reach a plateau much higher than its initial value at $t=0$.

If we switch on recombination processes ($B_r > 0$) (with a rate much larger than in the realistic system), the $\delta n - \delta p$ peak starts to decay even on the timescale of our simulation,  see the upper panel in Fig. \ref{fig:bending_largwin}(d). Now, the system will return  to the initial equilibrium state and we observe a continuous recovery of the BB, as shown by the green and red lines in  Fig. \ref{fig:bending_largwin}(a) and also in the upper panel of Fig. \ref{fig:bending_largwin}(b). The higher the recombination rate, the faster the return to the initial state. \\

These simulations present a good qualitative agreement with our \hbox{Tr-ARPES} measurements. Very quickly after the system is excited by the pump pulse, the photogenerated charge carriers are redistributed in the SCR under the effect of the pre-existing electric field and diffusion. This charge redistribution leads to a transient compensation of the electric field at the surface, and as a consequence to a compensation of the initial BB. As experimentally extracted from the transient evolution of the VB (see Fig. \ref{fig3}(f)), the SPV builds up in the first ps. According to our quench calculations, it happens on the $0.1$ fs timescale (see Fig. \ref{fig:bending_largwin}(b)). There are two reasons for this discrepancy. Firstly, in the experiment the photoexcitation occurs with a finite duration of about 50 fs and the total time resolution of our experiment is 100 fs, well above this timescale.  Secondly, it has been demonstrated in the pristine BP case that a fast transient broadening of the VB happens in the first ps after the photoexcitation \cite{chen2018band}. This effect is still present in the K doped case\cite{hedayat2020non} and explain the very fast increase of intensity in Fig. \ref{fig3}(f), which is quickly vanishing and counterbalanced by the depletion caused by the SPV. Consequently, limited pulses duration and transient broadening explain why the time trace in Fig. \ref{fig3}(f) is broadened and why the measured time constant of the SPV build up has a finite duration $>$ 500 fs.

Then, the delayed recombination (the plateau on \hbox{Fig. \ref{fig4}(d)}) regime is experimentally seen, but not captured by our theoretical model. Such features have already been observed in other systems such as silicon oxide\cite{widdra2003time} or GaAs\cite{tokudomi2008ultrafast} but, to the best of our knowledge, no model has been able to explain it. We speculate that the origin of this delayed recombination could be due to an additional effect, not captured by our theoretical model, that prevents the recombination of electrons and holes immediately after the pump, like spatial separation or the trapping by defects.  Another possibility is that, in this temporal regime, recombination takes place but is exactly compensated by another effect which is unknown to us. We leave the explanation of the origin of this feature for future works.

Finally, the calculations unambiguously demonstrate that the relaxation regime can only be explained by taking into account a recombination term of the photoexcited charge carriers, otherwise the system would remain in a photo-doped nonequilibrium state with almost zero BB indefinitely, which is not observed experimentally.

Overall, our dynamic simulations based on the solution of the DDEs equations satisfactorily capture the physics of the problem (as expected in Fig. \ref{fig4}(d)), although simulations with realistically slow recombination times were not possible. Since simulations up to ns timescales are beyond our capabilities we had to choose extremely fast recombination times (a few fs against ns). Still these calculations demonstrate that the relaxation of the band structure on a timescale is controlled by the electron-hole recombination.

%%%%%%%%%%%%%%%%%%%%%%%%%%%%CONCLUSIONS%%%%%%%%%%%%%%%%%%%%%%%%%%%%%%

\bigskip

\begin{center}
\textbf{IV. CONCLUSIONS}
\end{center}

To summarize, the ultrafast dynamics of the SPV in photoexcited K doped BP has been measured with Tr-ARPES and compared to numerical simulations of the DDEs. We analysed the detailed temporal evolution of the SPV and the band structure near the Fermi level, from a few ps before to a few hundreds ps after the photoexcitation. Our results demonstrate that the timescales of the SPV and the spectral shape modication of the VB are correlated, and support the scenario where the transient screening of  $\phi_{BB}$ is responsible for the apparent reopening of the band gap. Our simulations confirm the ultrafast transient evolution of $\phi_{BB}$ (on the timescale of the pump pulse) and the importance of the electron-hole recombination processes for the long-time evolution. Without recombination, the system would remain trapped in a nonequilibrium steady state with almost no BB. 

Our experimental observations and our theoretical approach are very general and should have implications also for other Tr-ARPES measurements of 2D semiconducting materials with significant BB. In particular, care has to be taken in the  interpretation of transient spectral shape modications, since these may reflect the evolution of the SPV rather than qualitative changes in the band structure.

\bigskip

%%%%%%%%%%%%%%%%%% REMERCIEMENTS %%%%%%%%%%%%%%%%%%%%%

\begin{center}
\textbf{ACKNOWLEDGMENTS}
\end{center}

This project was supported by the Swiss National Science Foundation (SNSF) Grants \hbox{No. P00P2$\_$170597}, \hbox{No. 200021-165539} and \hbox{No. 200021-196966.} The calculations have been performed on the Beo05 cluster at the University of Fribourg. We are very grateful to P. Aebi for sharing with us his photoemission setup. Skillful technical assistance was provided by J.L. Andrey, M. Andrey, F. Bourqui, B. Hediger and O. Raetzo.

%%%%%%%%%%%%%%%%%% APPENDIX %%%%%%%%%%%%%%%%%%%%%

\bigskip

\begin{center}
\textbf{APPENDIX}
\end{center}

%%%%%%%%%%%%%%%%%%%%%%%%%%%%%%%%%%%%%%%%%%%%%%%%%%%%%%%%%%

\begin{center}
\textbf{1. SPV Characterization in Pristine Black Phosphorus}
\end{center}

Figure \ref{fig6} displays the Tr-ARPES measurements of BP at -1 ps pump-probe delay without (red) and with (blue) the pump. From this, it is possible to evaluate the magnitude and the sign of the SPV. We observe that the blue spectrum is shifted by 13 meV to low KE with respect to the red one.  Since the pristine BP surface is p doped due to the presence of surface acceptor states, an upward BB of the VB is expected as illustrated in the middle panel of Fig. \ref{fig7}(a). This is confirmed by the negative value of the SPV which is compensating the upward BB. 

\bigskip

\begin{figure}
\begin{center}
\includegraphics[scale=0.125]{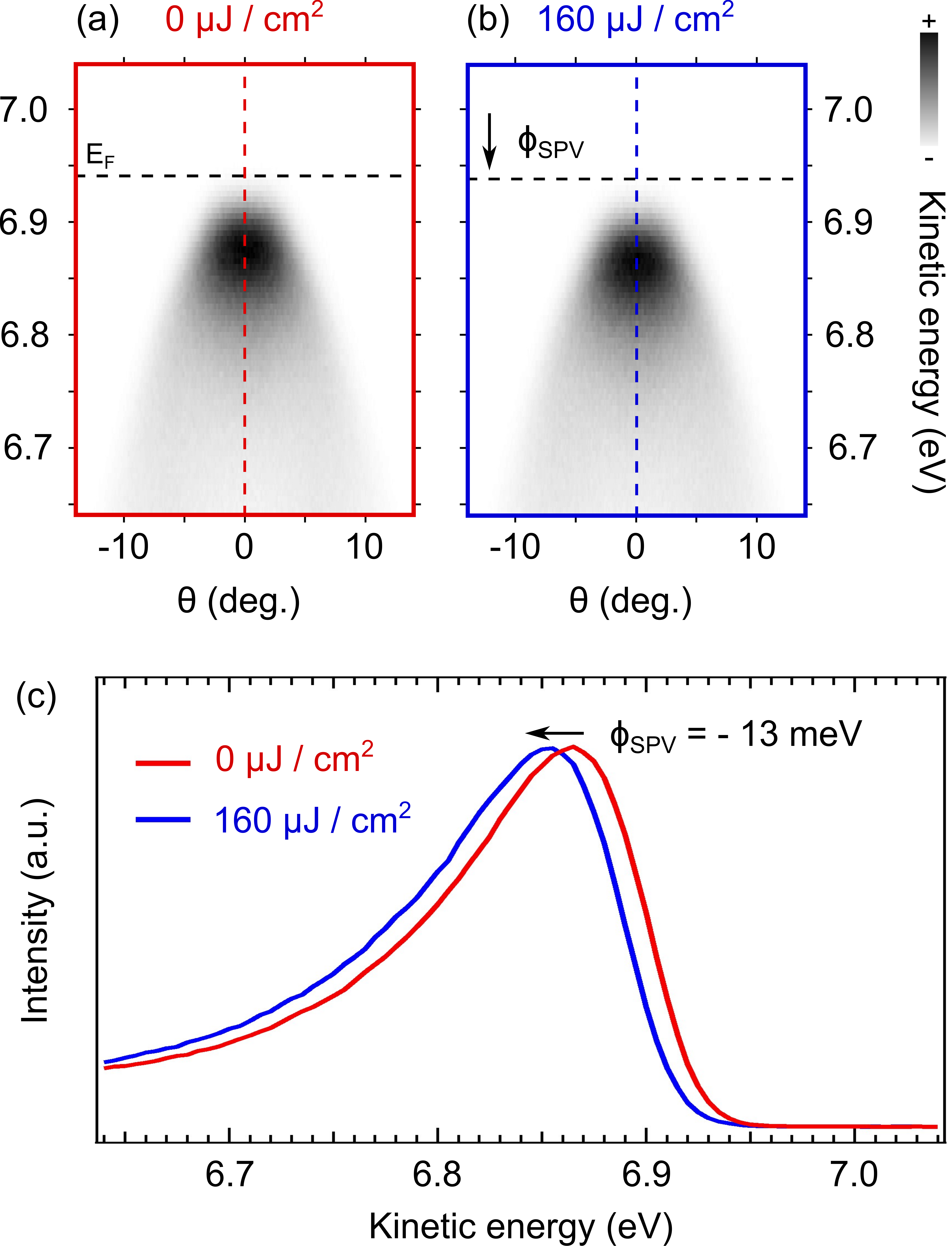} 
\end{center}
\caption{\label{fig6} Tr-ARPES measurements of BP along the ZZ direction at -1 ps pump-probe delay for a pump fluence of (a) 0 $\mu$J/cm$^{2}$ and (b) of 160 $\mu$J/cm$^{2}$. (c) Corresponding EDCs taken at normal emission. A negative SPV of -13 meV is observed due to the compensation of the upward $\phi_{BB}$ as depicted in Fig. \ref{fig7}(a).}
\end{figure}

\bigskip

\begin{center}
\textbf{2. SPV Characterization in K Doped Black Phosphorus}
\end{center}

As discussed above in the p-doped BP case, Fig. \ref{fig8} displays the Tr-ARPES measurements of K doped BP at -1 ps pump-probe delay without (red) and with (blue) the pump. In contrast to the pristine BP case, we observe that the blue spectrum is shifted by 200 meV to high KE with respect to the red one.  Due to the strong surface dipole, a long-range downward BB of both the CB and the VB is expected as illustrated in the middle panel of Fig. \ref{fig7}(b). This is confirmed by the large positive value of the SPV which is compensating this BB. 

\bigskip

\begin{figure}
\begin{center}
\includegraphics[scale=0.13]{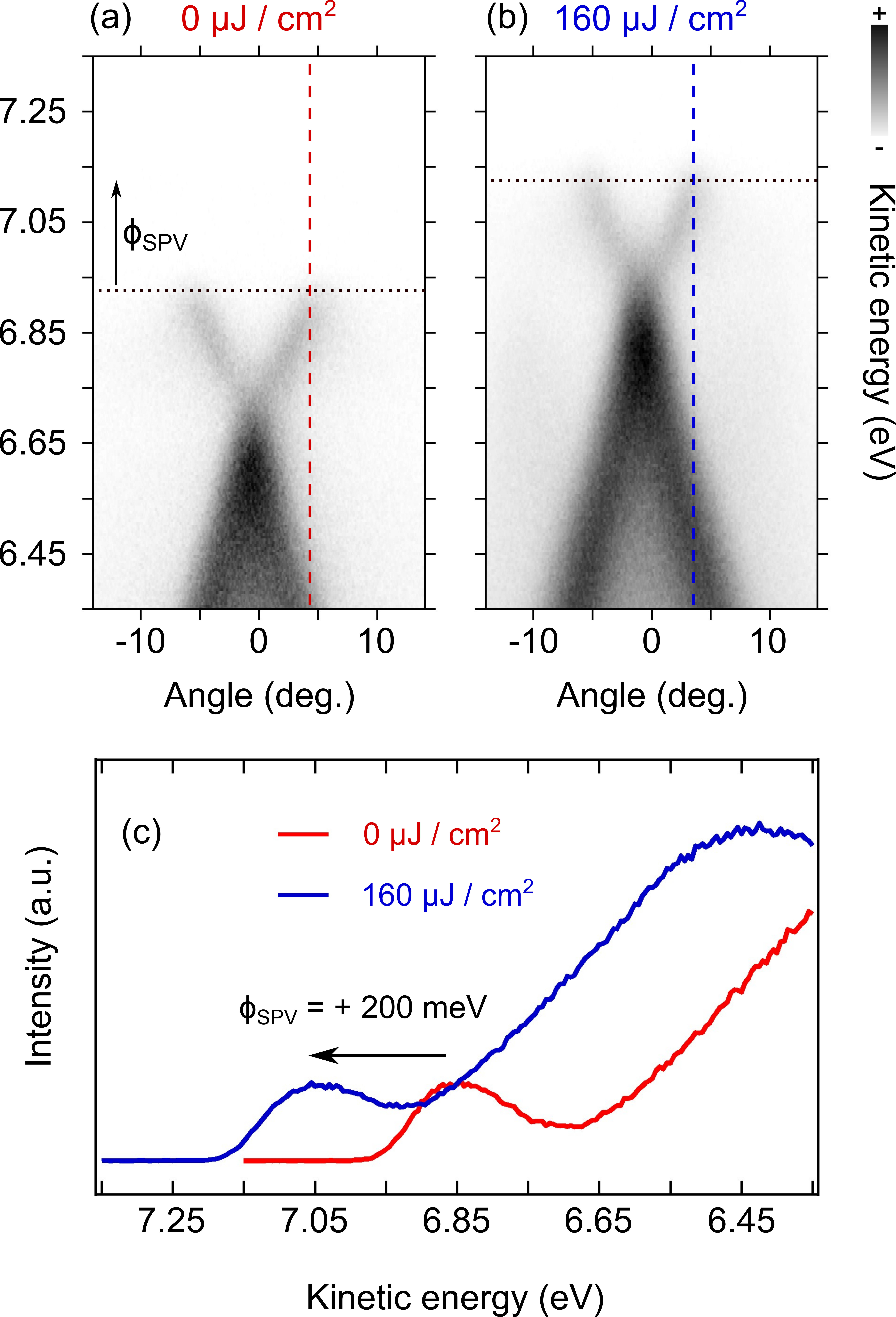} 
\end{center}
\caption{\label{fig8} Tr-ARPES measurements of K doped BP along the AC direction at \hbox{-1 ps} pump-probe delay for a pump fluence of (a) 0 $\mu$J/cm$^{2}$ and (b) of 160 $\mu$J/cm$^{2}$. (c) Corresponding EDCs taken four degrees off normal emission. A positive SPV of +200 meV is observed due to the compensation of the downward $\phi_{BB}$ as depicted in Fig. \ref{fig7}(b).}
\end{figure}

%%%%%%%%%%%%%%%%%%%%%%%%%%%%%%%%%%%%%%%%%%%%%%%%%%%%%%%%%%

\begin{center}
\textbf{3. Evidence of Spectral Shape Modification Along the Armchair Direction}
\end{center}

In Fig. \ref{fig9}, we show that the spectral shape modification of the VB is also visible along the AC direction. As discussed in the case of the ZZ direction in Fig. \ref{fig2} of the main text, we observe the same blue/white/red contrast in the difference plot (see Fig \ref{fig9}(c)) before and after the photoexcitation of the pump. The main difference with the ZZ direction is the absence of spectral weight at the bottom of CB, as already discussed in Ref. \onlinecite{chen2018band}.

\bigskip

\begin{figure}
\centering
\includegraphics[scale=0.225]{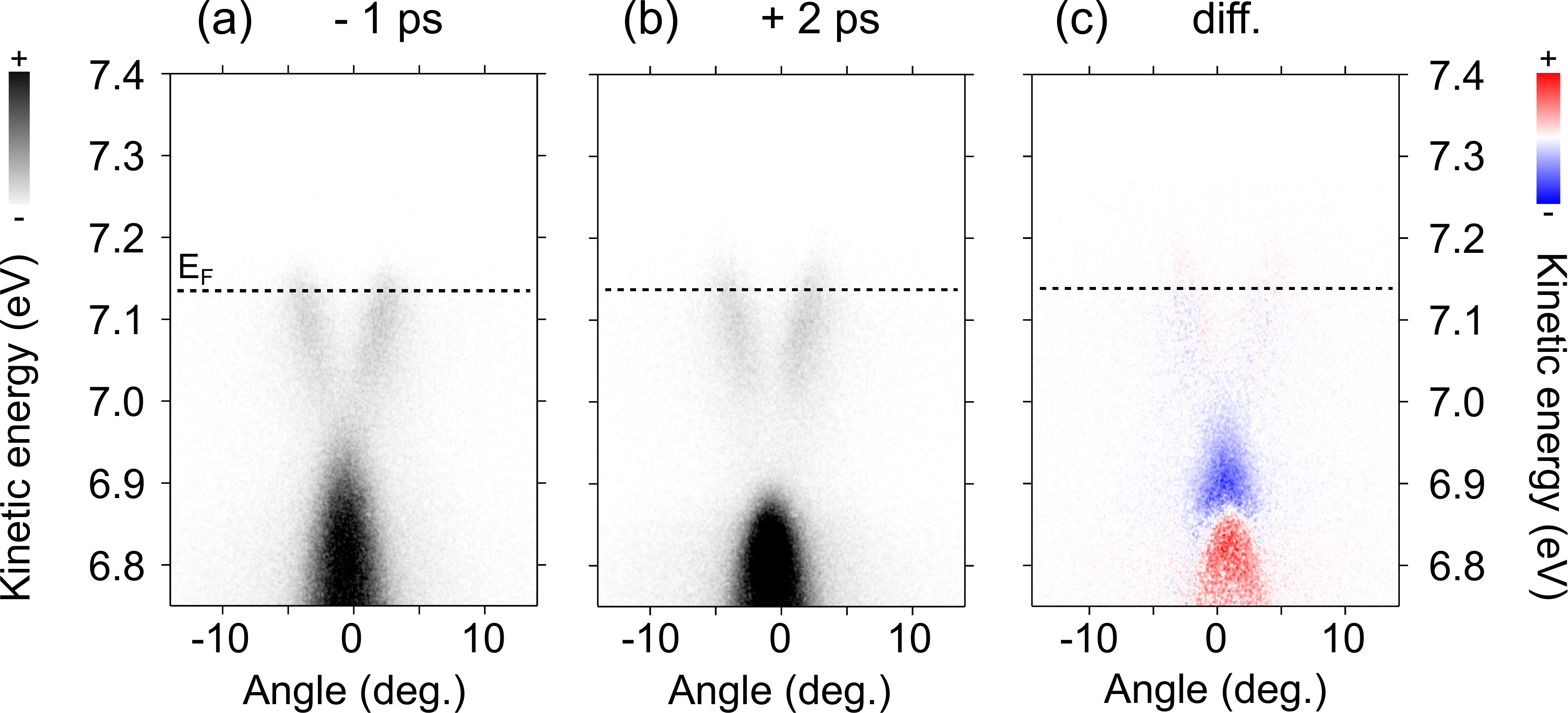}
\caption{\label{fig9} Tr-ARPES measurements of K doped BP along the AC direction using a pump fluence of 160 $\mu$J/cm$^{2}$ for pump-probe delays of (a) -1 ps and (b) +2 ps. (c) Intensity difference between the panels (b) and (a). Red and blue colors respectively correspond to an increase and a depletion of photoemission intensity. The transient dynamics is the same at the one observed in Fig. \ref{fig2} of the main text.}
\end{figure}

\bigskip

%%%%%%%%%%%%%%%%%%%%%%%%%%%%%%%%%%%%%%%%%%%%%%%%%%%%%%%%%%

\begin{center}
\textbf{4. Raws EDCs and Photoemission Cutoff Dynamics}
\end{center}

Figure \ref{fig10}(a) displays EDCs of the K doped BP sample taken at normal emission showing both a time dependent energy shift and VB spectral shape evolution. In contrast, the CB is only energy shifted. This is a selection of EDCs we used to produce Fig. \ref{fig3}(a,b). Figure \ref{fig10}(b) shows the temporal evolution of the low energy electrons cutoff position. Its shape is similar to the time dependent SPV of the VB and CB (see Fig. \ref{fig3}(a,b)).

\bigskip

\begin{figure}
\centering
\includegraphics[scale=0.135]{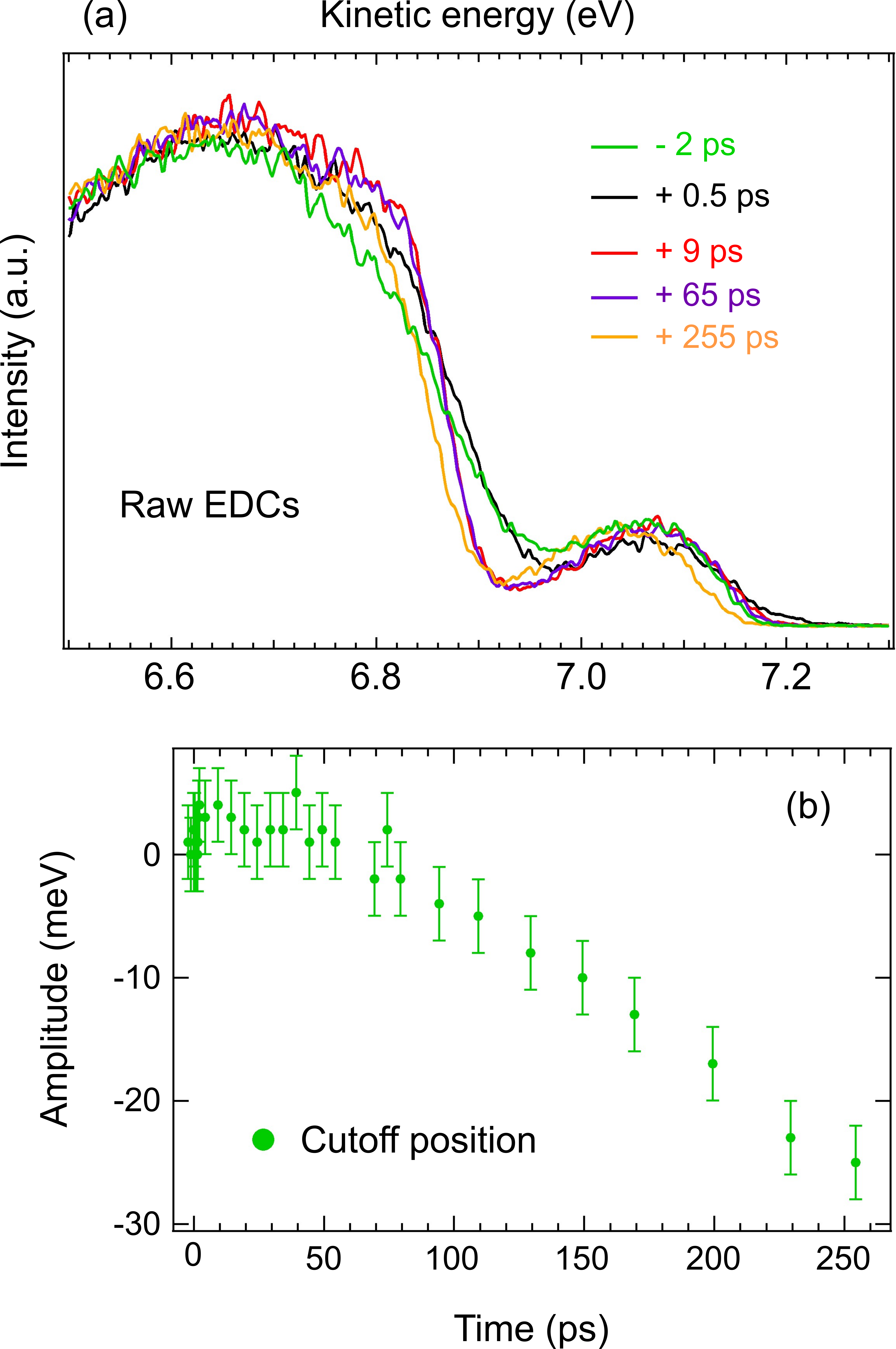}
\caption{\label{fig10} (a) EDCs of the K doped BP ARPES spectra taken at normal emission for a few pump-probe delays. These EDCs show both a time dependant energy shift and VB spectral shape evolution. (b) Temporal evolution of the low energy electrons cutoff position.}
\end{figure}

%%%%%%%%%%%%%%%%%%%%%%%%%%%%%%%%%%%%%%%%%%%%%%%%%%%%%%%%%%

\begin{center}
\textbf{5. Time Constants as Obtained by Exponential Fits}
\end{center}

In Fig. \ref{fig11}, we give a systematic fit of the pertinent time constants discussed in the main text, for both the VB and the CB of photoexcited K doped BP.  

\bigskip

\begin{figure*}
\begin{center}
\includegraphics[scale=0.115]{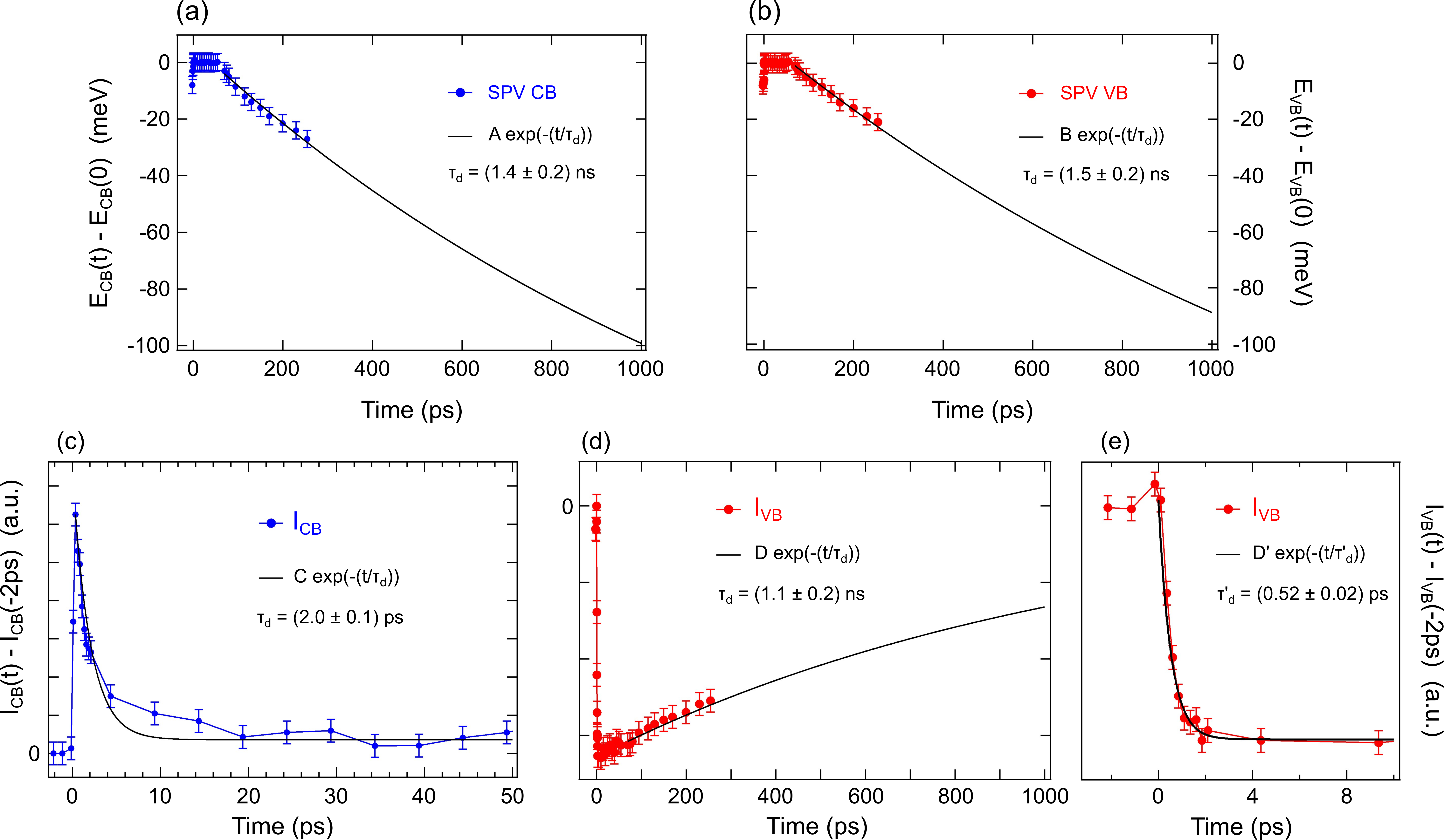} 
\end{center}
\caption{\label{fig11} Temporal evolution of the SPV at the surface of K doped BP for (a) the CB and (b) the VB. (c) Temporal evolution of I\textsubscript{CB} and (d,e) I\textsubscript{VB} as defined in the Fig. \ref{fig3} of the main text. For all these spectra, solid lines are the exponential fit used to extract the characteristic time constants discussed in the text.}
\end{figure*}

\bigskip

%%%%%%%%%%%%%%%%%%%%%%%%%%%%%%%%%%%%%%%%%%%%%%%%%%%%%%%%%%

\begin{center}
\textbf{6. Simulation of the Carrier Dynamics in Photoexcited Black Phosphorus}
\end{center}

\section{Model and parameters}
Pristine bulk BP is an intrinsically p-doped semiconductor with a small band gap of $E_g=0.3$ eV \cite{morita1986semiconducting,chen2018band}. The intrinsic carrier density $p_\mathrm{intr}$ can be as large as $O(10^{20})\mathrm{m^{-3}}$ at $T=200$ K \cite{Rad_Recom_BP_2016}, but at very low temperature, $p_\mathrm{intr}$ (and also $n_\mathrm{intr}$) decays exponentially with decreasing $T$. At the temperature of the experiment, $T=30K$, the density is as small as 0.016 $\mathrm{m^{-3}}$ and thus negligible. In order to electron dope the system, K atoms are deposited on the surface of BP. These K atoms release their $s$ electrons into the sample and create a 2D electron gas near the (positively charged) surface. The spatial separation between the ionized K atoms and the electron gas creates a strong electric field \cite{kim2015observation}. When electron-hole pairs are created by the pump, their dynamics is affected by this field and by the different mobilities in the VB and CB.  

The BB along the $z$ axis is mainly controlled by the carrier dynamics along the $z$ direction.
Assuming a homogeneous system in the $x$-$y$ plane, we consider a one-dimensional drift-diffusion problem along the $z$ axis. 
The mobilities of the electrons and holes in BP are taken from Ref. \onlinecite{JPSJ_1983}.
At $T=30$ K, $\mu_n=540$ $\mathrm{cm^2/Vs}$ and $\mu_p=6000$ $\mathrm{cm^2/Vs}$,
which makes the drift velocity $v=\mu E$ as large as $O(10)$ \AA$/\mathrm{fs}$. Thus the simulation needs a timestep of the order of attoseconds to resolve the ultrafast carrier dynamics.

In the simulation, we assume a sample thickness of $L=1000$ \AA, which is large compared to the atomic scales, but much smaller than the actual sample thickness ($\sim 1\mathrm{mm}$). There are two reasons for this choice.
On the one hand, the strong electric field requires a real-space grid spacing as small as $0.1$ \AA. Thus a realistic 
thickness requires huge matrices of dimension $O(10^8)$ which is computationally very demanding. 
On the other hand, the penetration depth of the pump pulse is only $O(10)$ $\mathrm{nm}$, and as we will show, the net difference in the electron and hole population is significant only in a narrow region near the surface. Hence, the relevant short-time carrier dynamics takes place near the surface. 
 
Furthermore, we assume that the photoexcited electron-hole pairs, with densities $\delta n(x,t=0_+)$ and $\delta p(x,t=0_+)$,
are generated instantaneously at time $t=0_+$, with an exponential profile  (see Fig. \ref{fig:static})
\begin{equation}
\delta n(x,t=0_+)=\delta p(x,t=0_+)=\gamma_0 e^{-x/\lambda_{ph}},
\label{eq:exp_pump}
\end{equation}
where the surface photoexcited carrier density $\gamma_0=5.7\cdot 10^{25}$ $\mathrm{m^{-3}}$ and the penetration depth 
$\lambda_{ph}=500$ \AA \mbox{}  are chosen in accordance with the experiment. The total density (per area) of photoexcited electron-hole pairs is denoted by 
$\sigma_p=\int_0^L \delta n(x,t=0_+)\mathrm{d}x=\int_0^L \delta p(x,t=0_+)\mathrm{d}x$. 
The K atoms are assumed to be deposited near the surface with a distribution that decays exponentially into the bulk, 
$N_{d}(x)=n_{u}\lambda_{d}e^{-\lambda_{d}\frac{x}{l_{0}}}$, where
$n_{u}=\sigma_{0}/l_{0}$ is the unit of density and $l_{0}/\lambda_{d}\approx 5$ \AA \mbox{}  represents the doping depth, while  
$\sigma_0$ is the total K coverage per unit area. In the case of a $\delta$-function distribution of K ions, the thickness of the 2D electron gas is less than 1 \AA, which is unrealistic. A more accurate quantum mechanical treatment based on a Schr\"odinger-Poisson solver would yield a broader charge distribution, and the above exponential profile for $N_d(x)$ is a simple way of mimicking this effect.  

As for the carrier densities $n$ and $p$, we neglect the intrinsic densities \cite{LowD_Semi}
\begin{align}
n_\mathrm{intr}=p_\mathrm{intr}&=\sqrt{\left(\frac{m_{e}^{*}m_{h}^{*}k_{B}^{2}T^{2}}{\pi^{2}\hbar^{4}}\right)^{3/2}}e^{-\frac{E_g}{2 k_B T}}\nonumber\\
&\approx 1.547\times10^{21}T^{3/2}e^{-\frac{1740.7}{T}}\mathrm{m^{-3}}
\label{eq:n_intrinsic}
\end{align}
where the averaged effective masses are $m_{e}^{*} \approx 0.413$ $m_e$ and $m_{h}^{*}\approx 0.335$ $m_e$ \cite{morita1986semiconducting} and the band gap 
is \hbox{$E_g\approx 0.3$ eV.} At temperature $T=30$K, 
$n_\mathrm{intr}\approx 0.016\mathrm{m^{-3}}$ which is much smaller than the excess carrier density \cite{Rad_Recom_BP_2016} (see also Fig.~3 of Ref. \onlinecite{Rad_Recom_BP_2016}. Even at $T=200$K, $n_\mathrm{intr}\sim O(10^{20})\mathrm{m^{-3}}\ll 10^{25}\mathrm{m^{-3}}$).
We only consider the photo-excited electron-hole densities ($\delta n$, $\delta p$) and the free electron density ($n_0(x)$) resulting from the fully ionized K atoms with surface density $\int_{0}^L n_0(x)\mathrm{d}x=\sigma_0$. Hence, the total electron and hole densities are   
\begin{equation}
n(x,t)=n_0(x)+\delta n(x,t)
\label{eq:nxt}
\end{equation}
and
\begin{equation}
p(x,t)=\delta p(x,t).
\label{eq:pxt}
\end{equation}
 
All the simulation parameters are listed in Tab.~\ref{tab:table_para}. 

\begin{table*}
\caption{\label{tab:table_para}
Parameters used in the simulation. The diffusion coefficients are obtained by the Einstein relation $D_{n,p}=\frac{\mu_{n,p}k_{B}T}{\mathrm{e}}$.  The values of $\mu_n$, $\mu_p$, $\varepsilon_r$ are from Ref. \onlinecite{JPSJ_1983}. The coefficient $\alpha$ in $B_r$ is a unit-less number.
}
\begin{ruledtabular}
\begin{tabular}{ccc}
Name & Symbol & Value\tabularnewline
\hline 
\hline 
Temperature & $T$ & 30K\tabularnewline
\hline 
Relative Permittivity & $\varepsilon_{r}$ & 8.3\tabularnewline
\hline 
Electron Mobility & $\mu_{n}$ & $540.0\times10^{-4}\mathrm{m^{2}/(Vs)}$\tabularnewline
\hline 
Electron Diffusion Coefficient & $D_{n}$ & $1.396\times10^{-4}\mathrm{m^{2}/s}$\tabularnewline
\hline 
Hole Mobility & $\mu_{p}$ & $6000\times10^{-4}\mathrm{m^{2}/(Vs)}$\tabularnewline
\hline 
Hole Diffusion Coefficient & $D_{p}$ & $15.51\times10^{-4}\mathrm{m^{2}/s}$\tabularnewline
\hline 
Potassium Coverage & $\sigma_{0}$ & $6.3\times10^{17}\mathrm{m}^{-2}$\tabularnewline
\hline
Photoexcited Carrier Volume & $\sigma_p$ & $\sim 4\sigma_0$\tabularnewline
\hline 
Length Unit & $l_{0}$ & $10^{-10}\mathrm{m}\  (1\text{Å})$\tabularnewline
\hline
Sample Length & L & $1000\cdot l_0$ \tabularnewline
\hline 
Number of $x$-grid points & N & 20001 \tabularnewline
\hline 
Electric Field Unit & $E_{u}=\frac{\sigma_{0}e}{2\varepsilon_{r}\varepsilon_{0}}$ & $6.867\times10^8\mathrm{V/m}$\tabularnewline
\hline 
Density Unit & $n_{u}=\frac{\sigma_0}{l_0}$ & $6.3\times10^{27} \mathrm{m^{-3}}$\tabularnewline
\hline 
Radiative Recombination Rate & $B_r=\frac{\alpha}{0.01\mathrm{fs}\cdot n_{u}}$ & $ 1.587\alpha\times 10^{-9}\mathrm{m^3/s}$ \tabularnewline
\end{tabular}
\end{ruledtabular}
\end{table*}

\section{Numerical Simulation}  
\subsection{Drift-Diffusion Equations}
The continuity equations for the electrons and holes read 
\begin{equation}
-\nabla\cdot\boldsymbol{J}_{n}+G_{n}-R_{n}=\frac{\partial n}{\partial t},
\label{eq:cont_n}
\end{equation}
and 
\begin{equation}
-\nabla\cdot\boldsymbol{J}_{p}+G_{p}-R_{p}=\frac{\partial p}{\partial t},
\label{eq:cont_p}
\end{equation}
respectively. Here, the particle flux densities are  
\begin{equation}
\boldsymbol{J}_{n}=-D_{n}\nabla n-n\mu_{n}\boldsymbol{E}
\label{eq:Jn}
\end{equation}
for the electrons and 
\begin{equation}
\boldsymbol{J}_{p}=-D_{p}\nabla p+p\mu_{p}\boldsymbol{E}
\label{eq:Jp}
\end{equation}
for the holes. $\mu_{n,p}$ are the drift mobilities and $D_{n,p}$ the diffusion coefficients which are related to $\mu_{n,p}$
by the Einstein relations $D_{n,p}=\frac{\mu_{n,p}k_B T}{e}$. $G$ and $R$ denote the generation rate and recombination rate, respectively. To illustrate the effect of recombination, we make the simple {\it ad hoc} approximation  
\begin{equation}
G-R \approx B_{r} n \delta p.
\label{eq:genrecom}
\end{equation}
%where $\delta n=n(x,t)-n(x,t<0)$ and $\delta p=p(x,t)-p(x,t<0)$. 
Here we neglect the thermal generations and recombinations 
and consider only inter-band recombinations \cite{Semi_cond_elect_book}.

Combining \cref{eq:nxt,eq:pxt,eq:cont_n,eq:cont_p,eq:Jn,eq:Jp,eq:genrecom}, we arrive at the drift-diffusion equations 
\begin{equation}
D_{n}\nabla^{2}n+\mu_{n}(\boldsymbol{E}\cdot\nabla n+n\nabla\cdot\boldsymbol{E})-B_r n\delta p=\frac{\partial \delta n}{\partial t}, 
\label{eq:dde_n}
\end{equation}
\begin{equation}
D_{p}\nabla^{2}\delta p-\mu_{p}(\boldsymbol{E}\cdot\nabla \delta p+\delta p\nabla\cdot\boldsymbol{E})-B_r n\delta p=\frac{\partial \delta p}{\partial t}. 
\label{eq:dde_p}
\end{equation}
Furthermore, the total electric field $E$ and total charge density satisfy the Poisson equation 
\begin{equation}
\nabla\cdot\boldsymbol{E}=\frac{e}{\kappa}(\delta p-n_0-\delta n +N_{d}).
\label{eq:Poiss}
\end{equation}
where $\kappa \equiv \varepsilon_{r}\varepsilon_{0}$ with 
$\varepsilon_{r}$ the relative permittivity and $\varepsilon_{0}$
the vacuum permittivity. 

For the stability of the simulation, it is advantageous to reformulate these equations  
in terms of $\delta n $ and $\delta p$. The free electron distribution $n_0(x)$ 
is obtained by solving the static (equilibrium) problem before the pump ($\delta n=\delta p=0$ for $t<0$)
\begin{equation}
D_{n}\nabla^{2}n_0+\mu_{n}(\boldsymbol{E}_0\cdot\nabla n_0+n\nabla\cdot\boldsymbol{E_0})=0,
\end{equation}
\begin{equation}
\nabla\cdot\boldsymbol{E}_0=\frac{e}{\kappa}(-n_0 +N_{d}).
\end{equation}
The numerical method used for computing the static $n_0$ and $\boldsymbol{E}_0$ is described in \cref{sec:steady}, 
where we have to set $\sigma_p=0$. Equation \eqref{eq:dde_n} can then be rewritten as 
\begin{align}
\frac{\partial\delta n}{\partial t}  & =  D_{n}\nabla^{2}\delta n+\mu_{n}(\delta\boldsymbol{E}\cdot \nabla n_{0}+\boldsymbol{E}\cdot\nabla\delta n)\nonumber\\
 & \quad +\mu_{n}(n_{0}\nabla\cdot \delta\boldsymbol{E}+\delta n\nabla\cdot \boldsymbol{E})-B_{r} n\delta p,
\label{eq:dde_n_new}
\end{align}
where $\boldsymbol E=\boldsymbol{E}_0+\delta\boldsymbol E$ and 
\begin{equation}
\nabla\cdot\delta\boldsymbol{E}=\frac{e}{\kappa}(\delta p-\delta n).
\label{eq:Poiss_dndp}
\end{equation}
For given $\delta n$ and $\delta p$, we calculate $\delta E$ and $E$ as
\begin{align}
\delta E(x)&=\frac{e}{\kappa}\int_{0}^{x}[\delta p-\delta n](x^{\prime})\mathrm{d}x^{\prime},\nonumber\\
E(x)&=\frac{e}{\kappa}\int_{0}^{x}[\delta p-\delta n-n_{0}+N_{d}](x^{\prime})\mathrm{d}x^{\prime}.
\label{eq:inte_E} 
\end{align}

\subsubsection{Discretization}
The density and electric field are continuous functions of the coordinate $x$ and time $t$. 
Numerically, we discretize $x$ and $t$ as $\{0=x_0<x_1<\cdots<x_N=L\}$ and
$\{0=t_0<t_1<\cdots<t_M=T\}$, and define $\tau_m\equiv t_{m+1}-t_m$.
Let $f_i^m$ denote the value of the function $f(x,t)$ at $x_i$ and $t_m$, and 
$\tau_m \equiv t_{m+1}-t_m$, $\boldsymbol{f}^m=[f_0^m,f_1^m,\cdots,f_N^m]$.
The first- and second-order derivative with respect to $x$ of a given function $f(x)$ 
can be represented in a matrix-vector multiplication form, 
$f^\prime_i\approx\sum_j D_{1;i,j}f_j$ and $f^{\prime\prime}_i \approx \sum_j D_{2;i,j}f_j$,
where $D_1$ and $D_2$ denote the finite difference matrices for the first and second-order derivative, respectively.
Here, we will use the central difference approximation to $D_1$ and $D_2$, 
\begin{align}
f^{\prime}_i & \approx \frac{f_{i+1}-f_{i-1}}{2\mathrm{d}x},\quad %\nonumber\\
f^{\prime\prime}_i  \approx \frac{f_{i+1}-2f_i+f_{i-1}}{{\mathrm{d}x}^2}.
\end{align}
One needs to properly treat the zero-flux boundary conditions $\boldsymbol{J}_{n,p}|_{x=0,L}=0$ (\cref{eq:Jn,eq:Jp}) 
which are equivalent to 
\begin{equation}
\nabla n|_{x=0,L}=\nabla p|_{x=0,L}=0,
\label{eq:zero_flux}
\end{equation}
since $\boldsymbol{E}|_{x=0,L}=\boldsymbol 0$ due to charge neutrality.
By introducing two auxiliary points $x_{-1}$ and $x_{N+1}$, we can write (using $n$ as an example)
\begin{align}
n^\prime_0 &\approx \frac{n_1-n_{-1}}{2\mathrm{d}x}=0,\quad %\nonumber\\
n^\prime_N \approx \frac{n_{N+1}-n_{N-1}}{2\mathrm{d}x}=0,
\end{align}
which gives $n_{-1}=n_{1}$ and $n_{N+1}=n_{N-1}$. Thus 
\begin{align}
n^{\prime\prime}_0 & \approx \frac{n_{1}-2n_0+n_{-1}}{{\mathrm{d}x}^2} \approx \frac{2n_{1}-2n_0}{{\mathrm{d}x}^2},\nonumber\\
n^{\prime\prime}_N & \approx \frac{n_{N+1}-2n_N+n_{N-1}}{{\mathrm{d}x}^2} \approx \frac{2n_{N-1}-2n_N}{{\mathrm{d}x}^2}.
\end{align}
Hence, we can write the finite-difference matrices for the density as 
\begin{equation}
D_{1}=\frac{1}{\mathrm{d}x}\left[\begin{array}{ccccc}
0 & 0\\
-\frac{1}{2} & 0 & \frac{1}{2}\\
 & \ddots & \ddots & \ddots\\
 &  & -\frac{1}{2} & 0 & \frac{1}{2}\\
 &  &  & 0 & 0
\end{array}\right],
\label{eq:D1_zeroflux}
\end{equation}
and 
\begin{equation}
D_{2}=\frac{1}{\mathrm{d}x^{2}}\left[\begin{array}{ccccc}
-2 & 2\\
1 & -2 & 1\\
 & \ddots & \ddots & \ddots\\
 &  & 1 & -2 & 1\\
 &  &  & 2 & -2
\end{array}\right].
\label{eq:D2_zeroflux}
\end{equation}

\subsubsection{Semi-implicit Euler Method}
The coupled \cref{eq:dde_n_new,eq:dde_p,eq:Poiss,eq:Poiss_dndp} are solved by using a semi-implicit Euler method:
\begin{align}
\frac{\delta\boldsymbol{n}^{m+1}-\delta\boldsymbol{n}^{m}}{\tau_{m}} & \approx D_{n}D_{2}\delta\boldsymbol{n}^{m+1}-B_{r}\delta\boldsymbol{p}_{\mathrm{diag}}^{m}\boldsymbol{n}^{m}\nonumber\\
 & +\mu_{n}[\delta\boldsymbol{E}_{\mathrm{diag}}^{m}D_{1}\boldsymbol{n}_{0}+\boldsymbol{E}^{m}_{\mathrm{diag}}D_{1}\delta\boldsymbol{n}^{m+1}]\nonumber\\
 & +\mu_{n}[(\nabla\cdot\delta\boldsymbol{E})^{m}_{\mathrm{diag}}\boldsymbol{n}_{0}+(\nabla\cdot\boldsymbol{E})_{\mathrm{diag}}^{m}\delta\boldsymbol{n}^{m+1}],
\label{eq:impl_euler_dn}
\end{align} 
and 
\begin{align}
\frac{\delta\boldsymbol{p}^{m+1}-\delta\boldsymbol{p}^{m}}{\tau_{m}} & \approx D_{p}D_{2}\delta\boldsymbol{p}^{m+1}-B_{r}\boldsymbol{n}_{\mathrm{diag}}^{m}\delta\boldsymbol{p}^{m}\nonumber\\
 & -\mu_{p}[\boldsymbol{E}_{\mathrm{diag}}^{m}D_{1}\delta\boldsymbol{p}^{m+1}+(\nabla\cdot\boldsymbol{E})_{\mathrm{diag}}^{m}\delta\boldsymbol{p}^{m+1}]
\label{eq:impl_euler_dp}.
\end{align} 
Here, $[\boldsymbol f]_{\mathrm{diag}}$ represents a diagonal matrix with the elements of the vector $\boldsymbol f$.
$\delta\boldsymbol{E}^{m}_\mathrm{diag}$ and $\boldsymbol{E}^{m}_{\mathrm{diag}}$ are calculated from \cref{eq:inte_E} using the density at $t_m$ and similarly for
$\nabla\delta\boldsymbol{E}^{m}$ and $\boldsymbol{E}^{m}$.

Eqs.~(\ref{eq:impl_euler_dn}) and (\ref{eq:impl_euler_dp}) can be reduced to the linear equations
\begin{align}
&\left[\mathrm{\mathbb{I}}-\tau_{m}D_{n}D_{2}-\tau_{m}\mu_{n}\left(\boldsymbol{E}_{\mathrm{diag}}^{m}D_{1}+(\nabla\cdot\boldsymbol{E})_{\mathrm{diag}}^{m}\right)\right]\delta\boldsymbol{n}^{m+1} \nonumber\\ 
&\quad\approx \delta \boldsymbol{n}^{m}-\tau_{m}B_{r}\delta\boldsymbol{p}_{\mathrm{diag}}^{m}\boldsymbol{n}^{m}\nonumber\\
&\hspace{8mm}+
 \tau_{m}\mu_{n}[\delta\boldsymbol{E}_{\mathrm{diag}}^{m}D_{1}\boldsymbol{n}_{0}+
(\nabla\cdot\delta\boldsymbol{E})^{m}_{\mathrm{diag}}\boldsymbol{n}_{0}]
\label{eq:dn_t_next}
\end{align}
and
\begin{align}
&\left[\mathbb{I}-\tau_{m}D_{p}D_{2}+\tau_{m}\mu_{p}\left(\boldsymbol{E}_{\mathrm{diag}}^{m}D_{1}+(\nabla\cdot\boldsymbol{E})_{\mathrm{diag}}^{m}\right)\right]\delta\boldsymbol{p}^{m+1} \nonumber\\ 
&\quad\approx \left[\mathbb{I}- \tau_{m}B_{r}\boldsymbol{n}_{\mathrm{diag}}^{m}\right]\delta\boldsymbol{p}^{m}.
\label{eq:dp_t_next}
\end{align} 
The numerical simulation consists of the following steps:
\begin{enumerate}
\item Find the static solution ($\boldsymbol{n}_0$ and $\boldsymbol{E}_0$) before the pump (see \cref{sec:steady}); turn on the pump $\delta \boldsymbol{n}^{m=0}=\delta\boldsymbol{p}^{m=0}$ as in \cref{eq:exp_pump} at 
$t=0_+$ ($m=0$). 
\item Update the electric field $\delta\boldsymbol{E}^{m}_\mathrm{diag}$ and $\boldsymbol{E}^{m}_{\mathrm{diag}}$ by \cref{eq:inte_E} using the density at $t_m$; update $(\nabla\cdot\boldsymbol{E})^{m}$ and $(\nabla\cdot\delta\boldsymbol{E})^{m}$ by \cref{eq:Poiss,eq:Poiss_dndp} using the density at $t_m$. 
\item Update the densities $\delta\boldsymbol{n}^{m+1}$ and $\delta\boldsymbol{p}^{m+1}$ at $t_{m+1}$ by solving the linear \cref{eq:dn_t_next,eq:dp_t_next}. 
\item Repeat steps 2 and 3 until $t=t_N$.
\end{enumerate}

For any intermediate time $t_m$, we can calculate the potential profile as 
\begin{equation} 
\phi(x,t_m)=\phi(L,t_m)+\int_{x}^{L}\boldsymbol{E}(x^{\prime},t_{m})\mathrm{d}x^{\prime},
\end{equation} 
where the potential reference is chosen as $\phi(L,t_m)=0$. The band bending profile is given by 
\begin{equation} 
-e\phi(x,t_m)=-e\int_{x}^{L}\boldsymbol{E}(x^{\prime},t_{m})\mathrm{d}x^{\prime},
\end{equation} 
and the surface photovoltage is given by $-e\phi(0,t_m)$.

\begin{figure*}
\begin{center}
\includegraphics[scale=0.4]{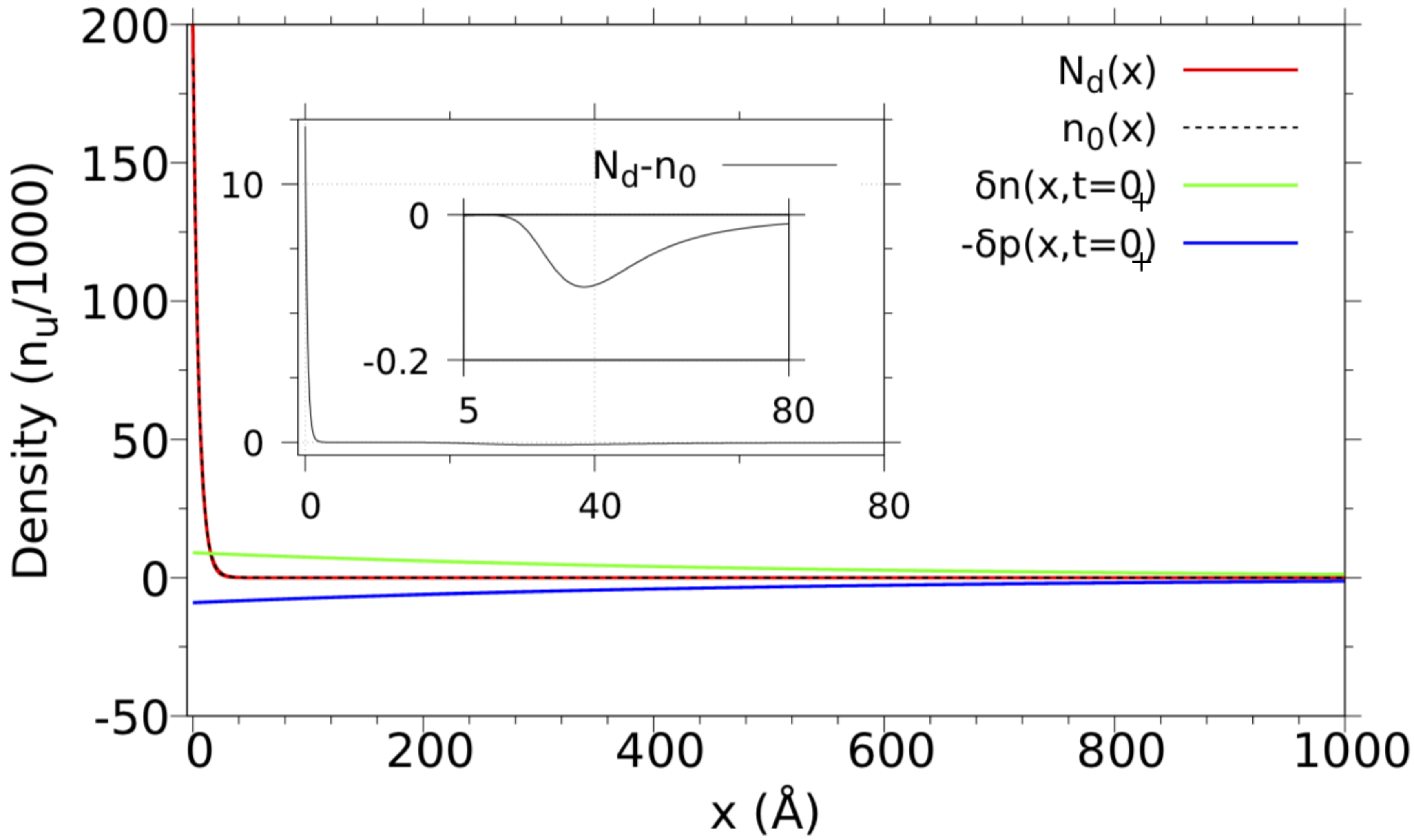} 
\end{center}
\caption{\label{fig:static} The static equilibrium state solution before the laser pump and the photoexcited carriers distribution at \hbox{$t=0_+$.}}
\end{figure*}

\subsection{Steady state solution}
\label{sec:steady}
Without recombination term in \cref{eq:dde_n,eq:dde_p},
the photoexcited carriers drift and diffuse inside the material
and eventually reach a steady state.
In this steady state, the current is zero everywhere 
\begin{equation}
0=\boldsymbol{J}_{n}=-D_{n}\nabla n-n\mu_{n}\boldsymbol{E},
\label{eq:Jneq0}
\end{equation}
\begin{equation}
0=\boldsymbol{J}_{p}=-D_{p}\nabla p+p\mu_{p}\boldsymbol{E}.
\label{eq:Jpeq0}
\end{equation}
To find the steady state solution, we need to solve the coupled \cref{eq:Poiss,eq:Jneq0,eq:Jpeq0}.

\subsubsection{Dimensionless formulation}
For simplicity, we split $\boldsymbol{E}$ into three components,
$E=E_{n}+E_{p}+E_{d}$, with $E_{n}^{\prime}=-\frac{e}{\kappa}n$,
$E_{p}^{\prime}=\frac{e}{\kappa}\delta p$ and $E_{d}^{\prime}=\frac{e}{\kappa}N_{d}$.
Furthermore, we introduce the unit $E_{u}=\frac{\sigma_{0}e}{2\kappa}$
for the electric field and $n_{u}=\frac{\sigma_{0}}{l_{0}}$ for the density,
and introduce the unit-less functions $y$, $h$ and $f$ for $E_{n}$,
$E_{p}$ and $E_{d}$, respectively. $E_{d}(x)$ can be easily calculated
from the known $N_{d}(x)$ as $E_{d}(x)=E_{u}f(x)$, with $f(x)=1-2e^{-\lambda_{d}\frac{x}{l_{0}}}$.
We can thus simplify \cref{eq:Jneq0,eq:Jpeq0} as
\begin{equation}
D_{n}E_{u}y^{\prime\prime}(x)+\mu_{n}E_{u}y^{\prime}(x)\cdot E_u [y(x)+h(x)+f(x)]=0,
\label{eq:steady_tmp_eq1}
\end{equation}
and
\begin{equation}
-D_{p}E_{u}h^{\prime\prime}(x)+\mu_{p}E_{u}h^{\prime}(x)\cdot E_u [y(x)+h(x)+f(x)]=0.
\label{eq:steady_tmp_eq2}
\end{equation}
Now we introduce the unit-less coefficient
\begin{equation}
cl_{0}=\frac{\mu_{n}E_{u}l_{0}}{D_{n}}=\frac{\mu_{p}E_{u}l_{0}}{D_{p}}=\frac{e}{k_{B}T}\cdot\frac{\sigma_{0}e}{2\kappa}\cdot l_{0}
\end{equation}
which allows to simplify \cref{eq:steady_tmp_eq1,eq:steady_tmp_eq2} as 
\begin{equation}
y^{\prime\prime}(x)+\frac{cl_{0}}{l_{0}}y^{\prime}(x)[f(x)+y(x)+h(x)]=0,
\label{eq:steady_tmp_eq3}
\end{equation}
and
\begin{equation}
-h^{\prime\prime}(x)+\frac{cl_{0}}{l_{0}}h^{\prime}(x)[f(x)+y(x)+h(x)]=0.
\label{eq:steady_tmp_eq4}
\end{equation}
We furthermore introduce the two auxiliary functions $P=y+h$, $Q=y-h$ which 
simplify the implementation of the boundary conditions, as shown below. The sum of Eq.~(\ref{eq:steady_tmp_eq3}) and Eq.~(\ref{eq:steady_tmp_eq4})
gives
\begin{equation}
P^{\prime\prime}+\frac{cl_{0}}{l_{0}}Q^{\prime}[f+P]=0,
\label{eq:steady_eq1}
\end{equation}
while the difference of Eq.~(\ref{eq:steady_tmp_eq3}) and Eq.~(\ref{eq:steady_tmp_eq4}) gives
\begin{equation}
Q^{\prime\prime}+\frac{cl_{0}}{l_{0}}P^{\prime}[f+P]=0.
\label{eq:steady_eq2}
\end{equation}

In the solution of the coupled second-order non-linear differential \cref{eq:steady_eq1,eq:steady_eq2} 
on the discretized grid points $0=x_0<x_1<x_2<\cdots x_{N-1}<x_N=L$, we need to impose the proper boundary conditions for $P$ and $Q$. 
In the following, we represent a function on the grid points by a vector $\boldsymbol{f}$ with the $i$-th element labeled as
$f_i=f(x_i)$ for simplicity. The conditions to be imposed are the following:
\begin{enumerate}
\item Charge neutrality requires that the electric field vanishes at the boundaries, i.e., $E|_{x=0}=E|_{x=L}=0$. Since $E=E_u(P+f)$, we thus have $P_0=-f_0=1.0$ and $P_N=-f_N\approx -1.0$ 
if $L\gg l_0$. 
\item The electric field component arising from $\delta p(x)$,
i.e. $E_{p}(x)=E_{u}h(x)$, satisfies $E_{p}|_{x=L}=E_{p}|_{x=0}+\frac{e}{\kappa}\int_{0}^{L}\delta p(x^{\prime})\mathrm{d}x^{\prime}=\frac{e}{\kappa}\sigma_{p}=E_{u}\frac{2\sigma_{p}}{\sigma_{0}}$.  Furthermore, the Gauss law requires that $E_{p}|_{x=0}=-E_{p}|_{x=L}$.
Thus we have $h_0=-\frac{\sigma_{p}}{\sigma_{0}}=-h_N$, which gives $Q_0=y_0-h_0=-f_0-2h_0$ and $Q_N=y_N-h_N=-f_N-2h_N$. 
\item $P_0^{\prime\prime}=Q_0^{\prime\prime}=P_N^{\prime\prime}=Q_N^{\prime\prime}=0$, since $E\sim f+P$ is zero at the boundaries.
\end{enumerate}

\subsubsection{Newton-Raphson Iterative Method}
Solving \cref{eq:steady_eq1,eq:steady_eq2} for $P$ and $Q$ is equivalent to solving a multivariable root problem 
$\boldsymbol{F}(\boldsymbol X)=0$. The Newton-Raphson iterative method \cite{Num_Recipe} is based on the first order Taylor expansion of $\boldsymbol{F}(\boldsymbol X)$
as $\boldsymbol{F}(\boldsymbol X+\delta \boldsymbol{X})\approx\boldsymbol{F}(\boldsymbol X)+\boldsymbol{\mathcal J}\cdot\delta\boldsymbol X$, 
where $\boldsymbol{\mathcal J}$ denotes the Jacobian matrix with elements $\mathcal J_{i,j}=\frac{\partial F_i}{\partial X_j}$. By setting 
$\boldsymbol{F}(\boldsymbol X+\delta \boldsymbol{X})=0$, we obtain a set of linear equations for $\delta \boldsymbol X$:
\begin{equation}
\boldsymbol{\mathcal J} \cdot \delta \boldsymbol X =-\boldsymbol F. 
\label{eq:Newton}
\end{equation}
$\delta \boldsymbol X$ is the correction to $\boldsymbol X$ and we update the solution as 
\begin{equation}
\boldsymbol{X}_{\mathrm{new}}=\boldsymbol{X}_{\mathrm{old}}+\delta \boldsymbol X 
\label{eq:update}
\end{equation}
until convergence.

To utilize the Newton-Raphson method, we need to combine \cref{eq:steady_eq1,eq:steady_eq2} into a matrix form. 
Since $P_{0,N}$, $P^{\prime\prime}_{0,N}$, 
$Q_{0,N}$, and $Q^{\prime\prime}_{0,N}$ are constants given by the boundary conditions, we only need to define the finite-difference matrices
$D_{1,2}$ on the internal grid points $\{x_1, x_2, \cdots, x_{N-1}\}$ with
\begin{equation}
D_{1}=\frac{1}{\mathrm{d}x}\left[\begin{array}{ccccc}
0 & \frac{1}{2}\\
-\frac{1}{2} & 0 & \frac{1}{2}\\
 & \ddots & \ddots & \ddots\\
 &  & -\frac{1}{2} & 0 & \frac{1}{2}\\
 &  &  & -\frac{1}{2} & 0
\end{array}\right],
\end{equation}
and 
\begin{equation}
D_{2}=\frac{1}{\mathrm{d}x^{2}}\left[\begin{array}{ccccc}
-2 & 1\\
1 & -2 & 1\\
 & \ddots & \ddots & \ddots\\
 &  & 1 & -2 & 1\\
 &  &  & 1 & -2
\end{array}\right].
\end{equation}
Let's introduce the vectors
\begin{align}
\vec{P}^T&=\left[P_1,P_2,\cdots,P_{N-1}\right],\nonumber\\
\vec{Q}^T&=\left[Q_1,Q_2,\cdots,Q_{N-1}\right],
\end{align}
and 
\begin{align}
\vec{v}_1^T&=\frac{1}{{\mathrm{d}x}^2}\left[P_0,0,\cdots,0,P_N\right],\nonumber\\
\vec{v}_2^T&=\frac{1}{{\mathrm{d}x}}\left[-\tfrac{1}{2}P_0,0,\cdots,0,\tfrac{1}{2}P_N\right],\nonumber\\
\vec{v}_3^T&=\frac{1}{{\mathrm{d}x}^2}\left[Q_0,0,\cdots,0,Q_N\right],\nonumber\\
\vec{v}_4^T&=\frac{1}{{\mathrm{d}x}}\left[-\tfrac{1}{2}Q_0,0,\cdots,0,\tfrac{1}{2}Q_N\right].
\end{align}
($\forall i$, $\dim \vec{v}_{i}^T =N-1$). 
The first and second derivatives are given by 
\begin{align}
\vec{P}^{\prime\prime}&=D_{2}\vec{P}+\vec{v}_{1},\nonumber\\
\vec{P}^{\prime}&=D_{1}\vec{P}+\vec{v}_{2},\nonumber\\
\vec{Q}^{\prime\prime}&=D_{2}\vec{Q}+\vec{v}_{3},\nonumber\\
\vec{Q}^{\prime}&=D_{1}\vec{Q}+\vec{v}_{4}.
\end{align}
Now we define the vector $\boldsymbol{F}$ from \cref{eq:steady_eq1,eq:steady_eq2} as
\begin{align}
\boldsymbol{F} & =\left(\begin{array}{c}
\boldsymbol{F}_{1}\\
\boldsymbol{F}_{2}
\end{array}\right)=\left(\begin{array}{c}
\vec{P}^{\prime\prime}+c[\vec{f}+\vec{P}]_{\mathrm{diag}}\vec{Q}^{\prime}\\
\vec{Q}^{\prime\prime}+c[\vec{f}+\vec{P}]_{\mathrm{diag}}\vec{P}^{\prime}
\end{array}\right)\nonumber\\
 & =\left(\begin{array}{c}
D_{2}\vec{P}+\vec{v}_{1}+c[\vec{f}+\vec{P}]_{\mathrm{diag}}\left(D_{1}\vec{Q}+\vec{v}_{4}\right)\\
D_{2}\vec{Q}+\vec{v}_{3}+c[\vec{f}+\vec{P}]_{\mathrm{diag}}\left(D_{1}\vec{P}+\vec{v}_{2}\right)
\end{array}\right),
\end{align}
The Jacobian matrix reads
\begin{align}
\boldsymbol{\mathcal J}&=\left[\begin{array}{cc}
\frac{\partial\boldsymbol{F}_{1}}{\partial\boldsymbol{\vec{P}}} & \frac{\partial\boldsymbol{F}_{1}}{\partial\boldsymbol{\vec{Q}}}\\
\frac{\partial\boldsymbol{F}_{2}}{\partial\boldsymbol{\vec{P}}} & \frac{\partial\boldsymbol{F}_{2}}{\partial\boldsymbol{\vec{Q}}}
\end{array}\right]\nonumber\\
&=\left[\begin{array}{cc} D_{2}+c\left(D_{1}\vec{Q}+\vec{v}_{4}\right)_{\mathrm{diag}} & c[\vec{f}+\vec{P}]_{\mathrm{diag}}D_{1}\\
c\left(D_{1}\vec{P}+\vec{v}_{2}\right)_{\mathrm{diag}}+c[\vec{f}+\vec{P}]_{\mathrm{diag}}D_{1} & D_{2}
\end{array}\right].
\end{align}
According to \cref{eq:update}, $\vec{P}$ and $\vec{Q}$ are updated
as
\begin{equation}
\left(\begin{array}{c}
\vec{P}\\
\vec{Q}
\end{array}\right)_{\mathrm{new}}=
\left(\begin{array}{c}
\vec{P}\\
\vec{Q}
\end{array}\right)_{\mathrm{old}}+\left(\begin{array}{c}
\delta\vec{P}\\
\delta\vec{Q}
\end{array}\right)
\end{equation}
after solving the linear equation
\begin{equation}
\boldsymbol{\mathcal J}\left(\begin{array}{c}
\delta\vec{P}\\
\delta\vec{Q}
\end{array}\right)=-\boldsymbol{F}.
\end{equation}

\subsection{Static State}

The static equilibrium state before the pump can be obtained in a similar manner as described in \cref{sec:steady}, 
by setting the parameter $\sigma_p$ to zero. In Fig. \ref{fig:static}, the distribution of $N_d(x)$ is shown 
as the red line, and the distribution of free electrons from ionized K is shown as the black dashed line. The inset shows 
the difference between $N_d(x)$ and $n_0(x)$ near the surface region. The sharp surface peak and the small broad peak in $N_d-n_0$
at $x\approx 40$ \AA \thinspace imply a spatial separation between the K ions and the free electrons, 
and this gives rise to a strong internal electric field
in this region, as shown by the green line ($t=0$) in Fig. \ref{fig:bending_largwin}(c).

\bibliography{biblio}

\begin{thebibliography}{68}%
\makeatletter
\providecommand \@ifxundefined [1]{%
 \@ifx{#1\undefined}
}%
\providecommand \@ifnum [1]{%
 \ifnum #1\expandafter \@firstoftwo
 \else \expandafter \@secondoftwo
 \fi
}%
\providecommand \@ifx [1]{%
 \ifx #1\expandafter \@firstoftwo
 \else \expandafter \@secondoftwo
 \fi
}%
\providecommand \natexlab [1]{#1}%
\providecommand \enquote  [1]{``#1''}%
\providecommand \bibnamefont  [1]{#1}%
\providecommand \bibfnamefont [1]{#1}%
\providecommand \citenamefont [1]{#1}%
\providecommand \href@noop [0]{\@secondoftwo}%
\providecommand \href [0]{\begingroup \@sanitize@url \@href}%
\providecommand \@href[1]{\@@startlink{#1}\@@href}%
\providecommand \@@href[1]{\endgroup#1\@@endlink}%
\providecommand \@sanitize@url [0]{\catcode `\\12\catcode `\$12\catcode
  `\&12\catcode `\#12\catcode `\^12\catcode `\_12\catcode `\%12\relax}%
\providecommand \@@startlink[1]{}%
\providecommand \@@endlink[0]{}%
\providecommand \url  [0]{\begingroup\@sanitize@url \@url }%
\providecommand \@url [1]{\endgroup\@href {#1}{\urlprefix }}%
\providecommand \urlprefix  [0]{URL }%
\providecommand \Eprint [0]{\href }%
\providecommand \doibase [0]{http://dx.doi.org/}%
\providecommand \selectlanguage [0]{\@gobble}%
\providecommand \bibinfo  [0]{\@secondoftwo}%
\providecommand \bibfield  [0]{\@secondoftwo}%
\providecommand \translation [1]{[#1]}%
\providecommand \BibitemOpen [0]{}%
\providecommand \bibitemStop [0]{}%
\providecommand \bibitemNoStop [0]{.\EOS\space}%
\providecommand \EOS [0]{\spacefactor3000\relax}%
\providecommand \BibitemShut  [1]{\csname bibitem#1\endcsname}%
\let\auto@bib@innerbib\@empty
%</preamble>
\bibitem [{\citenamefont {Chaves}\ \emph {et~al.}(2020)\citenamefont {Chaves},
  \citenamefont {Azadani}, \citenamefont {Alsalman}, \citenamefont {da~Costa},
  \citenamefont {Frisenda}, \citenamefont {Chaves}, \citenamefont {Song},
  \citenamefont {Kim}, \citenamefont {He}, \citenamefont {Zhou} \emph
  {et~al.}}]{chaves2020bandgap}%
  \BibitemOpen
  \bibfield  {author} {\bibinfo {author} {\bibfnamefont {A.}~\bibnamefont
  {Chaves}}, \bibinfo {author} {\bibfnamefont {J.}~\bibnamefont {Azadani}},
  \bibinfo {author} {\bibfnamefont {H.}~\bibnamefont {Alsalman}}, \bibinfo
  {author} {\bibfnamefont {D.}~\bibnamefont {da~Costa}}, \bibinfo {author}
  {\bibfnamefont {R.}~\bibnamefont {Frisenda}}, \bibinfo {author}
  {\bibfnamefont {A.}~\bibnamefont {Chaves}}, \bibinfo {author} {\bibfnamefont
  {S.~H.}\ \bibnamefont {Song}}, \bibinfo {author} {\bibfnamefont
  {Y.}~\bibnamefont {Kim}}, \bibinfo {author} {\bibfnamefont {D.}~\bibnamefont
  {He}}, \bibinfo {author} {\bibfnamefont {J.}~\bibnamefont {Zhou}},  \emph
  {et~al.},\ }\href@noop {} {\bibfield  {journal} {\bibinfo  {journal} {Npj 2D
  Mater. Appl.}\ }\textbf {\bibinfo {volume} {4}},\ \bibinfo {pages} {1}
  (\bibinfo {year} {2020})}\BibitemShut {NoStop}%
\bibitem [{\citenamefont {Ak}\ and\ \citenamefont
  {Novoselov}(2007)}]{ak2007rise}%
  \BibitemOpen
  \bibfield  {author} {\bibinfo {author} {\bibfnamefont {G.}~\bibnamefont
  {Ak}}\ and\ \bibinfo {author} {\bibfnamefont {K.}~\bibnamefont {Novoselov}},\
  }\href@noop {} {\bibfield  {journal} {\bibinfo  {journal} {Nat Mater}\
  }\textbf {\bibinfo {volume} {6}},\ \bibinfo {pages} {183} (\bibinfo {year}
  {2007})}\BibitemShut {NoStop}%
\bibitem [{\citenamefont {Vogt}\ \emph {et~al.}(2012)\citenamefont {Vogt},
  \citenamefont {De~Padova}, \citenamefont {Quaresima}, \citenamefont {Avila},
  \citenamefont {Frantzeskakis}, \citenamefont {Asensio}, \citenamefont
  {Resta}, \citenamefont {Ealet},\ and\ \citenamefont
  {Le~Lay}}]{vogt2012silicene}%
  \BibitemOpen
  \bibfield  {author} {\bibinfo {author} {\bibfnamefont {P.}~\bibnamefont
  {Vogt}}, \bibinfo {author} {\bibfnamefont {P.}~\bibnamefont {De~Padova}},
  \bibinfo {author} {\bibfnamefont {C.}~\bibnamefont {Quaresima}}, \bibinfo
  {author} {\bibfnamefont {J.}~\bibnamefont {Avila}}, \bibinfo {author}
  {\bibfnamefont {E.}~\bibnamefont {Frantzeskakis}}, \bibinfo {author}
  {\bibfnamefont {M.~C.}\ \bibnamefont {Asensio}}, \bibinfo {author}
  {\bibfnamefont {A.}~\bibnamefont {Resta}}, \bibinfo {author} {\bibfnamefont
  {B.}~\bibnamefont {Ealet}}, \ and\ \bibinfo {author} {\bibfnamefont
  {G.}~\bibnamefont {Le~Lay}},\ }\href@noop {} {\bibfield  {journal} {\bibinfo
  {journal} {Phys. Rev. Lett.}\ }\textbf {\bibinfo {volume} {108}},\ \bibinfo
  {pages} {155501} (\bibinfo {year} {2012})}\BibitemShut {NoStop}%
\bibitem [{\citenamefont {Li}\ \emph {et~al.}(2014{\natexlab{a}})\citenamefont
  {Li}, \citenamefont {Lu}, \citenamefont {Pan}, \citenamefont {Qin},
  \citenamefont {Wang}, \citenamefont {Wang}, \citenamefont {Cao},
  \citenamefont {Du},\ and\ \citenamefont {Gao}}]{li2014buckled}%
  \BibitemOpen
  \bibfield  {author} {\bibinfo {author} {\bibfnamefont {L.}~\bibnamefont
  {Li}}, \bibinfo {author} {\bibfnamefont {S.-z.}\ \bibnamefont {Lu}}, \bibinfo
  {author} {\bibfnamefont {J.}~\bibnamefont {Pan}}, \bibinfo {author}
  {\bibfnamefont {Z.}~\bibnamefont {Qin}}, \bibinfo {author} {\bibfnamefont
  {Y.-q.}\ \bibnamefont {Wang}}, \bibinfo {author} {\bibfnamefont
  {Y.}~\bibnamefont {Wang}}, \bibinfo {author} {\bibfnamefont {G.-y.}\
  \bibnamefont {Cao}}, \bibinfo {author} {\bibfnamefont {S.}~\bibnamefont
  {Du}}, \ and\ \bibinfo {author} {\bibfnamefont {H.-J.}\ \bibnamefont {Gao}},\
  }\href@noop {} {\bibfield  {journal} {\bibinfo  {journal} {Adv. Mater.}\
  }\textbf {\bibinfo {volume} {26}},\ \bibinfo {pages} {4820} (\bibinfo {year}
  {2014}{\natexlab{a}})}\BibitemShut {NoStop}%
\bibitem [{\citenamefont {Ji}\ \emph {et~al.}(2016)\citenamefont {Ji},
  \citenamefont {Song}, \citenamefont {Liu}, \citenamefont {Yan}, \citenamefont
  {Huo}, \citenamefont {Zhang}, \citenamefont {Su}, \citenamefont {Liao},
  \citenamefont {Wang}, \citenamefont {Ni} \emph {et~al.}}]{ji2016two}%
  \BibitemOpen
  \bibfield  {author} {\bibinfo {author} {\bibfnamefont {J.}~\bibnamefont
  {Ji}}, \bibinfo {author} {\bibfnamefont {X.}~\bibnamefont {Song}}, \bibinfo
  {author} {\bibfnamefont {J.}~\bibnamefont {Liu}}, \bibinfo {author}
  {\bibfnamefont {Z.}~\bibnamefont {Yan}}, \bibinfo {author} {\bibfnamefont
  {C.}~\bibnamefont {Huo}}, \bibinfo {author} {\bibfnamefont {S.}~\bibnamefont
  {Zhang}}, \bibinfo {author} {\bibfnamefont {M.}~\bibnamefont {Su}}, \bibinfo
  {author} {\bibfnamefont {L.}~\bibnamefont {Liao}}, \bibinfo {author}
  {\bibfnamefont {W.}~\bibnamefont {Wang}}, \bibinfo {author} {\bibfnamefont
  {Z.}~\bibnamefont {Ni}},  \emph {et~al.},\ }\href@noop {} {\bibfield
  {journal} {\bibinfo  {journal} {Nat. Commun.}\ }\textbf {\bibinfo {volume}
  {7}},\ \bibinfo {pages} {1} (\bibinfo {year} {2016})}\BibitemShut {NoStop}%
\bibitem [{\citenamefont {Zhu}\ \emph {et~al.}(2017)\citenamefont {Zhu},
  \citenamefont {Cai}, \citenamefont {Yi}, \citenamefont {Chen}, \citenamefont
  {Dai}, \citenamefont {Niu}, \citenamefont {Guo}, \citenamefont {Xie},
  \citenamefont {Liu}, \citenamefont {Cho} \emph
  {et~al.}}]{zhu2017multivalency}%
  \BibitemOpen
  \bibfield  {author} {\bibinfo {author} {\bibfnamefont {Z.}~\bibnamefont
  {Zhu}}, \bibinfo {author} {\bibfnamefont {X.}~\bibnamefont {Cai}}, \bibinfo
  {author} {\bibfnamefont {S.}~\bibnamefont {Yi}}, \bibinfo {author}
  {\bibfnamefont {J.}~\bibnamefont {Chen}}, \bibinfo {author} {\bibfnamefont
  {Y.}~\bibnamefont {Dai}}, \bibinfo {author} {\bibfnamefont {C.}~\bibnamefont
  {Niu}}, \bibinfo {author} {\bibfnamefont {Z.}~\bibnamefont {Guo}}, \bibinfo
  {author} {\bibfnamefont {M.}~\bibnamefont {Xie}}, \bibinfo {author}
  {\bibfnamefont {F.}~\bibnamefont {Liu}}, \bibinfo {author} {\bibfnamefont
  {J.-H.}\ \bibnamefont {Cho}},  \emph {et~al.},\ }\href@noop {} {\bibfield
  {journal} {\bibinfo  {journal} {Phys. Rev. Lett.}\ }\textbf {\bibinfo
  {volume} {119}},\ \bibinfo {pages} {106101} (\bibinfo {year}
  {2017})}\BibitemShut {NoStop}%
\bibitem [{\citenamefont {Kochat}\ \emph {et~al.}(2018)\citenamefont {Kochat},
  \citenamefont {Samanta}, \citenamefont {Zhang}, \citenamefont {Bhowmick},
  \citenamefont {Manimunda}, \citenamefont {Asif}, \citenamefont {Stender},
  \citenamefont {Vajtai}, \citenamefont {Singh}, \citenamefont {Tiwary} \emph
  {et~al.}}]{kochat2018atomically}%
  \BibitemOpen
  \bibfield  {author} {\bibinfo {author} {\bibfnamefont {V.}~\bibnamefont
  {Kochat}}, \bibinfo {author} {\bibfnamefont {A.}~\bibnamefont {Samanta}},
  \bibinfo {author} {\bibfnamefont {Y.}~\bibnamefont {Zhang}}, \bibinfo
  {author} {\bibfnamefont {S.}~\bibnamefont {Bhowmick}}, \bibinfo {author}
  {\bibfnamefont {P.}~\bibnamefont {Manimunda}}, \bibinfo {author}
  {\bibfnamefont {S.~A.~S.}\ \bibnamefont {Asif}}, \bibinfo {author}
  {\bibfnamefont {A.~S.}\ \bibnamefont {Stender}}, \bibinfo {author}
  {\bibfnamefont {R.}~\bibnamefont {Vajtai}}, \bibinfo {author} {\bibfnamefont
  {A.~K.}\ \bibnamefont {Singh}}, \bibinfo {author} {\bibfnamefont {C.~S.}\
  \bibnamefont {Tiwary}},  \emph {et~al.},\ }\href@noop {} {\bibfield
  {journal} {\bibinfo  {journal} {Sci. Adv.}\ }\textbf {\bibinfo {volume}
  {4}},\ \bibinfo {pages} {e1701373} (\bibinfo {year} {2018})}\BibitemShut
  {NoStop}%
\bibitem [{\citenamefont {Deng}\ \emph {et~al.}(2018)\citenamefont {Deng},
  \citenamefont {Xia}, \citenamefont {Ma}, \citenamefont {Chen}, \citenamefont
  {Shan}, \citenamefont {Zhai}, \citenamefont {Li}, \citenamefont {Zhao},
  \citenamefont {Xu}, \citenamefont {Duan} \emph {et~al.}}]{deng2018epitaxial}%
  \BibitemOpen
  \bibfield  {author} {\bibinfo {author} {\bibfnamefont {J.}~\bibnamefont
  {Deng}}, \bibinfo {author} {\bibfnamefont {B.}~\bibnamefont {Xia}}, \bibinfo
  {author} {\bibfnamefont {X.}~\bibnamefont {Ma}}, \bibinfo {author}
  {\bibfnamefont {H.}~\bibnamefont {Chen}}, \bibinfo {author} {\bibfnamefont
  {H.}~\bibnamefont {Shan}}, \bibinfo {author} {\bibfnamefont {X.}~\bibnamefont
  {Zhai}}, \bibinfo {author} {\bibfnamefont {B.}~\bibnamefont {Li}}, \bibinfo
  {author} {\bibfnamefont {A.}~\bibnamefont {Zhao}}, \bibinfo {author}
  {\bibfnamefont {Y.}~\bibnamefont {Xu}}, \bibinfo {author} {\bibfnamefont
  {W.}~\bibnamefont {Duan}},  \emph {et~al.},\ }\href@noop {} {\bibfield
  {journal} {\bibinfo  {journal} {Nat. Mater.}\ }\textbf {\bibinfo {volume}
  {17}},\ \bibinfo {pages} {1081} (\bibinfo {year} {2018})}\BibitemShut
  {NoStop}%
\bibitem [{\citenamefont {Wu}\ \emph {et~al.}(2019)\citenamefont {Wu},
  \citenamefont {Drozdov}, \citenamefont {Eltinge}, \citenamefont {Zahl},
  \citenamefont {Ismail-Beigi}, \citenamefont {Bo{\v{z}}ovi{\'c}},\ and\
  \citenamefont {Gozar}}]{wu2019large}%
  \BibitemOpen
  \bibfield  {author} {\bibinfo {author} {\bibfnamefont {R.}~\bibnamefont
  {Wu}}, \bibinfo {author} {\bibfnamefont {I.~K.}\ \bibnamefont {Drozdov}},
  \bibinfo {author} {\bibfnamefont {S.}~\bibnamefont {Eltinge}}, \bibinfo
  {author} {\bibfnamefont {P.}~\bibnamefont {Zahl}}, \bibinfo {author}
  {\bibfnamefont {S.}~\bibnamefont {Ismail-Beigi}}, \bibinfo {author}
  {\bibfnamefont {I.}~\bibnamefont {Bo{\v{z}}ovi{\'c}}}, \ and\ \bibinfo
  {author} {\bibfnamefont {A.}~\bibnamefont {Gozar}},\ }\href@noop {}
  {\bibfield  {journal} {\bibinfo  {journal} {Nat. Nanotechnol.}\ }\textbf
  {\bibinfo {volume} {14}},\ \bibinfo {pages} {44} (\bibinfo {year}
  {2019})}\BibitemShut {NoStop}%
\bibitem [{\citenamefont {Bihlmayer}\ \emph {et~al.}(2020)\citenamefont
  {Bihlmayer}, \citenamefont {Sassmannshausen}, \citenamefont {Kubetzka},
  \citenamefont {Bl{\"u}gel}, \citenamefont {von Bergmann},\ and\ \citenamefont
  {Wiesendanger}}]{bihlmayer2020plumbene}%
  \BibitemOpen
  \bibfield  {author} {\bibinfo {author} {\bibfnamefont {G.}~\bibnamefont
  {Bihlmayer}}, \bibinfo {author} {\bibfnamefont {J.}~\bibnamefont
  {Sassmannshausen}}, \bibinfo {author} {\bibfnamefont {A.}~\bibnamefont
  {Kubetzka}}, \bibinfo {author} {\bibfnamefont {S.}~\bibnamefont
  {Bl{\"u}gel}}, \bibinfo {author} {\bibfnamefont {K.}~\bibnamefont {von
  Bergmann}}, \ and\ \bibinfo {author} {\bibfnamefont {R.}~\bibnamefont
  {Wiesendanger}},\ }\href@noop {} {\bibfield  {journal} {\bibinfo  {journal}
  {Phys. Rev. Lett.}\ }\textbf {\bibinfo {volume} {124}},\ \bibinfo {pages}
  {126401} (\bibinfo {year} {2020})}\BibitemShut {NoStop}%
\bibitem [{\citenamefont {Manzeli}\ \emph {et~al.}(2017)\citenamefont
  {Manzeli}, \citenamefont {Ovchinnikov}, \citenamefont {Pasquier},
  \citenamefont {Yazyev},\ and\ \citenamefont {Kis}}]{manzeli20172d}%
  \BibitemOpen
  \bibfield  {author} {\bibinfo {author} {\bibfnamefont {S.}~\bibnamefont
  {Manzeli}}, \bibinfo {author} {\bibfnamefont {D.}~\bibnamefont
  {Ovchinnikov}}, \bibinfo {author} {\bibfnamefont {D.}~\bibnamefont
  {Pasquier}}, \bibinfo {author} {\bibfnamefont {O.~V.}\ \bibnamefont
  {Yazyev}}, \ and\ \bibinfo {author} {\bibfnamefont {A.}~\bibnamefont {Kis}},\
  }\href@noop {} {\bibfield  {journal} {\bibinfo  {journal} {Nat. Rev. Mater.}\
  }\textbf {\bibinfo {volume} {2}},\ \bibinfo {pages} {17033} (\bibinfo {year}
  {2017})}\BibitemShut {NoStop}%
\bibitem [{\citenamefont {Elias}\ \emph {et~al.}(2019)\citenamefont {Elias},
  \citenamefont {Valvin}, \citenamefont {Pelini}, \citenamefont {Summerfield},
  \citenamefont {Mellor}, \citenamefont {Cheng}, \citenamefont {Eaves},
  \citenamefont {Foxon}, \citenamefont {Beton}, \citenamefont {Novikov} \emph
  {et~al.}}]{elias2019direct}%
  \BibitemOpen
  \bibfield  {author} {\bibinfo {author} {\bibfnamefont {C.}~\bibnamefont
  {Elias}}, \bibinfo {author} {\bibfnamefont {P.}~\bibnamefont {Valvin}},
  \bibinfo {author} {\bibfnamefont {T.}~\bibnamefont {Pelini}}, \bibinfo
  {author} {\bibfnamefont {A.}~\bibnamefont {Summerfield}}, \bibinfo {author}
  {\bibfnamefont {C.}~\bibnamefont {Mellor}}, \bibinfo {author} {\bibfnamefont
  {T.}~\bibnamefont {Cheng}}, \bibinfo {author} {\bibfnamefont
  {L.}~\bibnamefont {Eaves}}, \bibinfo {author} {\bibfnamefont
  {C.}~\bibnamefont {Foxon}}, \bibinfo {author} {\bibfnamefont
  {P.}~\bibnamefont {Beton}}, \bibinfo {author} {\bibfnamefont
  {S.}~\bibnamefont {Novikov}},  \emph {et~al.},\ }\href@noop {} {\bibfield
  {journal} {\bibinfo  {journal} {Nat. Commun.}\ }\textbf {\bibinfo {volume}
  {10}},\ \bibinfo {pages} {1} (\bibinfo {year} {2019})}\BibitemShut {NoStop}%
\bibitem [{\citenamefont {Zhang}\ \emph {et~al.}(2021)\citenamefont {Zhang},
  \citenamefont {Holbrook}, \citenamefont {Cheng}, \citenamefont {Nam},
  \citenamefont {Liu}, \citenamefont {Pan}, \citenamefont {West}, \citenamefont
  {Zhang}, \citenamefont {Chou},\ and\ \citenamefont
  {Shih}}]{zhang2020epitaxial}%
  \BibitemOpen
  \bibfield  {author} {\bibinfo {author} {\bibfnamefont {H.}~\bibnamefont
  {Zhang}}, \bibinfo {author} {\bibfnamefont {M.}~\bibnamefont {Holbrook}},
  \bibinfo {author} {\bibfnamefont {F.}~\bibnamefont {Cheng}}, \bibinfo
  {author} {\bibfnamefont {H.}~\bibnamefont {Nam}}, \bibinfo {author}
  {\bibfnamefont {M.}~\bibnamefont {Liu}}, \bibinfo {author} {\bibfnamefont
  {C.-R.}\ \bibnamefont {Pan}}, \bibinfo {author} {\bibfnamefont
  {D.}~\bibnamefont {West}}, \bibinfo {author} {\bibfnamefont {S.}~\bibnamefont
  {Zhang}}, \bibinfo {author} {\bibfnamefont {M.-Y.}\ \bibnamefont {Chou}}, \
  and\ \bibinfo {author} {\bibfnamefont {C.-K.}\ \bibnamefont {Shih}},\
  }\href@noop {} {\bibfield  {journal} {\bibinfo  {journal} {ACS Nano}\
  }\textbf {\bibinfo {volume} {15}},\ \bibinfo {pages} {2497} (\bibinfo {year}
  {2021})}\BibitemShut {NoStop}%
\bibitem [{\citenamefont {Lichtenstein}\ \emph {et~al.}(2012)\citenamefont
  {Lichtenstein}, \citenamefont {Heyde}, \citenamefont {Ulrich}, \citenamefont
  {Nilius},\ and\ \citenamefont {Freund}}]{lichtenstein2012probing}%
  \BibitemOpen
  \bibfield  {author} {\bibinfo {author} {\bibfnamefont {L.}~\bibnamefont
  {Lichtenstein}}, \bibinfo {author} {\bibfnamefont {M.}~\bibnamefont {Heyde}},
  \bibinfo {author} {\bibfnamefont {S.}~\bibnamefont {Ulrich}}, \bibinfo
  {author} {\bibfnamefont {N.}~\bibnamefont {Nilius}}, \ and\ \bibinfo {author}
  {\bibfnamefont {H.-J.}\ \bibnamefont {Freund}},\ }\href@noop {} {\bibfield
  {journal} {\bibinfo  {journal} {J. Phys. Condens. Mat.}\ }\textbf {\bibinfo
  {volume} {24}},\ \bibinfo {pages} {354010} (\bibinfo {year}
  {2012})}\BibitemShut {NoStop}%
\bibitem [{\citenamefont {Kremer}\ \emph {et~al.}(2019)\citenamefont {Kremer},
  \citenamefont {Alvarez~Quiceno}, \citenamefont {Lisi}, \citenamefont
  {Pierron}, \citenamefont {Gonz{\'a}lez}, \citenamefont {Sicot}, \citenamefont
  {Kierren}, \citenamefont {Malterre}, \citenamefont {Rault}, \citenamefont
  {Le~F{\`e}vre}, \citenamefont {Bertran}, \citenamefont {Dappe}, \citenamefont
  {Coraux}, \citenamefont {Pochet},\ and\ \citenamefont
  {Fagot-Revurat}}]{kremer2019electronic}%
  \BibitemOpen
  \bibfield  {author} {\bibinfo {author} {\bibfnamefont {G.}~\bibnamefont
  {Kremer}}, \bibinfo {author} {\bibfnamefont {J.~C.}\ \bibnamefont
  {Alvarez~Quiceno}}, \bibinfo {author} {\bibfnamefont {S.}~\bibnamefont
  {Lisi}}, \bibinfo {author} {\bibfnamefont {T.}~\bibnamefont {Pierron}},
  \bibinfo {author} {\bibfnamefont {C.}~\bibnamefont {Gonz{\'a}lez}}, \bibinfo
  {author} {\bibfnamefont {M.}~\bibnamefont {Sicot}}, \bibinfo {author}
  {\bibfnamefont {B.}~\bibnamefont {Kierren}}, \bibinfo {author} {\bibfnamefont
  {D.}~\bibnamefont {Malterre}}, \bibinfo {author} {\bibfnamefont {J.~E.}\
  \bibnamefont {Rault}}, \bibinfo {author} {\bibfnamefont {P.}~\bibnamefont
  {Le~F{\`e}vre}}, \bibinfo {author} {\bibfnamefont {F.}~\bibnamefont
  {Bertran}}, \bibinfo {author} {\bibfnamefont {Y.~J.}\ \bibnamefont {Dappe}},
  \bibinfo {author} {\bibfnamefont {J.}~\bibnamefont {Coraux}}, \bibinfo
  {author} {\bibfnamefont {P.}~\bibnamefont {Pochet}}, \ and\ \bibinfo {author}
  {\bibfnamefont {Y.}~\bibnamefont {Fagot-Revurat}},\ }\href@noop {} {\bibfield
   {journal} {\bibinfo  {journal} {ACS Nano}\ }\textbf {\bibinfo {volume}
  {13}},\ \bibinfo {pages} {4720} (\bibinfo {year} {2019})}\BibitemShut
  {NoStop}%
\bibitem [{\citenamefont {Kremer}\ \emph {et~al.}(2021)\citenamefont {Kremer},
  \citenamefont {Alvarez~Quiceno}, \citenamefont {Pierron}, \citenamefont
  {Gonz{\'a}lez}, \citenamefont {Sicot}, \citenamefont {Kierren}, \citenamefont
  {Moreau}, \citenamefont {Rault}, \citenamefont {Le~F{\`e}vre}, \citenamefont
  {Bertran}, \citenamefont {Dappe}, \citenamefont {Coraux}, \citenamefont
  {Pochet},\ and\ \citenamefont {Fagot-Revurat}}]{kremer2020dispersing}%
  \BibitemOpen
  \bibfield  {author} {\bibinfo {author} {\bibfnamefont {G.}~\bibnamefont
  {Kremer}}, \bibinfo {author} {\bibfnamefont {J.~C.}\ \bibnamefont
  {Alvarez~Quiceno}}, \bibinfo {author} {\bibfnamefont {T.}~\bibnamefont
  {Pierron}}, \bibinfo {author} {\bibfnamefont {C.}~\bibnamefont
  {Gonz{\'a}lez}}, \bibinfo {author} {\bibfnamefont {M.}~\bibnamefont {Sicot}},
  \bibinfo {author} {\bibfnamefont {B.}~\bibnamefont {Kierren}}, \bibinfo
  {author} {\bibfnamefont {L.}~\bibnamefont {Moreau}}, \bibinfo {author}
  {\bibfnamefont {J.~E.}\ \bibnamefont {Rault}}, \bibinfo {author}
  {\bibfnamefont {P.}~\bibnamefont {Le~F{\`e}vre}}, \bibinfo {author}
  {\bibfnamefont {F.}~\bibnamefont {Bertran}}, \bibinfo {author} {\bibfnamefont
  {Y.~J.}\ \bibnamefont {Dappe}}, \bibinfo {author} {\bibfnamefont
  {J.}~\bibnamefont {Coraux}}, \bibinfo {author} {\bibfnamefont
  {P.}~\bibnamefont {Pochet}}, \ and\ \bibinfo {author} {\bibfnamefont
  {Y.}~\bibnamefont {Fagot-Revurat}},\ }\href
  {http://iopscience.iop.org/article/10.1088/2053-1583/abf715} {\bibfield
  {journal} {\bibinfo  {journal} {2D Mater.}\ } (\bibinfo {year}
  {2021})}\BibitemShut {NoStop}%
\bibitem [{\citenamefont {Guo}\ \emph {et~al.}(2014)\citenamefont {Guo},
  \citenamefont {Lu}, \citenamefont {Wang}, \citenamefont {Wu},\ and\
  \citenamefont {Zeng}}]{guo2014tuning}%
  \BibitemOpen
  \bibfield  {author} {\bibinfo {author} {\bibfnamefont {H.}~\bibnamefont
  {Guo}}, \bibinfo {author} {\bibfnamefont {N.}~\bibnamefont {Lu}}, \bibinfo
  {author} {\bibfnamefont {L.}~\bibnamefont {Wang}}, \bibinfo {author}
  {\bibfnamefont {X.}~\bibnamefont {Wu}}, \ and\ \bibinfo {author}
  {\bibfnamefont {X.~C.}\ \bibnamefont {Zeng}},\ }\href@noop {} {\bibfield
  {journal} {\bibinfo  {journal} {J. Phys. Chem. C}\ }\textbf {\bibinfo
  {volume} {118}},\ \bibinfo {pages} {7242} (\bibinfo {year}
  {2014})}\BibitemShut {NoStop}%
\bibitem [{\citenamefont {Rodin}\ \emph {et~al.}(2014)\citenamefont {Rodin},
  \citenamefont {Carvalho},\ and\ \citenamefont
  {Castro~Neto}}]{rodin2014strain}%
  \BibitemOpen
  \bibfield  {author} {\bibinfo {author} {\bibfnamefont {A.~S.}\ \bibnamefont
  {Rodin}}, \bibinfo {author} {\bibfnamefont {A.}~\bibnamefont {Carvalho}}, \
  and\ \bibinfo {author} {\bibfnamefont {A.~H.}\ \bibnamefont {Castro~Neto}},\
  }\href@noop {} {\bibfield  {journal} {\bibinfo  {journal} {Phys. Rev. Lett.}\
  }\textbf {\bibinfo {volume} {112}},\ \bibinfo {pages} {176801} (\bibinfo
  {year} {2014})}\BibitemShut {NoStop}%
\bibitem [{\citenamefont {Ramasubramaniam}\ \emph {et~al.}(2011)\citenamefont
  {Ramasubramaniam}, \citenamefont {Naveh},\ and\ \citenamefont
  {Towe}}]{ramasubramaniam2011tunable}%
  \BibitemOpen
  \bibfield  {author} {\bibinfo {author} {\bibfnamefont {A.}~\bibnamefont
  {Ramasubramaniam}}, \bibinfo {author} {\bibfnamefont {D.}~\bibnamefont
  {Naveh}}, \ and\ \bibinfo {author} {\bibfnamefont {E.}~\bibnamefont {Towe}},\
  }\href@noop {} {\bibfield  {journal} {\bibinfo  {journal} {Phys. Rev. B}\
  }\textbf {\bibinfo {volume} {84}},\ \bibinfo {pages} {205325} (\bibinfo
  {year} {2011})}\BibitemShut {NoStop}%
\bibitem [{\citenamefont {Kang}\ \emph {et~al.}(2017)\citenamefont {Kang},
  \citenamefont {Kim}, \citenamefont {Ryu}, \citenamefont {Jung}, \citenamefont
  {Kim}, \citenamefont {Moreschini}, \citenamefont {Jozwiak}, \citenamefont
  {Rotenberg}, \citenamefont {Bostwick},\ and\ \citenamefont
  {Kim}}]{kang2017universal}%
  \BibitemOpen
  \bibfield  {author} {\bibinfo {author} {\bibfnamefont {M.}~\bibnamefont
  {Kang}}, \bibinfo {author} {\bibfnamefont {B.}~\bibnamefont {Kim}}, \bibinfo
  {author} {\bibfnamefont {S.~H.}\ \bibnamefont {Ryu}}, \bibinfo {author}
  {\bibfnamefont {S.~W.}\ \bibnamefont {Jung}}, \bibinfo {author}
  {\bibfnamefont {J.}~\bibnamefont {Kim}}, \bibinfo {author} {\bibfnamefont
  {L.}~\bibnamefont {Moreschini}}, \bibinfo {author} {\bibfnamefont
  {C.}~\bibnamefont {Jozwiak}}, \bibinfo {author} {\bibfnamefont
  {E.}~\bibnamefont {Rotenberg}}, \bibinfo {author} {\bibfnamefont
  {A.}~\bibnamefont {Bostwick}}, \ and\ \bibinfo {author} {\bibfnamefont
  {K.~S.}\ \bibnamefont {Kim}},\ }\href@noop {} {\bibfield  {journal} {\bibinfo
   {journal} {Nano Lett.}\ }\textbf {\bibinfo {volume} {17}},\ \bibinfo {pages}
  {1610} (\bibinfo {year} {2017})}\BibitemShut {NoStop}%
\bibitem [{\citenamefont {Gong}\ \emph {et~al.}(2014)\citenamefont {Gong},
  \citenamefont {Liu}, \citenamefont {Lupini}, \citenamefont {Shi},
  \citenamefont {Lin}, \citenamefont {Najmaei}, \citenamefont {Lin},
  \citenamefont {El{\'\i}as}, \citenamefont {Berkdemir}, \citenamefont {You}
  \emph {et~al.}}]{gong2014band}%
  \BibitemOpen
  \bibfield  {author} {\bibinfo {author} {\bibfnamefont {Y.}~\bibnamefont
  {Gong}}, \bibinfo {author} {\bibfnamefont {Z.}~\bibnamefont {Liu}}, \bibinfo
  {author} {\bibfnamefont {A.~R.}\ \bibnamefont {Lupini}}, \bibinfo {author}
  {\bibfnamefont {G.}~\bibnamefont {Shi}}, \bibinfo {author} {\bibfnamefont
  {J.}~\bibnamefont {Lin}}, \bibinfo {author} {\bibfnamefont {S.}~\bibnamefont
  {Najmaei}}, \bibinfo {author} {\bibfnamefont {Z.}~\bibnamefont {Lin}},
  \bibinfo {author} {\bibfnamefont {A.~L.}\ \bibnamefont {El{\'\i}as}},
  \bibinfo {author} {\bibfnamefont {A.}~\bibnamefont {Berkdemir}}, \bibinfo
  {author} {\bibfnamefont {G.}~\bibnamefont {You}},  \emph {et~al.},\
  }\href@noop {} {\bibfield  {journal} {\bibinfo  {journal} {Nano Lett.}\
  }\textbf {\bibinfo {volume} {14}},\ \bibinfo {pages} {442} (\bibinfo {year}
  {2014})}\BibitemShut {NoStop}%
\bibitem [{\citenamefont {Ma}\ \emph {et~al.}(2014)\citenamefont {Ma},
  \citenamefont {Isarraraz}, \citenamefont {Wang}, \citenamefont {Preciado},
  \citenamefont {Klee}, \citenamefont {Bobek}, \citenamefont {Yamaguchi},
  \citenamefont {Li}, \citenamefont {Odenthal}, \citenamefont {Nguyen} \emph
  {et~al.}}]{ma2014postgrowth}%
  \BibitemOpen
  \bibfield  {author} {\bibinfo {author} {\bibfnamefont {Q.}~\bibnamefont
  {Ma}}, \bibinfo {author} {\bibfnamefont {M.}~\bibnamefont {Isarraraz}},
  \bibinfo {author} {\bibfnamefont {C.~S.}\ \bibnamefont {Wang}}, \bibinfo
  {author} {\bibfnamefont {E.}~\bibnamefont {Preciado}}, \bibinfo {author}
  {\bibfnamefont {V.}~\bibnamefont {Klee}}, \bibinfo {author} {\bibfnamefont
  {S.}~\bibnamefont {Bobek}}, \bibinfo {author} {\bibfnamefont
  {K.}~\bibnamefont {Yamaguchi}}, \bibinfo {author} {\bibfnamefont
  {E.}~\bibnamefont {Li}}, \bibinfo {author} {\bibfnamefont {P.~M.}\
  \bibnamefont {Odenthal}}, \bibinfo {author} {\bibfnamefont {A.}~\bibnamefont
  {Nguyen}},  \emph {et~al.},\ }\href@noop {} {\bibfield  {journal} {\bibinfo
  {journal} {ACS Nano}\ }\textbf {\bibinfo {volume} {8}},\ \bibinfo {pages}
  {4672} (\bibinfo {year} {2014})}\BibitemShut {NoStop}%
\bibitem [{\citenamefont {Villaos}\ \emph {et~al.}(2019)\citenamefont
  {Villaos}, \citenamefont {Crisostomo}, \citenamefont {Huang}, \citenamefont
  {Huang}, \citenamefont {Padama}, \citenamefont {Albao}, \citenamefont {Lin},\
  and\ \citenamefont {Chuang}}]{villaos2019thickness}%
  \BibitemOpen
  \bibfield  {author} {\bibinfo {author} {\bibfnamefont {R.~A.~B.}\
  \bibnamefont {Villaos}}, \bibinfo {author} {\bibfnamefont {C.~P.}\
  \bibnamefont {Crisostomo}}, \bibinfo {author} {\bibfnamefont {Z.-Q.}\
  \bibnamefont {Huang}}, \bibinfo {author} {\bibfnamefont {S.-M.}\ \bibnamefont
  {Huang}}, \bibinfo {author} {\bibfnamefont {A.~A.~B.}\ \bibnamefont
  {Padama}}, \bibinfo {author} {\bibfnamefont {M.~A.}\ \bibnamefont {Albao}},
  \bibinfo {author} {\bibfnamefont {H.}~\bibnamefont {Lin}}, \ and\ \bibinfo
  {author} {\bibfnamefont {F.-C.}\ \bibnamefont {Chuang}},\ }\href@noop {}
  {\bibfield  {journal} {\bibinfo  {journal} {Npj 2D Mater. Appl.}\ }\textbf
  {\bibinfo {volume} {3}},\ \bibinfo {pages} {1} (\bibinfo {year}
  {2019})}\BibitemShut {NoStop}%
\bibitem [{\citenamefont {Lin}\ \emph {et~al.}(2020)\citenamefont {Lin},
  \citenamefont {Villaos}, \citenamefont {Hlevyack}, \citenamefont {Chen},
  \citenamefont {Liu}, \citenamefont {Hsu}, \citenamefont {Avila},
  \citenamefont {Mo}, \citenamefont {Chuang},\ and\ \citenamefont
  {Chiang}}]{lin2020dimensionality}%
  \BibitemOpen
  \bibfield  {author} {\bibinfo {author} {\bibfnamefont {M.-K.}\ \bibnamefont
  {Lin}}, \bibinfo {author} {\bibfnamefont {R.~A.~B.}\ \bibnamefont {Villaos}},
  \bibinfo {author} {\bibfnamefont {J.~A.}\ \bibnamefont {Hlevyack}}, \bibinfo
  {author} {\bibfnamefont {P.}~\bibnamefont {Chen}}, \bibinfo {author}
  {\bibfnamefont {R.-Y.}\ \bibnamefont {Liu}}, \bibinfo {author} {\bibfnamefont
  {C.-H.}\ \bibnamefont {Hsu}}, \bibinfo {author} {\bibfnamefont
  {J.}~\bibnamefont {Avila}}, \bibinfo {author} {\bibfnamefont {S.-K.}\
  \bibnamefont {Mo}}, \bibinfo {author} {\bibfnamefont {F.-C.}\ \bibnamefont
  {Chuang}}, \ and\ \bibinfo {author} {\bibfnamefont {T.-C.}\ \bibnamefont
  {Chiang}},\ }\href@noop {} {\bibfield  {journal} {\bibinfo  {journal} {Phys.
  Rev. Lett.}\ }\textbf {\bibinfo {volume} {124}},\ \bibinfo {pages} {036402}
  (\bibinfo {year} {2020})}\BibitemShut {NoStop}%
\bibitem [{\citenamefont {Novoselov}\ \emph {et~al.}(2016)\citenamefont
  {Novoselov}, \citenamefont {Mishchenko}, \citenamefont {Carvalho},\ and\
  \citenamefont {Neto}}]{novoselov20162d}%
  \BibitemOpen
  \bibfield  {author} {\bibinfo {author} {\bibfnamefont {K.}~\bibnamefont
  {Novoselov}}, \bibinfo {author} {\bibfnamefont {o.~A.}\ \bibnamefont
  {Mishchenko}}, \bibinfo {author} {\bibfnamefont {o.~A.}\ \bibnamefont
  {Carvalho}}, \ and\ \bibinfo {author} {\bibfnamefont {A.~C.}\ \bibnamefont
  {Neto}},\ }\href@noop {} {\bibfield  {journal} {\bibinfo  {journal}
  {Science}\ }\textbf {\bibinfo {volume} {353}} (\bibinfo {year}
  {2016})}\BibitemShut {NoStop}%
\bibitem [{\citenamefont {Morita}(1986)}]{morita1986semiconducting}%
  \BibitemOpen
  \bibfield  {author} {\bibinfo {author} {\bibfnamefont {A.}~\bibnamefont
  {Morita}},\ }\href@noop {} {\bibfield  {journal} {\bibinfo  {journal} {Appl.
  Phys. A}\ }\textbf {\bibinfo {volume} {39}},\ \bibinfo {pages} {227}
  (\bibinfo {year} {1986})}\BibitemShut {NoStop}%
\bibitem [{\citenamefont {Li}\ \emph {et~al.}(2014{\natexlab{b}})\citenamefont
  {Li}, \citenamefont {Yu}, \citenamefont {Ye}, \citenamefont {Ge},
  \citenamefont {Ou}, \citenamefont {Wu}, \citenamefont {Feng}, \citenamefont
  {Chen},\ and\ \citenamefont {Zhang}}]{li2014black}%
  \BibitemOpen
  \bibfield  {author} {\bibinfo {author} {\bibfnamefont {L.}~\bibnamefont
  {Li}}, \bibinfo {author} {\bibfnamefont {Y.}~\bibnamefont {Yu}}, \bibinfo
  {author} {\bibfnamefont {G.~J.}\ \bibnamefont {Ye}}, \bibinfo {author}
  {\bibfnamefont {Q.}~\bibnamefont {Ge}}, \bibinfo {author} {\bibfnamefont
  {X.}~\bibnamefont {Ou}}, \bibinfo {author} {\bibfnamefont {H.}~\bibnamefont
  {Wu}}, \bibinfo {author} {\bibfnamefont {D.}~\bibnamefont {Feng}}, \bibinfo
  {author} {\bibfnamefont {X.~H.}\ \bibnamefont {Chen}}, \ and\ \bibinfo
  {author} {\bibfnamefont {Y.}~\bibnamefont {Zhang}},\ }\href@noop {}
  {\bibfield  {journal} {\bibinfo  {journal} {Nature Nanotechnol.}\ }\textbf
  {\bibinfo {volume} {9}},\ \bibinfo {pages} {372} (\bibinfo {year}
  {2014}{\natexlab{b}})}\BibitemShut {NoStop}%
\bibitem [{\citenamefont {Ling}\ \emph {et~al.}(2015)\citenamefont {Ling},
  \citenamefont {Wang}, \citenamefont {Huang}, \citenamefont {Xia},\ and\
  \citenamefont {Dresselhaus}}]{ling2015renaissance}%
  \BibitemOpen
  \bibfield  {author} {\bibinfo {author} {\bibfnamefont {X.}~\bibnamefont
  {Ling}}, \bibinfo {author} {\bibfnamefont {H.}~\bibnamefont {Wang}}, \bibinfo
  {author} {\bibfnamefont {S.}~\bibnamefont {Huang}}, \bibinfo {author}
  {\bibfnamefont {F.}~\bibnamefont {Xia}}, \ and\ \bibinfo {author}
  {\bibfnamefont {M.~S.}\ \bibnamefont {Dresselhaus}},\ }\href@noop {}
  {\bibfield  {journal} {\bibinfo  {journal} {P. Natl. A. Sci.}\ }\textbf
  {\bibinfo {volume} {112}},\ \bibinfo {pages} {4523} (\bibinfo {year}
  {2015})}\BibitemShut {NoStop}%
\bibitem [{\citenamefont {Mu}\ \emph {et~al.}(2019)\citenamefont {Mu},
  \citenamefont {Wang},\ and\ \citenamefont {Sun}}]{mu2019two}%
  \BibitemOpen
  \bibfield  {author} {\bibinfo {author} {\bibfnamefont {X.}~\bibnamefont
  {Mu}}, \bibinfo {author} {\bibfnamefont {J.}~\bibnamefont {Wang}}, \ and\
  \bibinfo {author} {\bibfnamefont {M.}~\bibnamefont {Sun}},\ }\href@noop {}
  {\bibfield  {journal} {\bibinfo  {journal} {Mater. Today. Phys.}\ }\textbf
  {\bibinfo {volume} {8}},\ \bibinfo {pages} {92} (\bibinfo {year}
  {2019})}\BibitemShut {NoStop}%
\bibitem [{\citenamefont {Dolui}\ and\ \citenamefont
  {Quek}(2015)}]{dolui2015quantum}%
  \BibitemOpen
  \bibfield  {author} {\bibinfo {author} {\bibfnamefont {K.}~\bibnamefont
  {Dolui}}\ and\ \bibinfo {author} {\bibfnamefont {S.~Y.}\ \bibnamefont
  {Quek}},\ }\href@noop {} {\bibfield  {journal} {\bibinfo  {journal} {Sci.
  Rep.}\ }\textbf {\bibinfo {volume} {5}},\ \bibinfo {pages} {11699} (\bibinfo
  {year} {2015})}\BibitemShut {NoStop}%
\bibitem [{\citenamefont {Liu}\ \emph {et~al.}(2017)\citenamefont {Liu},
  \citenamefont {Qiu}, \citenamefont {Carvalho}, \citenamefont {Bao},
  \citenamefont {Xu}, \citenamefont {Tan}, \citenamefont {Liu}, \citenamefont
  {Castro~Neto}, \citenamefont {Loh},\ and\ \citenamefont {Lu}}]{liu2017gate}%
  \BibitemOpen
  \bibfield  {author} {\bibinfo {author} {\bibfnamefont {Y.}~\bibnamefont
  {Liu}}, \bibinfo {author} {\bibfnamefont {Z.}~\bibnamefont {Qiu}}, \bibinfo
  {author} {\bibfnamefont {A.}~\bibnamefont {Carvalho}}, \bibinfo {author}
  {\bibfnamefont {Y.}~\bibnamefont {Bao}}, \bibinfo {author} {\bibfnamefont
  {H.}~\bibnamefont {Xu}}, \bibinfo {author} {\bibfnamefont {S.~J.}\
  \bibnamefont {Tan}}, \bibinfo {author} {\bibfnamefont {W.}~\bibnamefont
  {Liu}}, \bibinfo {author} {\bibfnamefont {A.}~\bibnamefont {Castro~Neto}},
  \bibinfo {author} {\bibfnamefont {K.~P.}\ \bibnamefont {Loh}}, \ and\
  \bibinfo {author} {\bibfnamefont {J.}~\bibnamefont {Lu}},\ }\href@noop {}
  {\bibfield  {journal} {\bibinfo  {journal} {Nano Lett.}\ }\textbf {\bibinfo
  {volume} {17}},\ \bibinfo {pages} {1970} (\bibinfo {year}
  {2017})}\BibitemShut {NoStop}%
\bibitem [{\citenamefont {Kim}\ \emph {et~al.}(2015)\citenamefont {Kim},
  \citenamefont {Baik}, \citenamefont {Ryu}, \citenamefont {Sohn},
  \citenamefont {Park}, \citenamefont {Park}, \citenamefont {Denlinger},
  \citenamefont {Yi}, \citenamefont {Choi},\ and\ \citenamefont
  {Kim}}]{kim2015observation}%
  \BibitemOpen
  \bibfield  {author} {\bibinfo {author} {\bibfnamefont {J.}~\bibnamefont
  {Kim}}, \bibinfo {author} {\bibfnamefont {S.~S.}\ \bibnamefont {Baik}},
  \bibinfo {author} {\bibfnamefont {S.~H.}\ \bibnamefont {Ryu}}, \bibinfo
  {author} {\bibfnamefont {Y.}~\bibnamefont {Sohn}}, \bibinfo {author}
  {\bibfnamefont {S.}~\bibnamefont {Park}}, \bibinfo {author} {\bibfnamefont
  {B.-G.}\ \bibnamefont {Park}}, \bibinfo {author} {\bibfnamefont
  {J.}~\bibnamefont {Denlinger}}, \bibinfo {author} {\bibfnamefont
  {Y.}~\bibnamefont {Yi}}, \bibinfo {author} {\bibfnamefont {H.~J.}\
  \bibnamefont {Choi}}, \ and\ \bibinfo {author} {\bibfnamefont {K.~S.}\
  \bibnamefont {Kim}},\ }\href@noop {} {\bibfield  {journal} {\bibinfo
  {journal} {Science}\ }\textbf {\bibinfo {volume} {349}},\ \bibinfo {pages}
  {723} (\bibinfo {year} {2015})}\BibitemShut {NoStop}%
\bibitem [{\citenamefont {Kim}\ \emph {et~al.}(2017)\citenamefont {Kim},
  \citenamefont {Jung}, \citenamefont {Kim}, \citenamefont {Choi},
  \citenamefont {Wei},\ and\ \citenamefont {Cho}}]{kim2017microscopic}%
  \BibitemOpen
  \bibfield  {author} {\bibinfo {author} {\bibfnamefont {S.-W.}\ \bibnamefont
  {Kim}}, \bibinfo {author} {\bibfnamefont {H.}~\bibnamefont {Jung}}, \bibinfo
  {author} {\bibfnamefont {H.-J.}\ \bibnamefont {Kim}}, \bibinfo {author}
  {\bibfnamefont {J.-H.}\ \bibnamefont {Choi}}, \bibinfo {author}
  {\bibfnamefont {S.-H.}\ \bibnamefont {Wei}}, \ and\ \bibinfo {author}
  {\bibfnamefont {J.-H.}\ \bibnamefont {Cho}},\ }\href@noop {} {\bibfield
  {journal} {\bibinfo  {journal} {Phys. Rev. B}\ }\textbf {\bibinfo {volume}
  {96}},\ \bibinfo {pages} {075416} (\bibinfo {year} {2017})}\BibitemShut
  {NoStop}%
\bibitem [{\citenamefont {Zhang}\ and\ \citenamefont
  {Yates~Jr}(2012)}]{zhang2012band}%
  \BibitemOpen
  \bibfield  {author} {\bibinfo {author} {\bibfnamefont {Z.}~\bibnamefont
  {Zhang}}\ and\ \bibinfo {author} {\bibfnamefont {J.~T.}\ \bibnamefont
  {Yates~Jr}},\ }\href@noop {} {\bibfield  {journal} {\bibinfo  {journal}
  {Chem. Rev.}\ }\textbf {\bibinfo {volume} {112}},\ \bibinfo {pages} {5520}
  (\bibinfo {year} {2012})}\BibitemShut {NoStop}%
\bibitem [{\citenamefont {Kiraly}\ \emph {et~al.}(2019)\citenamefont {Kiraly},
  \citenamefont {Knol}, \citenamefont {Volckaert}, \citenamefont {Biswas},
  \citenamefont {Rudenko}, \citenamefont {Prishchenko}, \citenamefont
  {Mazurenko}, \citenamefont {Katsnelson}, \citenamefont {Hofmann},
  \citenamefont {Wegner} \emph {et~al.}}]{kiraly2019anisotropic}%
  \BibitemOpen
  \bibfield  {author} {\bibinfo {author} {\bibfnamefont {B.}~\bibnamefont
  {Kiraly}}, \bibinfo {author} {\bibfnamefont {E.~J.}\ \bibnamefont {Knol}},
  \bibinfo {author} {\bibfnamefont {K.}~\bibnamefont {Volckaert}}, \bibinfo
  {author} {\bibfnamefont {D.}~\bibnamefont {Biswas}}, \bibinfo {author}
  {\bibfnamefont {A.~N.}\ \bibnamefont {Rudenko}}, \bibinfo {author}
  {\bibfnamefont {D.~A.}\ \bibnamefont {Prishchenko}}, \bibinfo {author}
  {\bibfnamefont {V.~G.}\ \bibnamefont {Mazurenko}}, \bibinfo {author}
  {\bibfnamefont {M.~I.}\ \bibnamefont {Katsnelson}}, \bibinfo {author}
  {\bibfnamefont {P.}~\bibnamefont {Hofmann}}, \bibinfo {author} {\bibfnamefont
  {D.}~\bibnamefont {Wegner}},  \emph {et~al.},\ }\href@noop {} {\bibfield
  {journal} {\bibinfo  {journal} {Phys. Rev. Lett.}\ }\textbf {\bibinfo
  {volume} {123}},\ \bibinfo {pages} {216403} (\bibinfo {year}
  {2019})}\BibitemShut {NoStop}%
\bibitem [{\citenamefont {Chen}\ \emph {et~al.}(2020)\citenamefont {Chen},
  \citenamefont {Dong}, \citenamefont {Giorgetti}, \citenamefont {Papalazarou},
  \citenamefont {Marsi}, \citenamefont {Zhang}, \citenamefont {Tian},
  \citenamefont {Ma}, \citenamefont {Cheng}, \citenamefont {Rueff} \emph
  {et~al.}}]{chen2020spectroscopy}%
  \BibitemOpen
  \bibfield  {author} {\bibinfo {author} {\bibfnamefont {Z.}~\bibnamefont
  {Chen}}, \bibinfo {author} {\bibfnamefont {J.}~\bibnamefont {Dong}}, \bibinfo
  {author} {\bibfnamefont {C.}~\bibnamefont {Giorgetti}}, \bibinfo {author}
  {\bibfnamefont {E.}~\bibnamefont {Papalazarou}}, \bibinfo {author}
  {\bibfnamefont {M.}~\bibnamefont {Marsi}}, \bibinfo {author} {\bibfnamefont
  {Z.}~\bibnamefont {Zhang}}, \bibinfo {author} {\bibfnamefont
  {B.}~\bibnamefont {Tian}}, \bibinfo {author} {\bibfnamefont {Q.}~\bibnamefont
  {Ma}}, \bibinfo {author} {\bibfnamefont {Y.}~\bibnamefont {Cheng}}, \bibinfo
  {author} {\bibfnamefont {J.-P.}\ \bibnamefont {Rueff}},  \emph {et~al.},\
  }\href@noop {} {\bibfield  {journal} {\bibinfo  {journal} {2D Mater.}\
  }\textbf {\bibinfo {volume} {7}},\ \bibinfo {pages} {035027} (\bibinfo {year}
  {2020})}\BibitemShut {NoStop}%
\bibitem [{\citenamefont {Hedayat}\ \emph {et~al.}(2021)\citenamefont
  {Hedayat}, \citenamefont {Ceraso}, \citenamefont {Soavi}, \citenamefont
  {Akhavan}, \citenamefont {Cadore}, \citenamefont {Dallera}, \citenamefont
  {Cerullo}, \citenamefont {Ferrari},\ and\ \citenamefont
  {Carpene}}]{hedayat2020non}%
  \BibitemOpen
  \bibfield  {author} {\bibinfo {author} {\bibfnamefont {H.}~\bibnamefont
  {Hedayat}}, \bibinfo {author} {\bibfnamefont {A.}~\bibnamefont {Ceraso}},
  \bibinfo {author} {\bibfnamefont {G.}~\bibnamefont {Soavi}}, \bibinfo
  {author} {\bibfnamefont {S.}~\bibnamefont {Akhavan}}, \bibinfo {author}
  {\bibfnamefont {A.}~\bibnamefont {Cadore}}, \bibinfo {author} {\bibfnamefont
  {C.}~\bibnamefont {Dallera}}, \bibinfo {author} {\bibfnamefont
  {G.}~\bibnamefont {Cerullo}}, \bibinfo {author} {\bibfnamefont {A.~C.}\
  \bibnamefont {Ferrari}}, \ and\ \bibinfo {author} {\bibfnamefont
  {E.}~\bibnamefont {Carpene}},\ }\href@noop {} {\bibfield  {journal} {\bibinfo
   {journal} {2D Mater.}\ }\textbf {\bibinfo {volume} {8}},\ \bibinfo {pages}
  {025020} (\bibinfo {year} {2021})}\BibitemShut {NoStop}%
\bibitem [{\citenamefont {Bovensiepen}\ and\ \citenamefont
  {Kirchmann}(2012)}]{bovensiepen2012elementary}%
  \BibitemOpen
  \bibfield  {author} {\bibinfo {author} {\bibfnamefont {U.}~\bibnamefont
  {Bovensiepen}}\ and\ \bibinfo {author} {\bibfnamefont {P.~S.}\ \bibnamefont
  {Kirchmann}},\ }\href@noop {} {\bibfield  {journal} {\bibinfo  {journal}
  {Laser Photonics Rev.}\ }\textbf {\bibinfo {volume} {6}},\ \bibinfo {pages}
  {589} (\bibinfo {year} {2012})}\BibitemShut {NoStop}%
\bibitem [{\citenamefont {Smallwood}\ \emph {et~al.}(2016)\citenamefont
  {Smallwood}, \citenamefont {Kaindl},\ and\ \citenamefont
  {Lanzara}}]{smallwood2016ultrafast}%
  \BibitemOpen
  \bibfield  {author} {\bibinfo {author} {\bibfnamefont {C.~L.}\ \bibnamefont
  {Smallwood}}, \bibinfo {author} {\bibfnamefont {R.~A.}\ \bibnamefont
  {Kaindl}}, \ and\ \bibinfo {author} {\bibfnamefont {A.}~\bibnamefont
  {Lanzara}},\ }\href@noop {} {\bibfield  {journal} {\bibinfo  {journal}
  {Europhys. Lett.}\ }\textbf {\bibinfo {volume} {115}},\ \bibinfo {pages}
  {27001} (\bibinfo {year} {2016})}\BibitemShut {NoStop}%
\bibitem [{\citenamefont {Widdra}\ \emph {et~al.}(2003)\citenamefont {Widdra},
  \citenamefont {Br{\"o}cker}, \citenamefont {Gie{\ss}el}, \citenamefont
  {Hertel}, \citenamefont {Kr{\"u}ger}, \citenamefont {Liero}, \citenamefont
  {Noack}, \citenamefont {Petrov}, \citenamefont {Pop}, \citenamefont {Schmidt}
  \emph {et~al.}}]{widdra2003time}%
  \BibitemOpen
  \bibfield  {author} {\bibinfo {author} {\bibfnamefont {W.}~\bibnamefont
  {Widdra}}, \bibinfo {author} {\bibfnamefont {D.}~\bibnamefont {Br{\"o}cker}},
  \bibinfo {author} {\bibfnamefont {T.}~\bibnamefont {Gie{\ss}el}}, \bibinfo
  {author} {\bibfnamefont {I.}~\bibnamefont {Hertel}}, \bibinfo {author}
  {\bibfnamefont {W.}~\bibnamefont {Kr{\"u}ger}}, \bibinfo {author}
  {\bibfnamefont {A.}~\bibnamefont {Liero}}, \bibinfo {author} {\bibfnamefont
  {F.}~\bibnamefont {Noack}}, \bibinfo {author} {\bibfnamefont
  {V.}~\bibnamefont {Petrov}}, \bibinfo {author} {\bibfnamefont
  {D.}~\bibnamefont {Pop}}, \bibinfo {author} {\bibfnamefont {P.}~\bibnamefont
  {Schmidt}},  \emph {et~al.},\ }\href@noop {} {\bibfield  {journal} {\bibinfo
  {journal} {Surf. Sci.}\ }\textbf {\bibinfo {volume} {543}},\ \bibinfo {pages}
  {87} (\bibinfo {year} {2003})}\BibitemShut {NoStop}%
\bibitem [{\citenamefont {Tokudomi}\ \emph {et~al.}(2008)\citenamefont
  {Tokudomi}, \citenamefont {Azuma}, \citenamefont {Takahashi},\ and\
  \citenamefont {Kamada}}]{tokudomi2008ultrafast}%
  \BibitemOpen
  \bibfield  {author} {\bibinfo {author} {\bibfnamefont {S.}~\bibnamefont
  {Tokudomi}}, \bibinfo {author} {\bibfnamefont {J.}~\bibnamefont {Azuma}},
  \bibinfo {author} {\bibfnamefont {K.}~\bibnamefont {Takahashi}}, \ and\
  \bibinfo {author} {\bibfnamefont {M.}~\bibnamefont {Kamada}},\ }\href@noop {}
  {\bibfield  {journal} {\bibinfo  {journal} {J. Phys. Soc. Jpn.}\ }\textbf
  {\bibinfo {volume} {77}},\ \bibinfo {pages} {014711} (\bibinfo {year}
  {2008})}\BibitemShut {NoStop}%
\bibitem [{\citenamefont {Tanaka}(2012)}]{tanaka2012utility}%
  \BibitemOpen
  \bibfield  {author} {\bibinfo {author} {\bibfnamefont {S.-i.}\ \bibnamefont
  {Tanaka}},\ }\href@noop {} {\bibfield  {journal} {\bibinfo  {journal} {J.
  Electron Spectrosc.}\ }\textbf {\bibinfo {volume} {185}},\ \bibinfo {pages}
  {152} (\bibinfo {year} {2012})}\BibitemShut {NoStop}%
\bibitem [{\citenamefont {Yang}\ \emph {et~al.}(2014)\citenamefont {Yang},
  \citenamefont {Sobota}, \citenamefont {Kirchmann},\ and\ \citenamefont
  {Shen}}]{yang2014electron}%
  \BibitemOpen
  \bibfield  {author} {\bibinfo {author} {\bibfnamefont {S.-L.}\ \bibnamefont
  {Yang}}, \bibinfo {author} {\bibfnamefont {J.~A.}\ \bibnamefont {Sobota}},
  \bibinfo {author} {\bibfnamefont {P.~S.}\ \bibnamefont {Kirchmann}}, \ and\
  \bibinfo {author} {\bibfnamefont {Z.-X.}\ \bibnamefont {Shen}},\ }\href@noop
  {} {\bibfield  {journal} {\bibinfo  {journal} {Appl. Phys. A}\ }\textbf
  {\bibinfo {volume} {116}},\ \bibinfo {pages} {85} (\bibinfo {year}
  {2014})}\BibitemShut {NoStop}%
\bibitem [{\citenamefont {Lange}\ \emph {et~al.}(2007)\citenamefont {Lange},
  \citenamefont {Schmidt},\ and\ \citenamefont {Nilges}}]{lange2007au3snp7}%
  \BibitemOpen
  \bibfield  {author} {\bibinfo {author} {\bibfnamefont {S.}~\bibnamefont
  {Lange}}, \bibinfo {author} {\bibfnamefont {P.}~\bibnamefont {Schmidt}}, \
  and\ \bibinfo {author} {\bibfnamefont {T.}~\bibnamefont {Nilges}},\
  }\href@noop {} {\bibfield  {journal} {\bibinfo  {journal} {Inorg. Chem.}\
  }\textbf {\bibinfo {volume} {46}},\ \bibinfo {pages} {4028} (\bibinfo {year}
  {2007})}\BibitemShut {NoStop}%
\bibitem [{\citenamefont {Faure}\ \emph {et~al.}(2012)\citenamefont {Faure},
  \citenamefont {Mauchain}, \citenamefont {Papalazarou}, \citenamefont {Yan},
  \citenamefont {Pinon}, \citenamefont {Marsi},\ and\ \citenamefont
  {Perfetti}}]{faure2012full}%
  \BibitemOpen
  \bibfield  {author} {\bibinfo {author} {\bibfnamefont {J.}~\bibnamefont
  {Faure}}, \bibinfo {author} {\bibfnamefont {J.}~\bibnamefont {Mauchain}},
  \bibinfo {author} {\bibfnamefont {E.}~\bibnamefont {Papalazarou}}, \bibinfo
  {author} {\bibfnamefont {W.}~\bibnamefont {Yan}}, \bibinfo {author}
  {\bibfnamefont {J.}~\bibnamefont {Pinon}}, \bibinfo {author} {\bibfnamefont
  {M.}~\bibnamefont {Marsi}}, \ and\ \bibinfo {author} {\bibfnamefont
  {L.}~\bibnamefont {Perfetti}},\ }\href@noop {} {\bibfield  {journal}
  {\bibinfo  {journal} {Rev. Sci. Instrum.}\ }\textbf {\bibinfo {volume}
  {83}},\ \bibinfo {pages} {043109} (\bibinfo {year} {2012})}\BibitemShut
  {NoStop}%
\bibitem [{\citenamefont {Kresse}\ and\ \citenamefont
  {Hafner}(1993)}]{kresse1993}%
  \BibitemOpen
  \bibfield  {author} {\bibinfo {author} {\bibfnamefont {G.}~\bibnamefont
  {Kresse}}\ and\ \bibinfo {author} {\bibfnamefont {J.}~\bibnamefont
  {Hafner}},\ }\href@noop {} {\bibfield  {journal} {\bibinfo  {journal} {Phys.
  Rev. B}\ }\textbf {\bibinfo {volume} {47}},\ \bibinfo {pages} {558} (\bibinfo
  {year} {1993})}\BibitemShut {NoStop}%
\bibitem [{\citenamefont {Kresse}\ and\ \citenamefont
  {Hafner}(1994)}]{kresse1994}%
  \BibitemOpen
  \bibfield  {author} {\bibinfo {author} {\bibfnamefont {G.}~\bibnamefont
  {Kresse}}\ and\ \bibinfo {author} {\bibfnamefont {J.}~\bibnamefont
  {Hafner}},\ }\href@noop {} {\bibfield  {journal} {\bibinfo  {journal} {Phys.
  Rev. B}\ }\textbf {\bibinfo {volume} {49}},\ \bibinfo {pages} {14251}
  (\bibinfo {year} {1994})}\BibitemShut {NoStop}%
\bibitem [{\citenamefont {Kresse}\ and\ \citenamefont
  {Furthm\"uller}(1996{\natexlab{a}})}]{kresse1996a}%
  \BibitemOpen
  \bibfield  {author} {\bibinfo {author} {\bibfnamefont {G.}~\bibnamefont
  {Kresse}}\ and\ \bibinfo {author} {\bibfnamefont {J.}~\bibnamefont
  {Furthm\"uller}},\ }\href@noop {} {\bibfield  {journal} {\bibinfo  {journal}
  {Comput. Mater. Sci.}\ }\textbf {\bibinfo {volume} {6}},\ \bibinfo {pages}
  {15 } (\bibinfo {year} {1996}{\natexlab{a}})}\BibitemShut {NoStop}%
\bibitem [{\citenamefont {Kresse}\ and\ \citenamefont
  {Furthm\"uller}(1996{\natexlab{b}})}]{kresse1996b}%
  \BibitemOpen
  \bibfield  {author} {\bibinfo {author} {\bibfnamefont {G.}~\bibnamefont
  {Kresse}}\ and\ \bibinfo {author} {\bibfnamefont {J.}~\bibnamefont
  {Furthm\"uller}},\ }\href@noop {} {\bibfield  {journal} {\bibinfo  {journal}
  {Phys. Rev. B}\ }\textbf {\bibinfo {volume} {54}},\ \bibinfo {pages} {11169}
  (\bibinfo {year} {1996}{\natexlab{b}})}\BibitemShut {NoStop}%
\bibitem [{\citenamefont {Bl\"ochl}(1994)}]{blochl1994}%
  \BibitemOpen
  \bibfield  {author} {\bibinfo {author} {\bibfnamefont {P.~E.}\ \bibnamefont
  {Bl\"ochl}},\ }\href@noop {} {\bibfield  {journal} {\bibinfo  {journal}
  {Phys. Rev. B}\ }\textbf {\bibinfo {volume} {50}},\ \bibinfo {pages} {17953}
  (\bibinfo {year} {1994})}\BibitemShut {NoStop}%
\bibitem [{\citenamefont {Kresse}\ and\ \citenamefont
  {Joubert}(1999)}]{kresse1999}%
  \BibitemOpen
  \bibfield  {author} {\bibinfo {author} {\bibfnamefont {G.}~\bibnamefont
  {Kresse}}\ and\ \bibinfo {author} {\bibfnamefont {D.}~\bibnamefont
  {Joubert}},\ }\href@noop {} {\bibfield  {journal} {\bibinfo  {journal} {Phys.
  Rev. B}\ }\textbf {\bibinfo {volume} {59}},\ \bibinfo {pages} {1758}
  (\bibinfo {year} {1999})}\BibitemShut {NoStop}%
\bibitem [{\citenamefont {Sun}\ \emph {et~al.}(2015)\citenamefont {Sun},
  \citenamefont {Ruzsinszky},\ and\ \citenamefont {Perdew}}]{sun2015}%
  \BibitemOpen
  \bibfield  {author} {\bibinfo {author} {\bibfnamefont {J.}~\bibnamefont
  {Sun}}, \bibinfo {author} {\bibfnamefont {A.}~\bibnamefont {Ruzsinszky}}, \
  and\ \bibinfo {author} {\bibfnamefont {J.~P.}\ \bibnamefont {Perdew}},\
  }\href@noop {} {\bibfield  {journal} {\bibinfo  {journal} {Phys. Rev. Lett.}\
  }\textbf {\bibinfo {volume} {115}},\ \bibinfo {pages} {036402} (\bibinfo
  {year} {2015})}\BibitemShut {NoStop}%
\bibitem [{\citenamefont {Brown}\ and\ \citenamefont
  {Rundqvist}(1965)}]{brown1965refinement}%
  \BibitemOpen
  \bibfield  {author} {\bibinfo {author} {\bibfnamefont {A.}~\bibnamefont
  {Brown}}\ and\ \bibinfo {author} {\bibfnamefont {S.}~\bibnamefont
  {Rundqvist}},\ }\href@noop {} {\bibfield  {journal} {\bibinfo  {journal}
  {Acta Crystallogr.}\ }\textbf {\bibinfo {volume} {19}},\ \bibinfo {pages}
  {684} (\bibinfo {year} {1965})}\BibitemShut {NoStop}%
\bibitem [{\citenamefont {Momma}\ and\ \citenamefont
  {Izumi}(2011)}]{momma2011vesta}%
  \BibitemOpen
  \bibfield  {author} {\bibinfo {author} {\bibfnamefont {K.}~\bibnamefont
  {Momma}}\ and\ \bibinfo {author} {\bibfnamefont {F.}~\bibnamefont {Izumi}},\
  }\href@noop {} {\bibfield  {journal} {\bibinfo  {journal} {J. Appl.
  Crystallogr.}\ }\textbf {\bibinfo {volume} {44}},\ \bibinfo {pages} {1272}
  (\bibinfo {year} {2011})}\BibitemShut {NoStop}%
\bibitem [{\citenamefont {Takahashi}\ \emph {et~al.}(1986)\citenamefont
  {Takahashi}, \citenamefont {Gunasekara}, \citenamefont {Ohsawa},
  \citenamefont {Ishii}, \citenamefont {Kinoshita}, \citenamefont {Suzuki},
  \citenamefont {Sagawa}, \citenamefont {Kato}, \citenamefont {Miyahara},\ and\
  \citenamefont {Shirotani}}]{taka1986}%
  \BibitemOpen
  \bibfield  {author} {\bibinfo {author} {\bibfnamefont {T.}~\bibnamefont
  {Takahashi}}, \bibinfo {author} {\bibfnamefont {N.}~\bibnamefont
  {Gunasekara}}, \bibinfo {author} {\bibfnamefont {H.}~\bibnamefont {Ohsawa}},
  \bibinfo {author} {\bibfnamefont {H.}~\bibnamefont {Ishii}}, \bibinfo
  {author} {\bibfnamefont {T.}~\bibnamefont {Kinoshita}}, \bibinfo {author}
  {\bibfnamefont {S.}~\bibnamefont {Suzuki}}, \bibinfo {author} {\bibfnamefont
  {T.}~\bibnamefont {Sagawa}}, \bibinfo {author} {\bibfnamefont
  {H.}~\bibnamefont {Kato}}, \bibinfo {author} {\bibfnamefont {T.}~\bibnamefont
  {Miyahara}}, \ and\ \bibinfo {author} {\bibfnamefont {I.}~\bibnamefont
  {Shirotani}},\ }\href@noop {} {\bibfield  {journal} {\bibinfo  {journal}
  {Phys. Rev. B}\ }\textbf {\bibinfo {volume} {33}},\ \bibinfo {pages} {4324}
  (\bibinfo {year} {1986})}\BibitemShut {NoStop}%
\bibitem [{\citenamefont {Golias}\ \emph {et~al.}(2016)\citenamefont {Golias},
  \citenamefont {Krivenkov},\ and\ \citenamefont
  {S{\'a}nchez-Barriga}}]{golias2016disentangling}%
  \BibitemOpen
  \bibfield  {author} {\bibinfo {author} {\bibfnamefont {E.}~\bibnamefont
  {Golias}}, \bibinfo {author} {\bibfnamefont {M.}~\bibnamefont {Krivenkov}}, \
  and\ \bibinfo {author} {\bibfnamefont {J.}~\bibnamefont
  {S{\'a}nchez-Barriga}},\ }\href@noop {} {\bibfield  {journal} {\bibinfo
  {journal} {Phys. Rev. B}\ }\textbf {\bibinfo {volume} {93}},\ \bibinfo
  {pages} {075207} (\bibinfo {year} {2016})}\BibitemShut {NoStop}%
\bibitem [{\citenamefont {Kiraly}\ \emph {et~al.}(2017)\citenamefont {Kiraly},
  \citenamefont {Hauptmann}, \citenamefont {Rudenko}, \citenamefont
  {Katsnelson},\ and\ \citenamefont {Khajetoorians}}]{kiraly2017probing}%
  \BibitemOpen
  \bibfield  {author} {\bibinfo {author} {\bibfnamefont {B.}~\bibnamefont
  {Kiraly}}, \bibinfo {author} {\bibfnamefont {N.}~\bibnamefont {Hauptmann}},
  \bibinfo {author} {\bibfnamefont {A.~N.}\ \bibnamefont {Rudenko}}, \bibinfo
  {author} {\bibfnamefont {M.~I.}\ \bibnamefont {Katsnelson}}, \ and\ \bibinfo
  {author} {\bibfnamefont {A.~A.}\ \bibnamefont {Khajetoorians}},\ }\href@noop
  {} {\bibfield  {journal} {\bibinfo  {journal} {Nano Lett.}\ }\textbf
  {\bibinfo {volume} {17}},\ \bibinfo {pages} {3607} (\bibinfo {year}
  {2017})}\BibitemShut {NoStop}%
\bibitem [{\citenamefont {Qiu}\ \emph {et~al.}(2017)\citenamefont {Qiu},
  \citenamefont {Fang}, \citenamefont {Carvalho}, \citenamefont {Rodin},
  \citenamefont {Liu}, \citenamefont {Tan}, \citenamefont {Telychko},
  \citenamefont {Lv}, \citenamefont {Su}, \citenamefont {Wang} \emph
  {et~al.}}]{qiu2017resolving}%
  \BibitemOpen
  \bibfield  {author} {\bibinfo {author} {\bibfnamefont {Z.}~\bibnamefont
  {Qiu}}, \bibinfo {author} {\bibfnamefont {H.}~\bibnamefont {Fang}}, \bibinfo
  {author} {\bibfnamefont {A.}~\bibnamefont {Carvalho}}, \bibinfo {author}
  {\bibfnamefont {A.}~\bibnamefont {Rodin}}, \bibinfo {author} {\bibfnamefont
  {Y.}~\bibnamefont {Liu}}, \bibinfo {author} {\bibfnamefont {S.~J.}\
  \bibnamefont {Tan}}, \bibinfo {author} {\bibfnamefont {M.}~\bibnamefont
  {Telychko}}, \bibinfo {author} {\bibfnamefont {P.}~\bibnamefont {Lv}},
  \bibinfo {author} {\bibfnamefont {J.}~\bibnamefont {Su}}, \bibinfo {author}
  {\bibfnamefont {Y.}~\bibnamefont {Wang}},  \emph {et~al.},\ }\href@noop {}
  {\bibfield  {journal} {\bibinfo  {journal} {Nano Lett.}\ }\textbf {\bibinfo
  {volume} {17}},\ \bibinfo {pages} {6935} (\bibinfo {year}
  {2017})}\BibitemShut {NoStop}%
\bibitem [{\citenamefont {Papalazarou}\ \emph {et~al.}(2018)\citenamefont
  {Papalazarou}, \citenamefont {Khalil}, \citenamefont {Caputo}, \citenamefont
  {Perfetti}, \citenamefont {Nilforoushan}, \citenamefont {Deng}, \citenamefont
  {Chen}, \citenamefont {Zhao}, \citenamefont {Taleb-Ibrahimi}, \citenamefont
  {Konczykowski} \emph {et~al.}}]{papalazarou2018unraveling}%
  \BibitemOpen
  \bibfield  {author} {\bibinfo {author} {\bibfnamefont {E.}~\bibnamefont
  {Papalazarou}}, \bibinfo {author} {\bibfnamefont {L.}~\bibnamefont {Khalil}},
  \bibinfo {author} {\bibfnamefont {M.}~\bibnamefont {Caputo}}, \bibinfo
  {author} {\bibfnamefont {L.}~\bibnamefont {Perfetti}}, \bibinfo {author}
  {\bibfnamefont {N.}~\bibnamefont {Nilforoushan}}, \bibinfo {author}
  {\bibfnamefont {H.}~\bibnamefont {Deng}}, \bibinfo {author} {\bibfnamefont
  {Z.}~\bibnamefont {Chen}}, \bibinfo {author} {\bibfnamefont {S.}~\bibnamefont
  {Zhao}}, \bibinfo {author} {\bibfnamefont {A.}~\bibnamefont
  {Taleb-Ibrahimi}}, \bibinfo {author} {\bibfnamefont {M.}~\bibnamefont
  {Konczykowski}},  \emph {et~al.},\ }\href@noop {} {\bibfield  {journal}
  {\bibinfo  {journal} {Phys. Rev. Mat.}\ }\textbf {\bibinfo {volume} {2}},\
  \bibinfo {pages} {104202} (\bibinfo {year} {2018})}\BibitemShut {NoStop}%
\bibitem [{\citenamefont {Chen}\ \emph {et~al.}(2018)\citenamefont {Chen},
  \citenamefont {Dong}, \citenamefont {Papalazarou}, \citenamefont {Marsi},
  \citenamefont {Giorgetti}, \citenamefont {Zhang}, \citenamefont {Tian},
  \citenamefont {Rueff}, \citenamefont {Taleb-Ibrahimi},\ and\ \citenamefont
  {Perfetti}}]{chen2018band}%
  \BibitemOpen
  \bibfield  {author} {\bibinfo {author} {\bibfnamefont {Z.}~\bibnamefont
  {Chen}}, \bibinfo {author} {\bibfnamefont {J.}~\bibnamefont {Dong}}, \bibinfo
  {author} {\bibfnamefont {E.}~\bibnamefont {Papalazarou}}, \bibinfo {author}
  {\bibfnamefont {M.}~\bibnamefont {Marsi}}, \bibinfo {author} {\bibfnamefont
  {C.}~\bibnamefont {Giorgetti}}, \bibinfo {author} {\bibfnamefont
  {Z.}~\bibnamefont {Zhang}}, \bibinfo {author} {\bibfnamefont
  {B.}~\bibnamefont {Tian}}, \bibinfo {author} {\bibfnamefont {J.-P.}\
  \bibnamefont {Rueff}}, \bibinfo {author} {\bibfnamefont {A.}~\bibnamefont
  {Taleb-Ibrahimi}}, \ and\ \bibinfo {author} {\bibfnamefont {L.}~\bibnamefont
  {Perfetti}},\ }\href@noop {} {\bibfield  {journal} {\bibinfo  {journal} {Nano
  Lett.}\ }\textbf {\bibinfo {volume} {19}},\ \bibinfo {pages} {488} (\bibinfo
  {year} {2018})}\BibitemShut {NoStop}%
\bibitem [{\citenamefont {Seah}\ and\ \citenamefont
  {Dench}(1979)}]{seah1979quantitative}%
  \BibitemOpen
  \bibfield  {author} {\bibinfo {author} {\bibfnamefont {M.~P.}\ \bibnamefont
  {Seah}}\ and\ \bibinfo {author} {\bibfnamefont {W.}~\bibnamefont {Dench}},\
  }\href@noop {} {\bibfield  {journal} {\bibinfo  {journal} {Surf. Interface
  Anal.}\ }\textbf {\bibinfo {volume} {1}},\ \bibinfo {pages} {2} (\bibinfo
  {year} {1979})}\BibitemShut {NoStop}%
\bibitem [{\citenamefont {Sobota}\ \emph {et~al.}(2012)\citenamefont {Sobota},
  \citenamefont {Yang}, \citenamefont {Analytis}, \citenamefont {Chen},
  \citenamefont {Fisher}, \citenamefont {Kirchmann},\ and\ \citenamefont
  {Shen}}]{sobota2012ultrafast}%
  \BibitemOpen
  \bibfield  {author} {\bibinfo {author} {\bibfnamefont {J.~A.}\ \bibnamefont
  {Sobota}}, \bibinfo {author} {\bibfnamefont {S.}~\bibnamefont {Yang}},
  \bibinfo {author} {\bibfnamefont {J.~G.}\ \bibnamefont {Analytis}}, \bibinfo
  {author} {\bibfnamefont {Y.~L.}\ \bibnamefont {Chen}}, \bibinfo {author}
  {\bibfnamefont {I.~R.}\ \bibnamefont {Fisher}}, \bibinfo {author}
  {\bibfnamefont {P.~S.}\ \bibnamefont {Kirchmann}}, \ and\ \bibinfo {author}
  {\bibfnamefont {Z.-X.}\ \bibnamefont {Shen}},\ }\href@noop {} {\bibfield
  {journal} {\bibinfo  {journal} {Phys. Rev. Lett.}\ }\textbf {\bibinfo
  {volume} {108}},\ \bibinfo {pages} {117403} (\bibinfo {year}
  {2012})}\BibitemShut {NoStop}%
\bibitem [{\citenamefont {Li}(2016)}]{Semi_cond_elect_book}%
  \BibitemOpen
  \bibfield  {author} {\bibinfo {author} {\bibfnamefont {S.}~\bibnamefont
  {Li}},\ }\href@noop {} {\emph {\bibinfo {title} {Semiconductor Physical
  Electronics}}}\ (\bibinfo  {publisher} {Springer-Verlag New York},\ \bibinfo
  {year} {2016})\BibitemShut {NoStop}%
\bibitem [{Note1()}]{Note1}%
  \BibitemOpen
  \bibinfo {note} {We caution here that the sub-fs timescale is obtained by
  assuming that the experimental mobilities are still valid for such a very
  high electric field. In practice, this might not be the case, resulting
  potentially a longer rise-time. As long as it remains shorter than the pump
  duration, our conclusions are not qualitatively affected.}\BibitemShut
  {Stop}%
\bibitem [{\citenamefont {Bhaskar}\ \emph {et~al.}(2016)\citenamefont
  {Bhaskar}, \citenamefont {Achtstein}, \citenamefont {Vermeulen},\ and\
  \citenamefont {Siebbeles}}]{Rad_Recom_BP_2016}%
  \BibitemOpen
  \bibfield  {author} {\bibinfo {author} {\bibfnamefont {P.}~\bibnamefont
  {Bhaskar}}, \bibinfo {author} {\bibfnamefont {A.~W.}\ \bibnamefont
  {Achtstein}}, \bibinfo {author} {\bibfnamefont {M.~J.~W.}\ \bibnamefont
  {Vermeulen}}, \ and\ \bibinfo {author} {\bibfnamefont {L.~D.~A.}\
  \bibnamefont {Siebbeles}},\ }\href@noop {} {\bibfield  {journal} {\bibinfo
  {journal} {J. Phys. Chem. C}\ }\textbf {\bibinfo {volume} {120}},\ \bibinfo
  {pages} {13836} (\bibinfo {year} {2016})}\BibitemShut {NoStop}%
\bibitem [{\citenamefont {Akahama}\ \emph {et~al.}(1983)\citenamefont
  {Akahama}, \citenamefont {Endo},\ and\ \citenamefont {Narita}}]{JPSJ_1983}%
  \BibitemOpen
  \bibfield  {author} {\bibinfo {author} {\bibfnamefont {Y.}~\bibnamefont
  {Akahama}}, \bibinfo {author} {\bibfnamefont {S.}~\bibnamefont {Endo}}, \
  and\ \bibinfo {author} {\bibfnamefont {S.-i.}\ \bibnamefont {Narita}},\
  }\href@noop {} {\bibfield  {journal} {\bibinfo  {journal} {JPSJ}\ }\textbf
  {\bibinfo {volume} {52}},\ \bibinfo {pages} {2148} (\bibinfo {year}
  {1983})}\BibitemShut {NoStop}%
\bibitem [{\citenamefont {Davies}(1997)}]{LowD_Semi}%
  \BibitemOpen
  \bibfield  {author} {\bibinfo {author} {\bibfnamefont {J.~H.}\ \bibnamefont
  {Davies}},\ }\href@noop {} {\emph {\bibinfo {title} {The Physics of
  Low-Dimensional Semiconductors}}}\ (\bibinfo  {publisher} {Cambridge
  University Press},\ \bibinfo {year} {1997})\BibitemShut {NoStop}%
\bibitem [{\citenamefont {Press}\ \emph {et~al.}(2007)\citenamefont {Press},
  \citenamefont {Teukolsky}, \citenamefont {Vetterling},\ and\ \citenamefont
  {Flannery}}]{Num_Recipe}%
  \BibitemOpen
  \bibfield  {author} {\bibinfo {author} {\bibfnamefont {W.~H.}\ \bibnamefont
  {Press}}, \bibinfo {author} {\bibfnamefont {S.~A.}\ \bibnamefont
  {Teukolsky}}, \bibinfo {author} {\bibfnamefont {W.~T.}\ \bibnamefont
  {Vetterling}}, \ and\ \bibinfo {author} {\bibfnamefont {B.~P.}\ \bibnamefont
  {Flannery}},\ }\href@noop {} {\emph {\bibinfo {title} {NUMERICAL RECIPES, The
  Art of Scientific Computing}}},\ \bibinfo {edition} {third edition}\ ed.\
  (\bibinfo  {publisher} {Cambridge University Press},\ \bibinfo {year}
  {2007})\BibitemShut {NoStop}%
\end{thebibliography}%
\end{document}